\documentclass[onecolumn]{aastex6}
\usepackage{amsmath,amssymb}
\usepackage{graphicx,mathptm}
\usepackage{times}
\usepackage{natbib}
\usepackage{longtable}
\usepackage{multirow}
\usepackage{rotating}
\usepackage{color}
\usepackage{epstopdf}
\usepackage{bm}
\usepackage{indentfirst}
\usepackage{booktabs}

\newcommand{\be}{\begin{equation}}
\newcommand{\ee}{\end{equation}}
\newcommand{\bea}{\begin{eqnarray}}
\newcommand{\eea}{\end{eqnarray}}
\newcommand {\apgt} {\ {\raise-.5ex\hbox{$\buildrel>\over\sim$}}\ }
\newcommand {\aplt} {\ {\raise-.5ex\hbox{$\buildrel<\over\sim$}}\ }
\usepackage{threeparttable}
\usepackage{subfigure}
\usepackage{mathrsfs}
\usepackage{url}
\begin{document}
\title{The influence of magnetic fields on absorption and emission spectroscopies}
\author{Heshou Zhang\altaffilmark{1,2} , Huirong Yan\altaffilmark{1,2} \& Philipp Richter\altaffilmark{2,3}}
\altaffiltext{1}{Deutsches Elektronen-Synchrotron DESY, Platanenallee 6, D-15738 Zeuthen, Germany;}
\altaffiltext{2}{Institut für Physik und Astronomie, Universität Potsdam, Haus 28, Karl-Liebknecht-Str. 24/25, D-14476 Potsdam, Germany;}
\altaffiltext{3}{Leibniz-Institut für Astrophysik Potsdam(AIP), An der Sternwarte 16, D-14482 Potsdam, Germany;}
\begin{abstract}
Spectroscopic observations play a fundamental role in astrophysics. They are crutial to determine important physical parameters, provide information about the composition of various objects in the universe, as well as depict motions in the universe. However, spectroscopic studies often do not consider the influence of magnetic fields. In this paper, we explore the influence of magnetic fields on the spectroscopic observations using the concept of atomic alignment. Synthetic spectra are generated to show the measurable changes of the spectra due to atomic alignment. The influences of atomic alignment on absorption from DLAs, emission from H\,{\sc ii} Regions, submillimeter fine-structure lines from star forming regions are presented as examples to show this effect in diffuse gas. Furthermore, we demonstrate the influence of atomic alignment on physical parameters derived from atomic line ratios, such as the alpha-to-iron ratio([X/Fe]), interstellar temperature, and ionization rate. We conclude that Ground State Alignment (GSA) should be taken into consideration in the error budget of spectroscopic studies with high signal-to-noise(S/N) ratio.
\end{abstract}
\keywords
{ISM: magnetic fields--spectroscopy--(stars:) circumstellar matter--Galaxy}
\section{Introduction}

Atomic spectra play a crucial role in studying the universe. Atomic absorption spectroscopy proves to be a strong tool for the analysis of chemical abundances (e.g., \citealt{2013ApJ...772..110F,2013ApJ...772..111R,2012PASP..124..566W,2015ApJ...800...14M}) and chemical evolution of circumburst medium of GRB (e.g., \citealt{2006A&A...451L..47F}); atomic emission spectra help the modeling of stellar atmosphere (e.g., \citealt{2001ApJS..132..403P}); FIR submillimeter fine structure lines provide abundant information in star forming regions (e.g., \citealt{1988ApJ...332..379S,1996ApJ...471..400C}). However, astrophysical magnetic fields, which exist everywhere in the universe and play crucial roles in many astrophysical processes, are not often considered in previous spectroscopic studies. As demonstrated in \citet{YLfine,YLhyf,YLhanle}, magnetic fields redistribute the angular momenta of atoms of diffuse medium among different sublevels on the ground state (known as Ground State Alignment, hereby GSA), and hence, influence the polarization of emission and absorption from the medium.  In this paper, we will show that due to GSA, magnetic fields should be taken into consideration into the error budget when analysing astrophysical spectroscopies.

The basic for GSA is explicit. Alignment refers to the orientation of the angular momentum of atoms. Occupation of the atoms on different sublevels of the ground state will alter when they interact with anisotropic radiation. With the existence of magnetic fields, the angular momenta of the atoms will redistribute among the sublevels on the ground state due to fast magnetic precession, which is addressed as magnetic realignment (see \citealt{2012JQSRT.113.1409Y,2013arXiv1302.3264Y} for details). GSA has been a well-developed theory. Optical pumping was firstly proposed by \citet{KASTLER-1950-234250}, and then the atomic alignment in the presence of magnetic fields was studied in laboratory \citep[see][]{Hawkins:1955fv}. Toy models of this effect were proposed by \citet{Varshalovich:1968qc,Varshalovich:1971mw}. Then, emission for atoms with idealized fine structure given a particular geometry of magnetic fields and light beam was discussed in \citet{Landolfi:1986lh}.

Calculations for GSA with fine and hyperfine structure in astrophysical environment were provided in \citet{YLfine,YLhyf,YLhanle}. They demonstrated an exclusive feature of GSA that it reveals the 3D direction of magnetic fields. \citet{ZYD15}  proved that GSA is a powerful magnetic tracer in general radiation fields.

Despite the fact that atomic alignment is a good magnetic diagnostic, magnetic fields have a notable influence on atomic spectra as a result of GSA. As illustrated in \citet{2012JQSRT.113.1409Y}, GSA will be effective when $\nu_L$ (Larmor precession rate) $>\tau^{-1}_R$ (radiative pumping rate) $>\tau^{-1}_c$ (collisional transition rate), which means the magnetic field can be up to 1 Gauss and approximately down to the scale smaller than micro Gauss even when the diffuse medium is quite close to the pumping source (the distance with the scale of several Au). Even longer distance from the pumping source to the diffuse medium allow even smaller magnetic fields to align the medium (down to $10^{-15}$ Gauss, see \citealt{2012JQSRT.113.1409Y}). Since the main magnetic field in the galaxy spiral arm has the scale of $10^{-5}$ Gauss, spectra observed from the diffuse medium in the galaxy, e.g. DLAs, are safely within the GSA range. As long as the strength of magnetic field is within the range of GSA, the only thing that makes the magnetic fields alter the spectrum is the change of atomic angular momentum due to atomic alignment.

In this paper, the basic formulae of GSA are presented in \S 2 to demonstrate the influence of magnetic fields on atomic lines. In \S 3, we use several examples of diffuse gas in different astrophysical environment to illustrate the influence of GSA effect on different types of atomic spectra. Synthetic spectra for absorption lines are generated according to a scenario of DLAs in galaxy. The influence of GSA on different emission lines are compared according to a scenario of extragalactic H\,{\sc ii} Regions. Results on submillimeter fine-structure lines are presented in Star Forming Regions. In \S 4, the influences of GSA on different astrophysical properties derived from atomic line ratios are demonstrated in different subsections. Conclusions and discussions are presented in \S 5.

\section{Physics and formulae}

The theory of GSA has been comprehensively illustrated in \citet{YLfine,YLhyf,YLhanle}. Thus we present below briefly the main physics of magnetic realignment and its influence on atomic spectra. For results, please directly skip this section and go to \S 3.

Anisotropic radiation can excite atoms, which is known as photo-excitations and consequently results in spontaneous emissions. As is well illustrated in previous GSA papers, both the photo-excitation and magnetic precession decide the occupations among the sublevels of the ground state. The equations to describe the evolution of atom densities on both upper and ground states are \citep[see][]{landi2004}:
\begin{equation}\label{upperevol}
\begin{split}
\dot{\rho}_q^k(J_u)&+2\pi i\nu_L g_u q\rho_q^k(J_u)=\\
&-\sum_{\substack{J_l}}A(J_u\rightarrow J_l)\rho_q^k(J_u)+\sum_{\substack{J_{l}k'}} [J_l]\left[\delta_{kk'}p_{k'}(J_u,J_l)B_{lu}\bar{J}^0_0+\sum_{\substack{Qq'}}r_{kk'}(J_u,J_l,Q,q')B_{lu}\bar{J}^2_Q\right]\rho^{k'}_{-q'}(J_l),
\end{split}
\end{equation}
\begin{equation}\label{groundevol}
\begin{split}
\dot{\rho}_q^k(J_l)+&2\pi i\nu_L g_l q\rho_q^k(J_l)=\\
&\sum_{\substack{J_u}}p_k(J_u,J_l)[J_u]A(J_u\rightarrow J_l)\rho_q^k(J_u)-\sum_{\substack{J_{u}k'}}\left[\delta_{kk'}B_{lu}\bar{J}^0_0+\sum_{\substack{Qq'}}s_{kk'}(J_u,J_l,Q,q')B_{lu}\bar{J}^2_Q\right]\rho^{k'}_{-q'}(J_l),
\end{split}
\end{equation}
in which
\begin{equation}
p_k(J_u,J_l)=(-1)^{J_u+J_l+1}\left\{\begin{array}{ccc}J_l&J_l&k\\J_u&J_u&1\end{array}\right\}, p_0(J_u,J_l)=\frac{1}{\sqrt{[J_u,J_l]}},
\end{equation}
\begin{equation}
r_{kk'}(J_u,J_l,Q,q)=(3[k,k',2])^{\frac{1}{2}}\left\{\begin{array}{ccc}1&J_u&J_l\\1&J_u&J_l\\2&k&k'\end{array}\right\}\left(\begin{array}{ccc}k&k'&K\\q&q'&Q\end{array}\right),
\end{equation}
\begin{equation}
s_{kk'}(J_u,J_l,Q,q)=(-1)^{J_l-J_u+1}[J_l](3[k,k',K])^\frac{1}{2}\left(\begin{array}{ccc}k&k'&2\\q&q'&Q\end{array}\right) \left\{\begin{array}{ccc}1&1&2\\J_l&J_l&J_u\end{array}\right\}\left\{\begin{array}{ccc}k&k'&2\\J_l&J_l&J_l\end{array}\right\}.
\end{equation}
The evolution of the ground state $[\rho_q^k(J_l)]$ and the upper state $[\rho_q^k(J_u)]$ are illustrated in Eq.~\eqref{groundevol} and Eq.~\eqref{upperevol}, respectively. The quantities $J_{u}$ and $J_{l}$ are the total angular momentum quantum numbers for the upper and lower levels, respectively. The quantity $A$ is the Einstein spontaneous emission rate and the quantity $B$ is the Einstein coefficient for absorption and stimulated emission\footnote[4]{The data of Einstein coefficients used in the paper are taken from the Atomic Line List (\url{http://www.pa.uky.edu/~peter/atomic/}) and the NIST Atomic Spectra Database.}. The quantities $\rho_q^k$ and $\bar{J}_Q^K$ are irreducible density matrices for the atoms and the radiation field, respectively. $6-j$ and $9-j$ symbols are represented by the matrices with $"\{$ $\}"$, whereas $3-j$ symbols are indicated by the matrices with $"()"$ (see \citealt{1989PhT....42l..68Z} for details). The symbol $[j]$ represents the quantity $2j+1$, e.g., $[J_l]=2J_l+1$, $[J_l,J_u]=(2J_l+1)(2J_u+1)$, etc. The second terms on the left side of Eq.~\eqref{upperevol} and Eq.~\eqref{groundevol} stand for the magnetic realignment. Due to fast magnetic precession, magnetic realignment on the upper levels can be neglected. The two terms on the right side represent spontaneous emissions and the excitations from lower levels. All the transitions to different upper states and different sublevels of the ground state are taken into account by summing up $J_u$ and $J_l$. Note that the symmetric processes of spontaneous emission and magnetic realignment conserve $k$ and $q$. Hence, the steady state occupations of atoms on the ground state are obtained by setting the left side of Eq.~\eqref{upperevol} and Eq.~\eqref{groundevol} zero:
\begin{equation}\label{groundoccu}
\begin{split}
2\pi i\rho_q^k(J_l)qg_l\nu_L&-\sum_{\substack{J_uk'}}\bigg\{p_k(J_u,J_l)\frac{[J_u]}{\sum_{J''_{l}}A''/A+i\Gamma'q} \sum_{\substack{J'_l}}B_{lu}[J'_l]\left[\delta_{kk'}p_{k'}(J_u,J'_l)\bar{J}^0_0+\sum_{\substack{Qq'}}r_{kk'}(J_u,J'_l,Q,q')\bar{J}^2_Q\right]\\
&-\delta_{J_lJ'_l}\left[\delta_{kk'}B_{lu}\bar{J}^0_0+\sum_{\substack{Qq'}}B_{lu}s_{kk'}(J_u,J_l,Q,q')\bar{J}^2_Q\right]\bigg\}\rho^{k'}_{-q'}(J'_l)=0
\end{split}
\end{equation}
where $\Gamma'$ equals $2\pi\nu_L g_u/A$.

The absorption from the state where the atoms are aligned is polarized, due to a difference of absorption parallel and perpendicular to the direction of alignment. Stokes parameters for absorption lines with finite optical depth are:
\begin{equation}\label{Stokesgen}
\begin{split}
I&=(I_0+Q_0)e^{-\tau(1+\frac{\eta_1}{\eta_0})}+(I_0-Q_0)e^{-\tau(1-\frac{\eta_1}{\eta_0})},\\
Q&=(I_0+Q_0)e^{-\tau(1+\frac{\eta_1}{\eta_0})}-(I_0-Q_0)e^{-\tau(1-\frac{\eta_1}{\eta_0})},\\
U&=U_0e^{-\tau},V=V_0e^{-\tau},
\end{split}
\end{equation}
in which $I_0$, $Q_0$, $U_0$, $V_0$ are the Stokes parameters of the background radiation. $\eta_i(i=0\sim3)$ are the corresponding absorption coefficients, which are defined as (\citealt{Landi-DeglInnocenti:1984kl}, see also \citealt{YLfine}):
\begin{equation}\label{etai}
\eta_i(\nu,\Omega)=\frac{h\nu_0}{4\pi}Bn(J_l)\Psi(\nu-\nu_0)\sum_{\substack{KQ}}(-1)^K\omega^{K}_{J_lJ_u}\sigma^K_Q(J_l)\mathcal{J}^K_Q(i,\Omega),
\end{equation}
where the quantity $\sigma^K_Q\equiv\frac{\rho^K_Q}{\rho^0_0}$ is the alignment parameter, the quantity $\rho^K_Q$ is the density matrix \footnote[5]{For example, the irreducible tensor for $J/F=1$ is $\rho_0^2=[\rho(1,1)-2\rho(1,0)+\rho(1,-1)]$.}. As illustrated in \citep{YLfine}, the alignment parameter is related to $\theta_{br}$, the angle between the direction of radiation and magnetic direction. The total atomic population $n(J_l)$ on the lower level $J_l$ is defined as $n\sqrt{[J_L]}\rho^0_0(J_l)$. $\Psi(\nu_-\nu_0)$ is the line profile. The quantity $\mathcal{J}^K_Q(i, \Omega)$ is the irreducible radiation tensor of incoming light averaged over the whole solid angle and the line profile:
\begin{equation}\label{orintensor}
\begin{split}
\mathcal{J}_0^0(i,\Omega)=\left(
\begin{array}{c}
1\\
0\\
0\\
\end{array}\right),
&\mathcal{J}_0^2(i,\Omega)=\frac{1}{\sqrt{2}}\left[
\begin{array}{c}
1-1.5\sin^{2}\theta\\
-3/2\sin^{2}\theta\\
0\\
\end{array}\right],\\
\mathcal{J}_{\pm2}^{2}(i,\Omega)=\sqrt{3}e^{\pm 2i\phi}\left[
\begin{array}{c}
\sin^{2}\theta/4\\
-(1+\cos^{2}\theta)/4\\
\mp i\cos\theta/2\\
\end{array}\right],
&\mathcal{J}_{\pm1}^{2}(i,\Omega)=\sqrt{3}e^{\pm i\phi}\left(
\begin{array}{c}
\mp \sin 2\theta/4\\
\mp \sin 2\theta/4\\
-i \sin \theta/2\\
\end{array}\right).
\end{split}
\end{equation}
the quantity $\mathcal{J}^K_Q(i, \Omega)$ is decided by its direction $\Omega(\theta,\phi)$, where $\theta$ is defined as the angle between the line of sight and the magnetic direction. Additionally,
\begin{equation}\label{omegai}
\omega^k_{J_lJ_u}\equiv\left\{\begin{array}{ccc}1&1&K\\J_l&J_l&J_u\end{array}\right\}/\left\{\begin{array}{ccc}1&1&0\\J_l&J_l&J_u\end{array}\right\}.
\end{equation}

Moreover, the influence of magnetic fields on upper level is negligible when the magnetic precession rate is larger than the emission rate of upper level. The differential occupation the on ground state which results from the atomic alignment is transferred to the upper atomic states by radiative excitation (see \citealt{2012JQSRT.113.1409Y} for details). The emission coefficients $\epsilon_i$ demonstrating the polarized emission from the upper level are \citep[see][]{YLhyf}:
\begin{equation}\label{epsiloni}
\epsilon_i(\nu,\Omega)=\frac{h\nu_0}{4\pi}An(J_u,\theta_{br})\Psi(\nu-\nu_0)\sum_{\substack{KQ}}\omega^{K}_{J_uJ_l}\sigma^K_Q(J_u,\theta_{br})\mathcal{J}^K_Q(i,\Omega).
\end{equation}
where the intensity of emission line is when $i=0$. Throughout the paper, we consider the case when the magnetic field is parallel to the incidental radiation as 1 and use the ratio between the cases with the change of the direction of magnetic field and this case to demonstrate the variation of spectra with the influence of magnetic field due to GSA. When magnetic field is parallel to the incident radiation, only the anisotropy of the radiation field contribute to the alignment, thus we can consider there is no magnetic realignment in this case. Moreover, for the sake of observation, the change of magnetic fields is presented in the observational coordinate system ($\theta,\phi$). The reason for the choice of the coordinate system is discussed in Appendix A.

In the optical thin case, where $\tau$ is proportional to the column density of the absorbing species, the change in the absorption coefficient from the GSA is proportional to the absorption depth of the line. Thus, if several transitions from the same species (atom/ion) are available in a given line spectrum, the GSA effect can be measured from the {\it differential} changes in the absorption depths of these lines, because the individual lines are affected {\it differently}. However, since the effect is expected to be small (see \S 3), a high spectral resolution and a very high signal-to-noise (S/N) is required for such an experiment. In the following, we evaluate more quantitatively under what conditions a direct measurement of the GSA effect is possible with current observational data.

\begin{figure*}
\centering
 \subfigure[]{
\includegraphics[width=0.44\columnwidth,
 height=0.25\textheight]{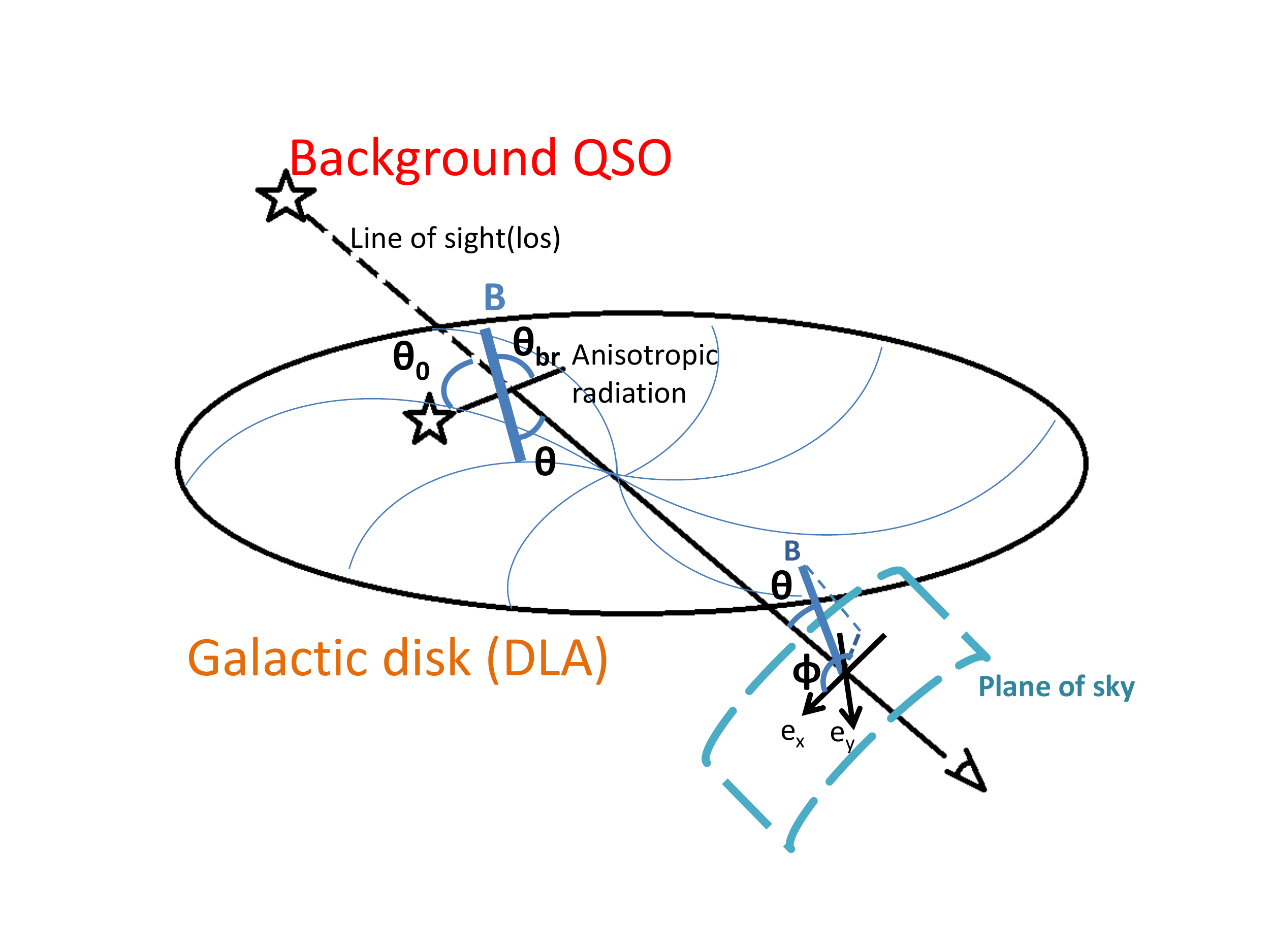}\label{scea}}
\subfigure[]{
 \includegraphics[width=0.41\columnwidth,
 height=0.25\textheight]{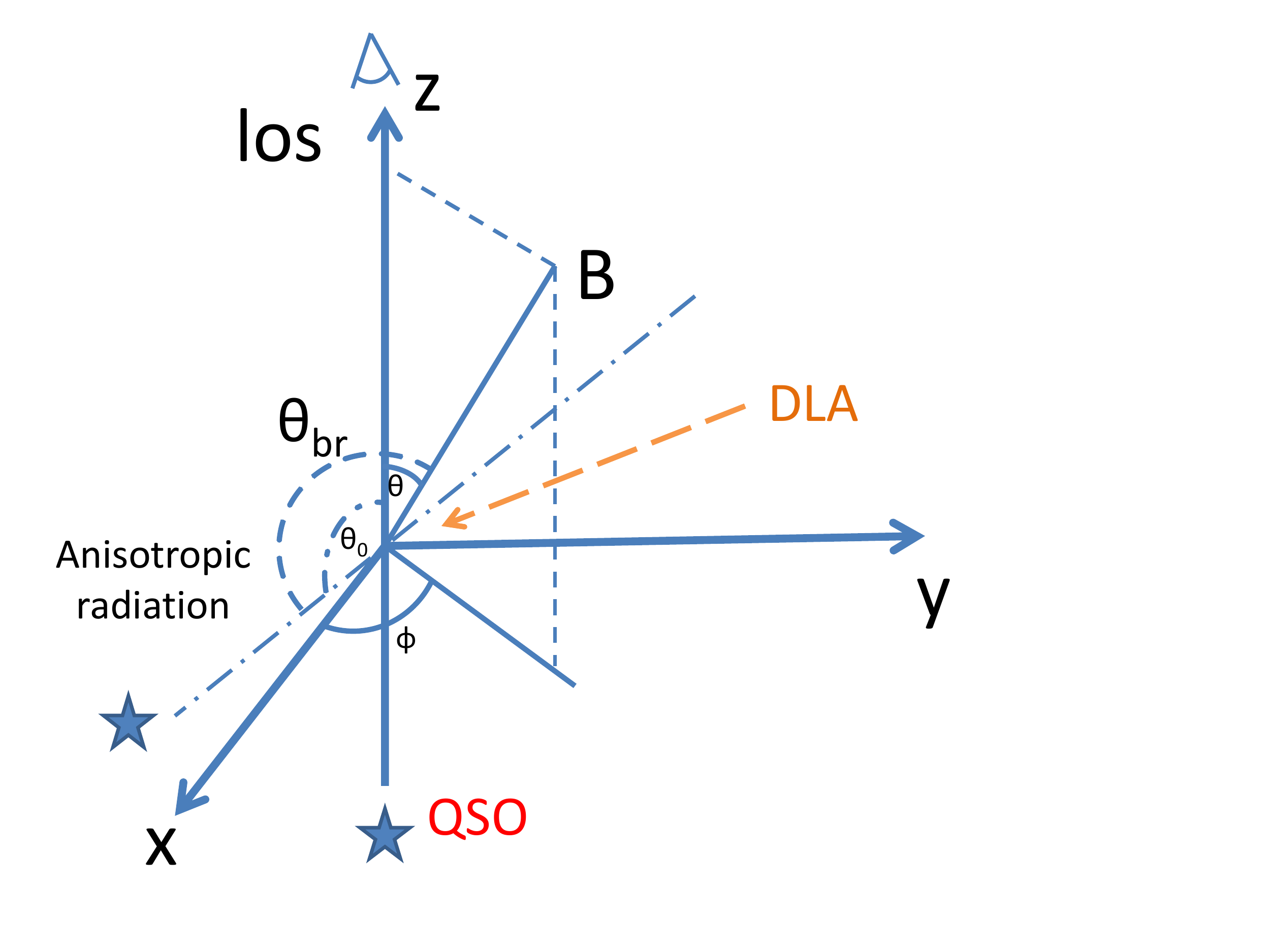}\label{sceb}}
\subfigure[]{
 \includegraphics[width=0.41\columnwidth,
 height=0.25\textheight]{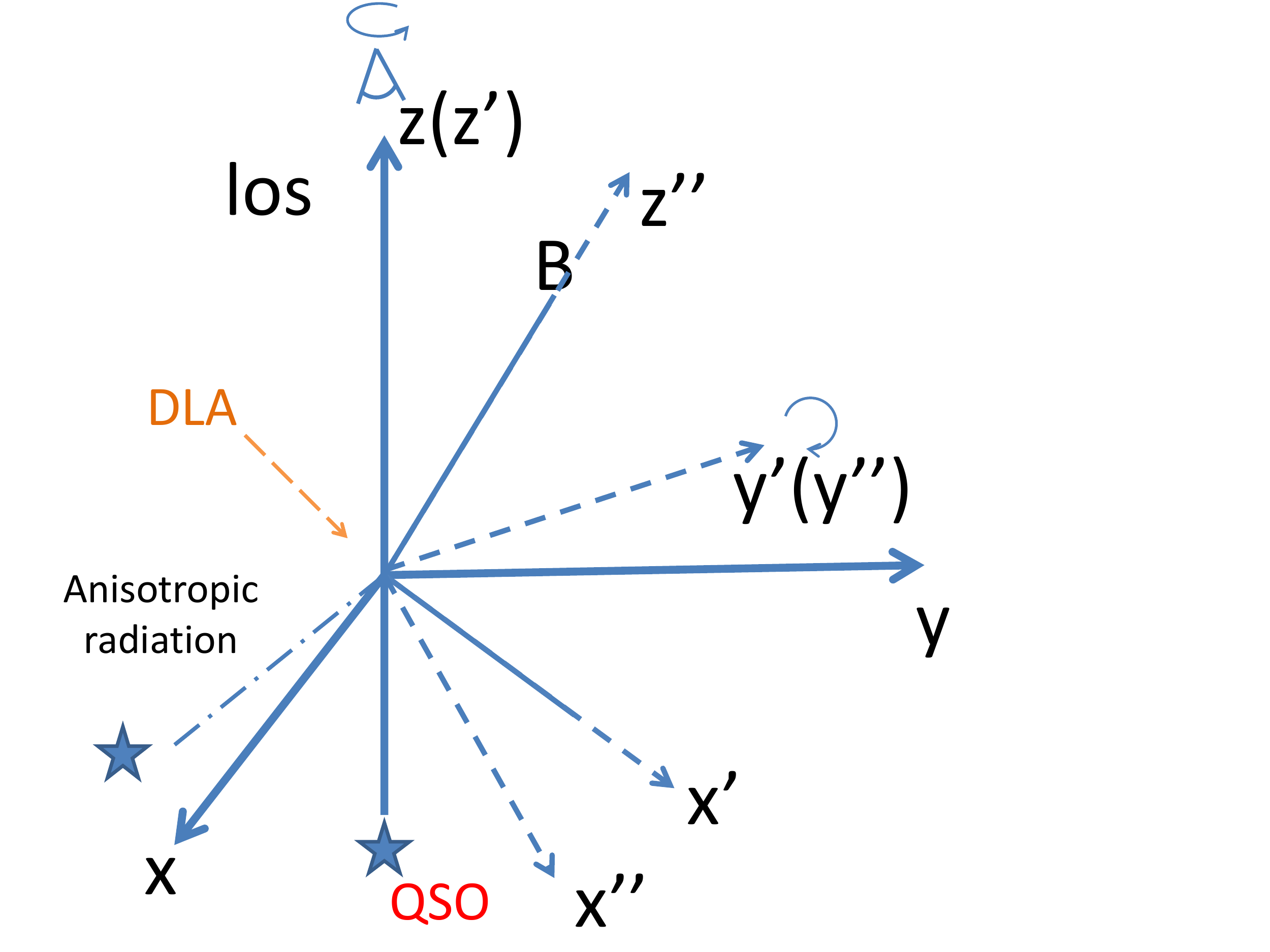}\label{scec}}
\subfigure[]{
 \includegraphics[width=0.41\columnwidth,
 height=0.25\textheight]{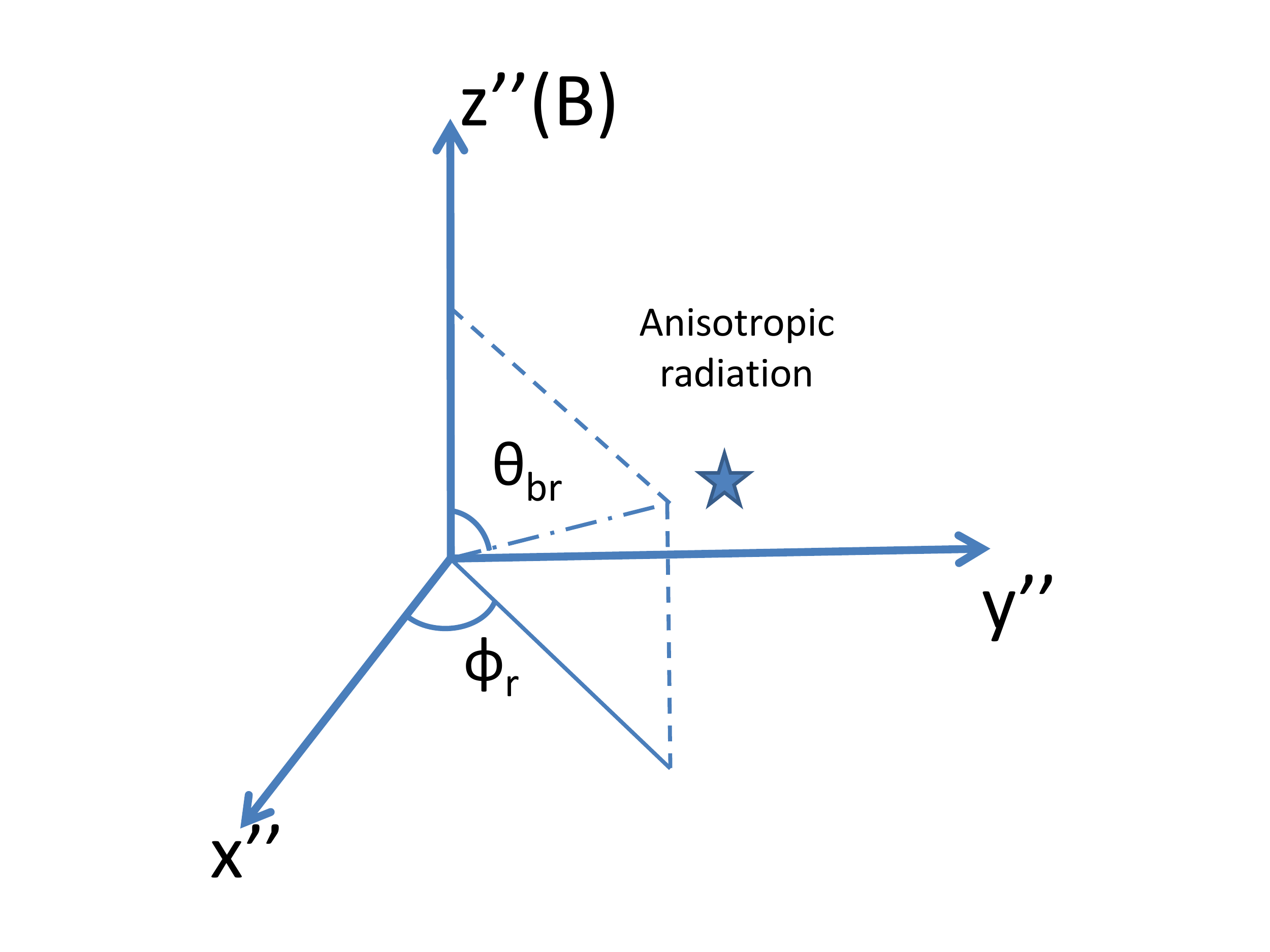}\label{sced}}
\caption{Scenario for atomic alignment in interstellar medium in galaxy.
(a) Typical environment where absorption line is polarized owing to GSA. Alignment is produced by a pumping source, which both influence the polarization of scattering(emission) lines and induce polarization on absorption of another background source from which the light passes through the aligned medium$^{\footnotesize 6}$. $\theta_{br},\theta$ are the polar angles of the pumping radiation and line of sight measured in reference to the magnetic field, respectively. $\theta_0$ is the angle between the anisotropic radiation and the line of sight;
(b) xyz-coordinate system with the line of sight being z-axis, which is the main coordinate system to compare the change of magnetic fields (see more details in Appendix A);
(c) coordinate system transformation: first rotate line of sight and make the magnetic field on x'z'-plane to form x'y'z
'-coordinate system, and then rotate y'-axis to make x''y''z''-coordinate system with the magnetic field being z''-axis;
(d) x''y''z''-coordinate system with the magnetic field being z''-axis.}\label{scena}
\end{figure*}
\footnotetext[6]{There could be degenerate case where the pumping source is the same as the background source \citep[see][]{YLfine}.}

\section{Influence of GSA on atomic spectra}

\subsection{Influence of GSA on absorption spectra}

\subsubsection{Scenario for absorption spectra}

As an example, we show in Fig.\ref{scena} a typical scenario, in which the effect of atomic alignment could lead to measurable effect in observational data. Given is a typical late-type galaxy with an extended neutral gas disk and an interstellar magnetic field. Star-formation is expected to take place in distinct regions in the disk (i.e., in spiral arms), so that at a given point in the disk the interstellar radiation field might be highly anisotropic. Imagine a bright background point source (e.g., a QSO) being located behind the gas disk. In the spectrum of the background QSO, the gas disk will leave its imprint through many absorption lines from neutral and ionized species, where most of the lines are located in the (restframe) UV. Such disk absorbers are known to contribute to the population of the so-called Damped Lyman $\alpha$ absorbers (DLAs) that are frequently observed at low and high redshift in QSO spectra. If in the region in which the sightlines pierces the disk the geometry of the ambient magnetic field and the anisotropy of the interstellar radiation field is fortunate, the atomic alignment will cause small but measurable changes in the central absorption depths of unsaturated absorption lines. Due to a small density in the interstellar medium, the collisional rate is much smaller than the Larmor precession rate. Hence, the magnetic realignment is effective. Given one specific object, the line of sight and the direction of radiation is fixed. Latter part of this section will present the variation of line intensity with the influence of magnetic fields.

\subsubsection{Influence absorption lines}

The alignment of singly-ionized sulfur S\,{\sc ii} absorption line in a beam of light is demonstrated in \citet{YLfine}. The ground state of S\,{\sc ii} is $4S_{\frac{3}{2}}^o$ ($J_l=\frac{3}{2}$) and the upper states are $4P_{\frac{1}{2}; \frac{3}{2}; \frac{5}{2}}$ ($J_u=\frac{1}{2}, \frac{3}{2}, \frac{5}{2}$). The transition wavelength from ground the level to three upper levels are 1250.58, 1253.81, 1259.52 $\mbox{\AA}$, respectively.

\begin{figure*}
\centering
 \subfigure[]{
\includegraphics[width=0.32\columnwidth,
 height=0.25\textheight]{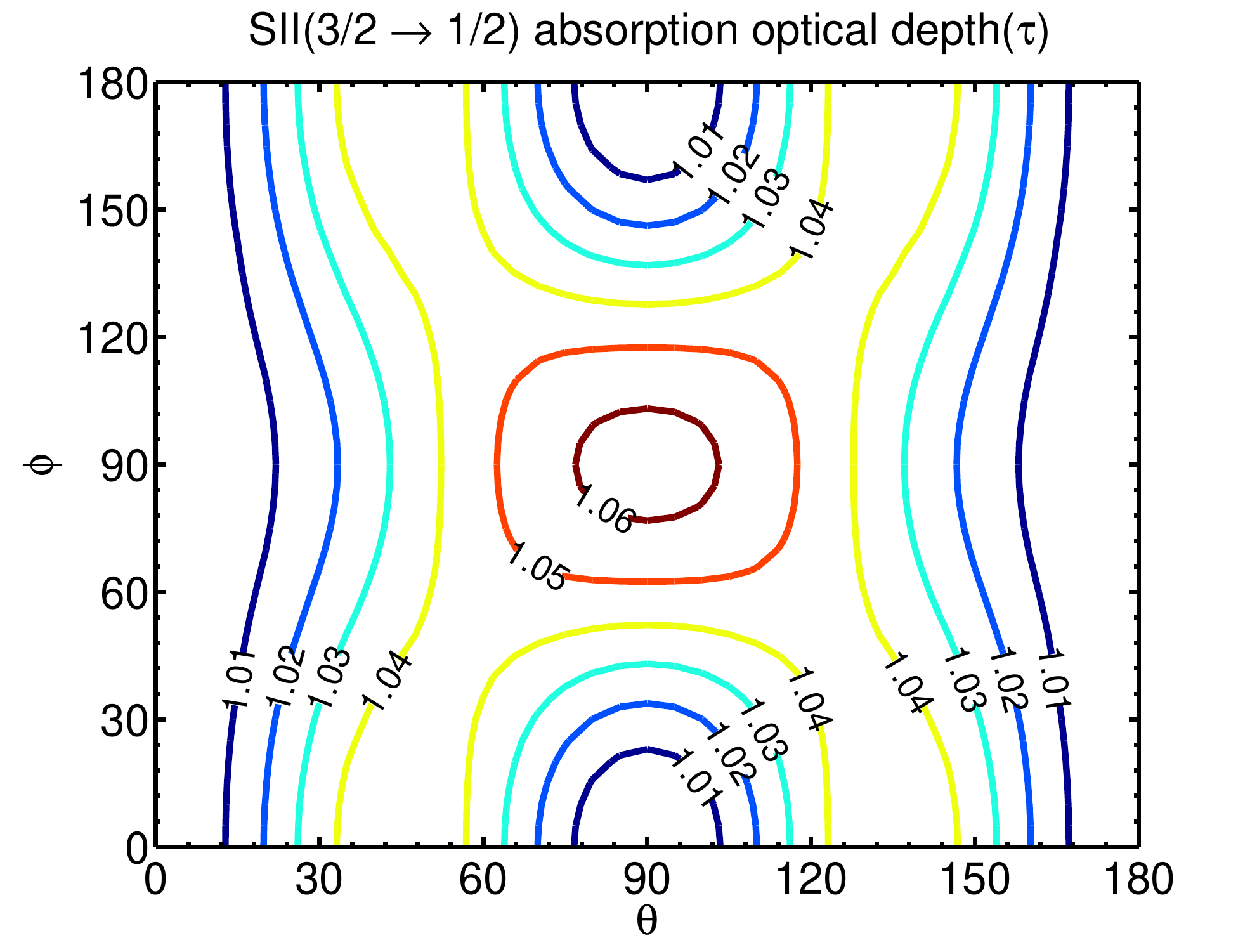}\label{S2bbrabv1c90}}
\subfigure[]{
\includegraphics[width=0.32\columnwidth,
 height=0.25\textheight]{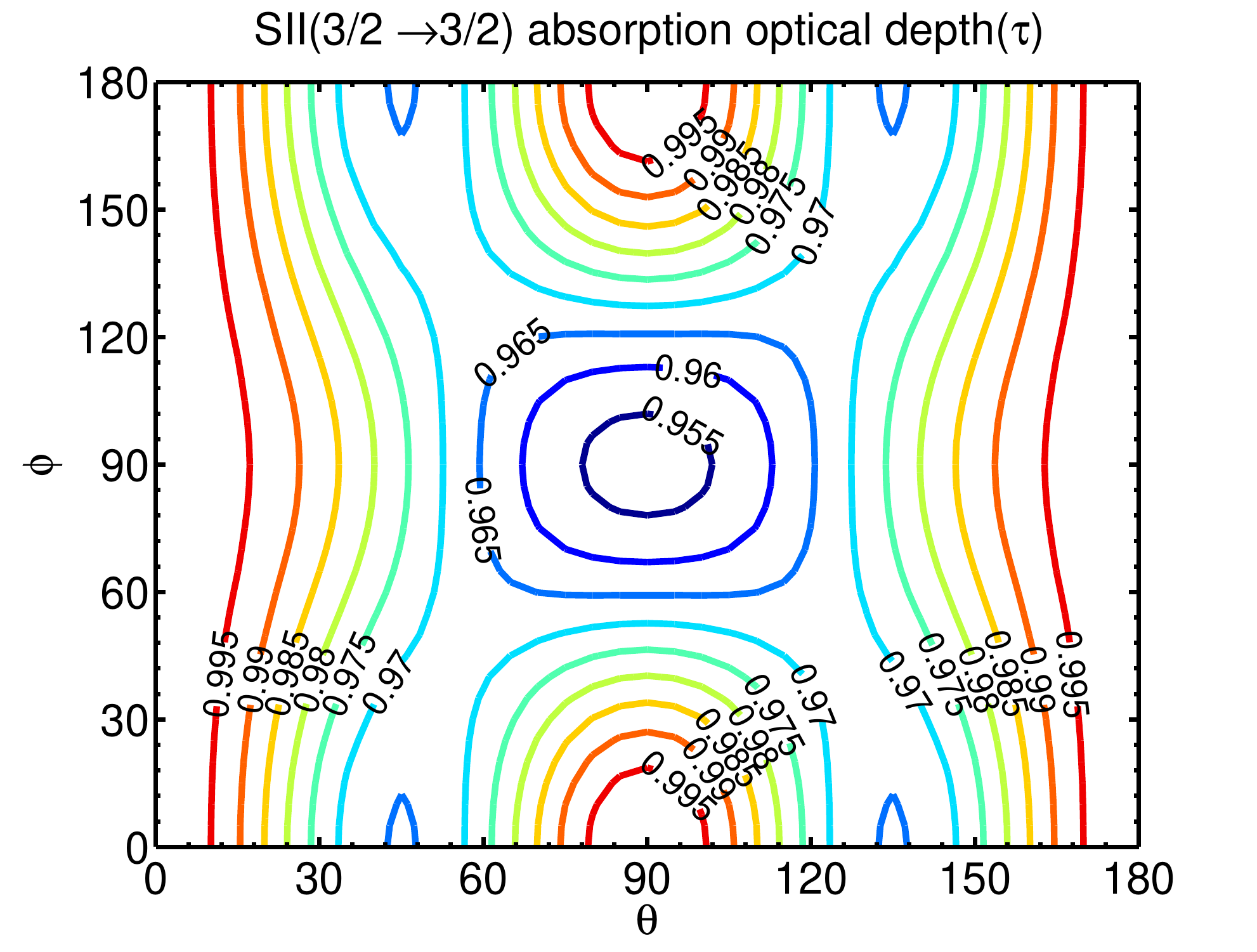}\label{S2bbrabv3c90}}
\subfigure[]{
\includegraphics[width=0.32\columnwidth,
 height=0.25\textheight]{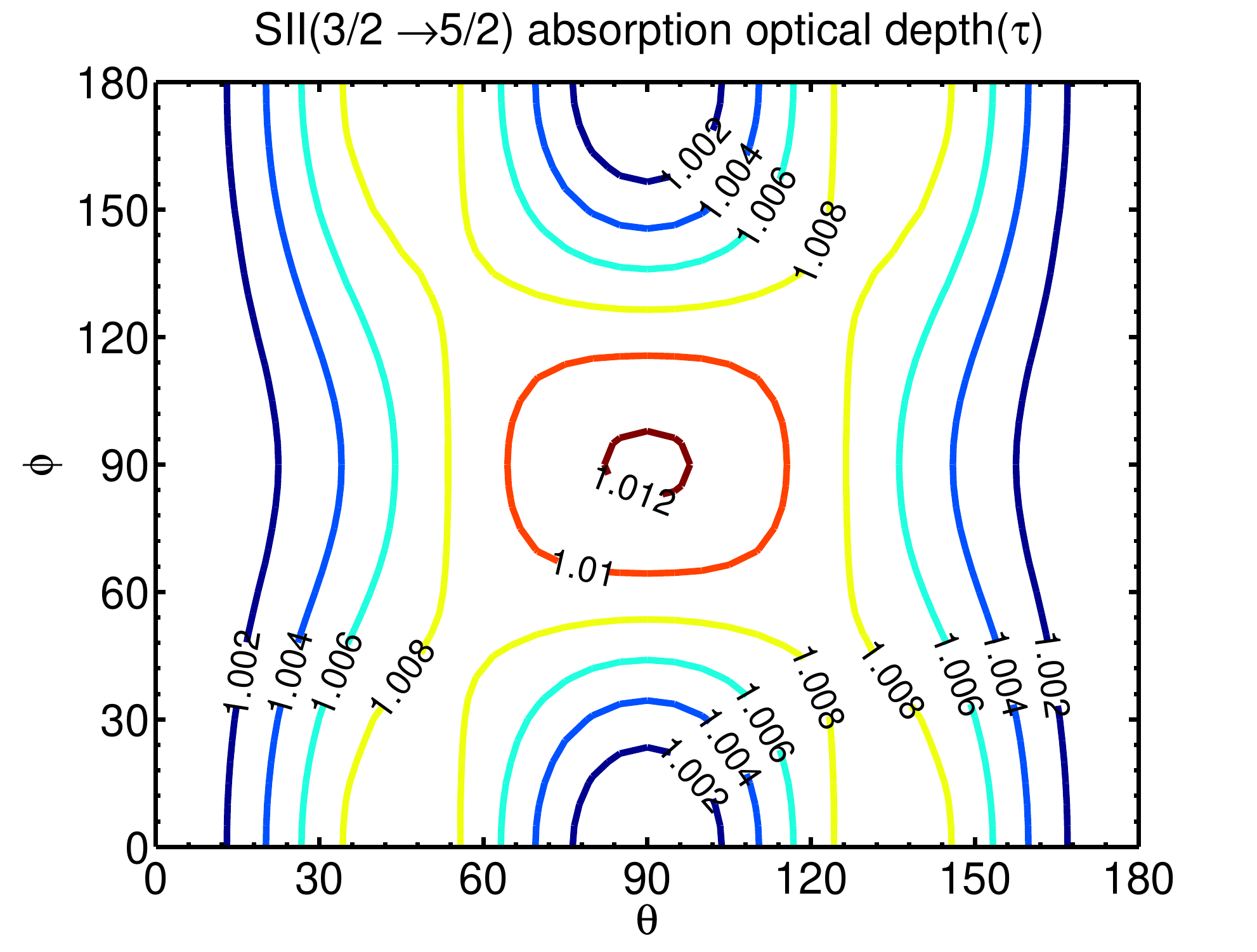}\label{S2bbrabv5c90}}
\caption{S\,{\sc ii} absorption line in blackbody radiation with $T_{source}=50000K$. The variation of line intensity with respect to the direction of the magnetic field is compared between the line (a)1250.58$\mbox{\AA}$, (b)1253.81$\mbox{\AA}$ and (c)1259.52$\mbox{\AA}$ in the case of line of sight vertical to radiation field($\theta_0=90^{\circ}$). $\theta$ and $\phi$ are the polar and azimuth angle of the magnetic field in line of sight coordinate defined in Fig. \ref{sceb}.\\}\label{S2bbrabcp}
\end{figure*}

\begin{figure*}
\centering
\includegraphics[width=0.9\columnwidth,
 height=0.75\textheight]{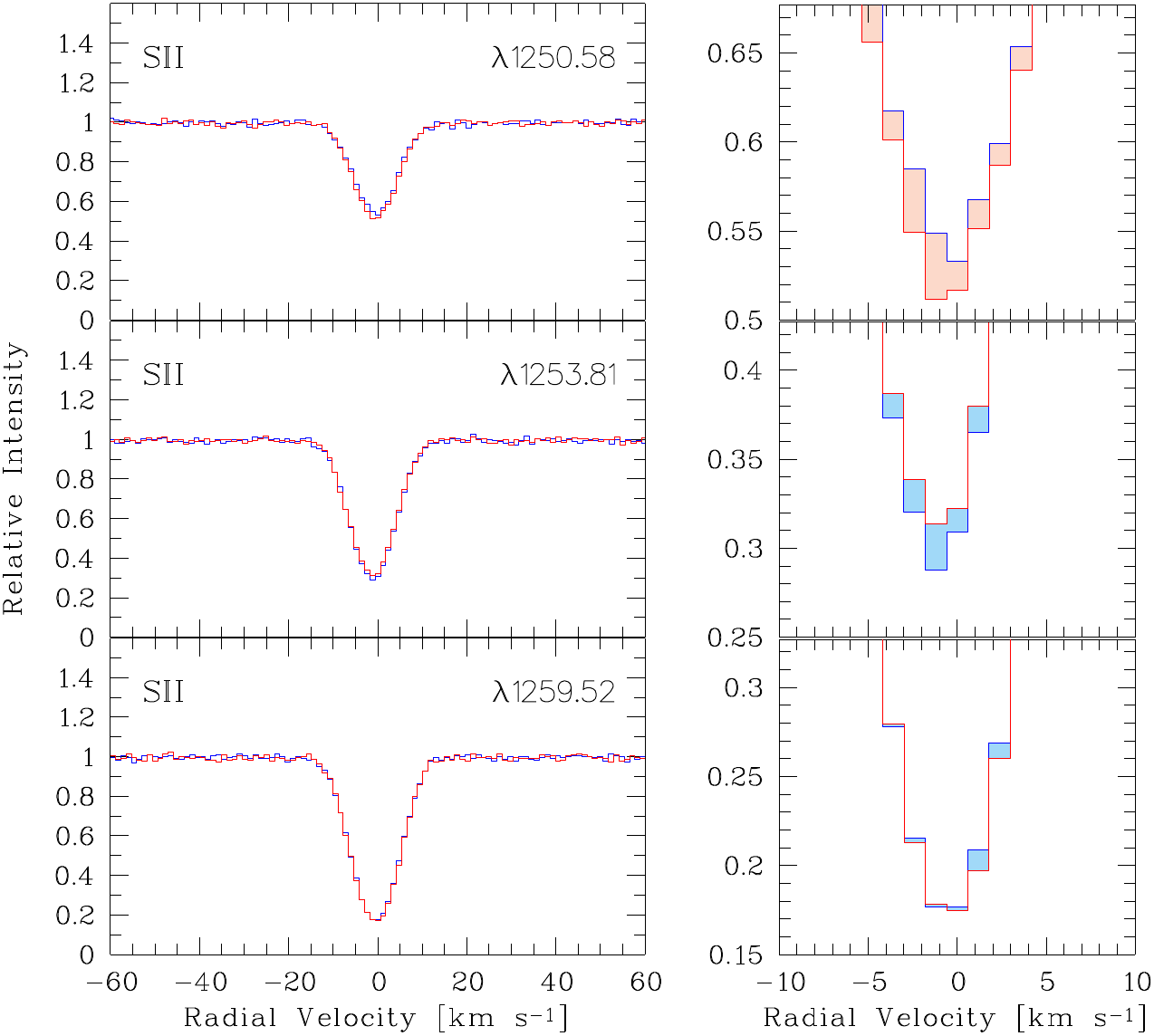}
\caption{Synthetic spectrum (in velocity scale) of a single-component S\,{\sc ii} $\lambda\lambda 1250.58,1253.81,1259.52$ absorption system with a S\,{\sc ii} column density of $N$(S\,{\sc ii}$)=4\times 10^{14}$ cm$^{-2}$, a Doppler paramter of $b=5$ km\,s$^{-1}$, a spectral resolution of $R=45,000$, and a S/N of 100 per resolution element. Such a S\,{\sc ii} column density would be expected for a DLA with log $N$(H\,{\sc i}$)=20.48$ and a sulfur abundance of $0.1$ solar assuming solar referene abundances of \citet{Asplund2009}. The synthetic spectra were generated using the {\tt FITLYMAN} routine \citep{Fontana1995} implemented in the ESO-MIDAS sofware package. Atomic data were taken from \citet{Morton2003}. The blue solid line shows the orginal specrum without the atomic alignment, the red solid lines shows the spectrum including the alignment effect, assuming $\theta_0=\theta=90^{\circ}$, $\phi=0$. The red-/blue-shaded areas indicate the excess/deficiency of absorption due to the alignment effect.
}\label{S2ob}
\end{figure*}

The three transitions of singly-ionized sulfur (S\,{\sc ii}; upper ionization potential 23.3 eV) at $\lambda\lambda 1250.58,1253.81,1259.52$ represent important tracers for neutral and weakly ionized gas in the local interstellar and circumgalactic medium and in gas in distant galaxies (e.g., \citealt{2014ApJ...780...76K,Welsh2012,Richter2001,2014A&A...572A.102F}). Sulfur is an $\alpha$ element, but shows only a weak depletion into dust grains (e.g.,\citealt{Savage1996}), so that the interstellar S abundance is often used as a proxy for the $\alpha$-abundance in the gas. The three lines mentioned above span a moderate oscillator strengths \citep{2014ApJ...780...76K,Morton2003} are all within 10 $\mbox{\AA}$, so that in an UV or optical (if absorption is redshifted) spectrum, these lines can be observed in the same wavelength region at an identical S/N. In addition, sulfur has a relatively low cosmic abundance \citep{Asplund2009}, so that under typical interstellar conditions (in particular in low-metallicity environments) these lines are not saturated. Thus, the S\,{\sc ii} $\lambda\lambda 1250.58,1253.81,1259.52$ represent the ideal transitions to study atomic alignment effects in interstellar environments with magnetic fields. Let's denote  the real intensity of absorption line as $I^{ab}_{re}$, which is proportional to real column density  $\tau_0$; and the observed intensity of absorption lines as $I^{ab}_{ob}$, which is proportional to the varied column density $\tau_0\pm\Delta\tau$. The ratio between observed absorption line intensity and real line intensity is defined by
\begin{equation}\label{ratioab1}
r^{ab}(\lambda,\theta_0,\theta_{br},\theta)=I^{ab}_{ob}/I^{ab}_{re}\propto\left(\rho^0_0(J_l)\sqrt{2}+\omega^2_{J_lJ_u}\rho^2_0(J_l)\left(1-1.5\sin^2\theta\right)\right),
\end{equation}
where $\lambda$ is the wavelength of the line, $\theta_0,\theta_{br},\theta$ are defined in Fig. \ref{scea}. The optical depth variation with respect to different magnetic field direction is presented in Fig. \ref{S2bbrabcp} given line of sight perpendicular to radiation direction($\theta_0=90^\circ$), radiation source temperature $T_{source}=50000K$. By comparing Fig. \ref{S2bbrabv1c90}, Fig. \ref{S2bbrabv3c90}, and Fig. \ref{S2bbrabv5c90}, it is obvious that with the same direction of radiation and line of sight, the influence of magnetic fields on different spectra is different.

Moreover, even for the same spectrum with the same radiation field observed from the same line of sight (hereby, the same $\theta_0$), the actual column density deduced from the spectrum can be different only due to the change of magnetic fields. For example, as demonstrated in Fig. \ref{S2bbrabv1c90}, the magnetic field does not influence the spectrum when the direction of the magnetic field is parallel to the line of sight ($\theta=0^{\circ}$), whereas enrich $7\%$ of the spectrum when the direction of the magnetic field is perpendicular to both the line of sight and incident radiation ($\theta=90^{\circ}, \phi=90^{\circ}$). Therefore, magnetic fields affect the column density deduced from the absorption line observed due to GSA.

In Fig.\ref{S2ob} we show the expected maximum effect of the atomic alignment for the three S\,{\sc ii} lines in an synthetic absorption spectrum of a DLA for $\theta_0=90^{\circ}$. To show the effect clearly, we zoom in the spectrum to the radial velocity range $[-10 km{\cdot}s^{-1}, 10 km{\cdot}s^{-1}]$. It is obvious that the alignment enriches $\lambda 1250.58$ whereas depletes $\lambda 1253.81$, and $\lambda 1259.52$ in different scales. Hence, magnetic fields have a small yet measurable effect on the spectra observed and should be taken into consideration for the error budget of spectroscopic studies.

\begin{table}[!hbp]
\centering
\caption{COLUMN DENSITY VARIATION FOR ABSORPTION LINE INDUCED BY GSA}\label{absorpretable}
\begin{tabular}{|c|c|c|c|c|c|c|}
\hline \hline
Species & Transition & Wavelength & $\Delta N/N_{min}$ & $\Delta N/N_{max}$ & $\Delta$ log$(N/N_0)_{min}$ & $\Delta$ log$(N/N_0)_{max}$ \\
\hline
O\,{\sc i} & $3P_{2}\rightarrow 3D^{\circ}$ & $1025.76\mbox{\AA}$ & $-6.37\%$ & $+0\%$ & $-2.86\times10^{-2}$ & $0$ \\
\hline
\multirow{3}{*}{S\,{\sc i}} & $3P_{2}\rightarrow 3D^{\circ}_{1}$ & $1474.57\mbox{\AA}$ & $-27.29\%$ & $+1.81\%$ & $-1.38\times10^{-1}$ & $7.79\times10^{-3}$ \\
\cline{2-7}
& $3P_{2}\rightarrow 3D^{\circ}_{2}$ & $1474.38\mbox{\AA}$ & $-22.82\%$ & $+2.62\%$ & $-1.13\times10^{-1}$ & $1.12\times10^{-2}$ \\
\cline{2-7}
& $3P_{2}\rightarrow 3D^{\circ}_{3}$ & $1473.99\mbox{\AA}$ & $-19.99\%$ & $+0.08\%$ & $9.69\times10^{-2}$ & $3.65\times10^{-4}$ \\
\hline
\multirow{3}{*}{S\,{\sc ii}} & $4S^{\circ}_{\frac{3}{2}}\rightarrow 4P_{\frac{1}{2}}$ & $1250.58\mbox{\AA}$ & $-7.56\%$ & $+6.33\%$ & $-3.41\times10^{-2}$ & $2.67\times10^{-2}$ \\
\cline{2-7}
& $4S^{\circ}_{\frac{3}{2}}\rightarrow 4P_{\frac{3}{2}}$ & $1253.81\mbox{\AA}$ & $-4.71\%$ & $+7.00\%$ & $-2.09\times10^{-2}$ & $2.94\times10^{-2}$ \\
\cline{2-7}
& $4S^{\circ}_{\frac{3}{2}}\rightarrow 4P_{\frac{5}{2}}$ & $1259.52\mbox{\AA}$ & $-1.61\%$ & $+1.22\%$ & $-7.04\times10^{-3}$ & $5.29\times10^{-3}$ \\
\hline
\multirow{3}{*}{Ti\,{\sc ii}} & $a4F_{\frac{3}{2}}\rightarrow z4D^{\circ}_{\frac{1}{2}}$ & $3072.97\mbox{\AA}$ & $-1.64\%$ & $+1.30\%$ & $-7.16\times10^{-3}$ & $5.62\times10^{-3}$ \\
\cline{2-7}
& $a4F_{\frac{3}{2}}\rightarrow z4F^{\circ}_{\frac{3}{2}}$ & $3241.98\mbox{\AA}$ & $-1.03\%$ & $+1.35\%$ & $-4.48\times10^{-3}$ & $5.81\times10^{-3}$ \\
\cline{2-7}
& $a4F_{\frac{3}{2}}\rightarrow z4F^{\circ}_{\frac{5}{2}}$ & $3229.19\mbox{\AA}$ & $-0.33\%$ & $+0.26\%$ & $-1.44\times10^{-3}$ & $1.12\times10^{-3}$ \\
\hline
\multirow{3}{*}{Fe\,{\sc ii}} & $a6D_{\frac{9}{2}}\rightarrow y6F^{\circ}_{\frac{7}{2}}$ & $1142.36\mbox{\AA}$ & $-1.69\%$ & $+4.42\%$ & $-7.40\times10^{-3}$ & $1.88\times10^{-2}$ \\
\cline{2-7}
& $a6D_{\frac{9}{2}}\rightarrow y6F^{\circ}_{\frac{9}{2}}$ & $1143.22\mbox{\AA}$ & $-2.86\%$ & $+7.94\%$ & $-1.26\times10^{-2}$ & $3.32\times10^{-2}$ \\
\cline{2-7}
& $a6D_{\frac{9}{2}}\rightarrow y6F^{\circ}_{\frac{11}{2}}$ & $1144.93\mbox{\AA}$ & $-0.46\%$ & $+3.21\%$ & $-2.02\times10^{-3}$ & $1.37\times10^{-2}$ \\
\hline
\hline
\multicolumn{7}{|c|}{Absorption from other sublevels of the ground level}\\
\hline
\multirow{3}{*}{C\,{\sc i}$^\ast$} & $3P_{1}\rightarrow 3P^{\circ}_{0}$ & $1260.93\mbox{\AA}$ & $-24.20\%$ & $+1.54\%$ & $-1.20\times10^{-1}$ & $6.64\times10^{-3}$ \\
\cline{2-7}
& $3P_{1}\rightarrow 3P^{\circ}_{1}$ & $1261.00\mbox{\AA}$ & $-17.45\%$ & $+0.46\%$ & $-8.33\times10^{-2}$ & $2\times10^{-3}$ \\
\cline{2-7}
& $3P_{1}\rightarrow 3P^{\circ}_{2}$ & $1261.12\mbox{\AA}$ & $-15.76\%$ & $+0\%$ & $-7.45\times10^{-2}$ & $0$ \\
\hline
\multirow{3}{*}{C\,{\sc ii}$^\ast$} & $2P^{\circ}_{\frac{3}{2}}\rightarrow 2S_{\frac{1}{2}}$ & $1037.02\mbox{\AA}$ & $-9.02\%$ & $+3.68\%$ & $-4.10\times10^{-2}$ & $1.57\times10^{-2}$ \\
\cline{2-7}
& $2P^{\circ}_{\frac{3}{2}}\rightarrow 2D_{\frac{3}{2}}$ & $1335.66\mbox{\AA}$ & $-10.34\%$ & $+1.29\%$ & $-4.74\times10^{-2}$ & $5.56\times10^{-3}$ \\
\cline{2-7}
& $2P^{\circ}_{\frac{3}{2}}\rightarrow 2D_{\frac{5}{2}}$ & $1335.71\mbox{\AA}$ & $-5.39\%$ & $+0.10\%$ & $2.41\times10^{-2}$ & $4.41\times10^{-4}$ \\
\hline
O\,{\sc i}$^\ast$ & $3P_{1}\rightarrow 3D^{\circ}$ & $1027.43\mbox{\AA}$ & $-9.57\%$ & $+0\%$ & $-4.37\times10^{-2}$ & $0$ \\
\hline
\multirow{2}{*}{Si\,{\sc ii}$^\ast$} & $2P^{\circ}_{\frac{3}{2}}\rightarrow 2D_{\frac{3}{2}}$ & $1264.74\mbox{\AA}$ & $-5.59\%$ & $+5.32\%$ & $2.50\times10^{-2}$ & $2.25\times10^{-2}$ \\
\cline{2-7}
& $2P^{\circ}_{\frac{3}{2}}\rightarrow 2D_{\frac{5}{2}}$ & $1265.00\mbox{\AA}$ & $-3.00\%$ & $+0.27\%$ & $-1.32\times10^{-2}$ & $1.18\times10^{-3}$ \\
\hline
\multirow{3}{*}{S\,{\sc i}$^\ast$} & $3P_{1}\rightarrow 3P^{\circ}_{0}$ & $1303.11\mbox{\AA}$ & $-21.05\%$ & $+1.72\%$ & $-1.03\times10^{-1}$ & $7.41\times10^{-3}$ \\
\cline{2-7}
& $3P_{1}\rightarrow 3P^{\circ}_{1}$ & $1302.86\mbox{\AA}$ & $-14.43\%$ & $+0.50\%$ & $-6.77\times10^{-2}$ & $2.18\times10^{-3}$ \\
\cline{2-7}
& $3P_{1}\rightarrow 3P^{\circ}_{2}$ & $1302.34\mbox{\AA}$ & $-12.74\%$ & $+0\%$ & $-5.92\times10^{-2}$ & $0$ \\
\hline
\multirow{3}{*}{S\,{\sc iii}$^\ast$} & $3P_{1}\rightarrow 3P^{\circ}_{0}$ & $1015.50\mbox{\AA}$ & $-24.67\%$ & $+1.53\%$ & $-1.23\times10^{-1}$ & $6.58\times10^{-3}$ \\
\cline{2-7}
& $3P_{1}\rightarrow 3P^{\circ}_{1}$ & $1015.57\mbox{\AA}$ & $-17.88\%$ & $+0.46\%$ & $-8.56\times10^{-2}$ & $1.99\times10^{-3}$ \\
\cline{2-7}
& $3P_{1}\rightarrow 3P^{\circ}_{2}$ & $1015.78\mbox{\AA}$ & $-16.19\%$ & $+0\%$ & $-7.67\times10^{-2}$ & $0$ \\
\hline
\multirow{2}{*}{S\,{\sc iv}$^\ast$} & $2P^{\circ}_{\frac{3}{2}}\rightarrow 2D_{\frac{3}{2}}$ & $1072.97\mbox{\AA}$ & $-1.20\%$ & $+2.20\%$ & $-5.25\times10^{-3}$ & $9.46\times10^{-3}$ \\
\cline{2-7}
& $2P^{\circ}_{\frac{3}{2}}\rightarrow 2D_{\frac{5}{2}}$ & $1073.52\mbox{\AA}$ & $-0.20\%$ & $+0.56\%$ & $-8.68\times10^{-4}$ & $2.43\times10^{-3}$ \\
\hline
\multirow{3}{*}{C\,{\sc i}$^{\ast\ast}$} & $3P_{2}\rightarrow 3D^{\circ}_{1}$ & $1193.65\mbox{\AA}$ & $-24.65\%$ & $+1.06\%$ & $-1.23\times10^{-1}$ & $4.60\times10^{-3}$ \\
\cline{2-7}
& $3P_{2}\rightarrow 3D^{\circ}_{2}$ & $1193.39\mbox{\AA}$ & $-20.93\%$ & $+1.51\%$ & $-1.02\times10^{-1}$ & $6.52\times10^{-3}$ \\
\cline{2-7}
& $3P_{2}\rightarrow 3D^{\circ}_{3}$ & $1193.24\mbox{\AA}$ & $-18.74\%$ & $+0\%$ & $-9.01\times10^{-2}$ & $0$ \\
\hline
\multirow{3}{*}{S\,{\sc iii}$^{\ast\ast}$}& $3P_{2}\rightarrow 3D^{\circ}_{1}$ & $1202.12\mbox{\AA}$ & $-24.20\%$ & $+0.95\%$ & $-1.20\times10^{-1}$ & $4.10\times10^{-3}$ \\
\cline{2-7}
& $3P_{2}\rightarrow 3D^{\circ}_{2}$ & $1201.73\mbox{\AA}$ & $-20.62\%$ & $+1.37\%$ & $-1.00\times10^{-1}$ & $5.90\times10^{-3}$ \\
\cline{2-7}
& $3P_{2}\rightarrow 3D^{\circ}_{3}$ & $1200.97\mbox{\AA}$ & $-18.53\%$ & $+0\%$ & $-8.90\times10^{-2}$ & $0$ \\
\hline
\end{tabular}
\end{table}

\begin{figure*}
\centering
\subfigure[]{
\includegraphics[width=0.47\columnwidth,
 height=0.35\textheight]{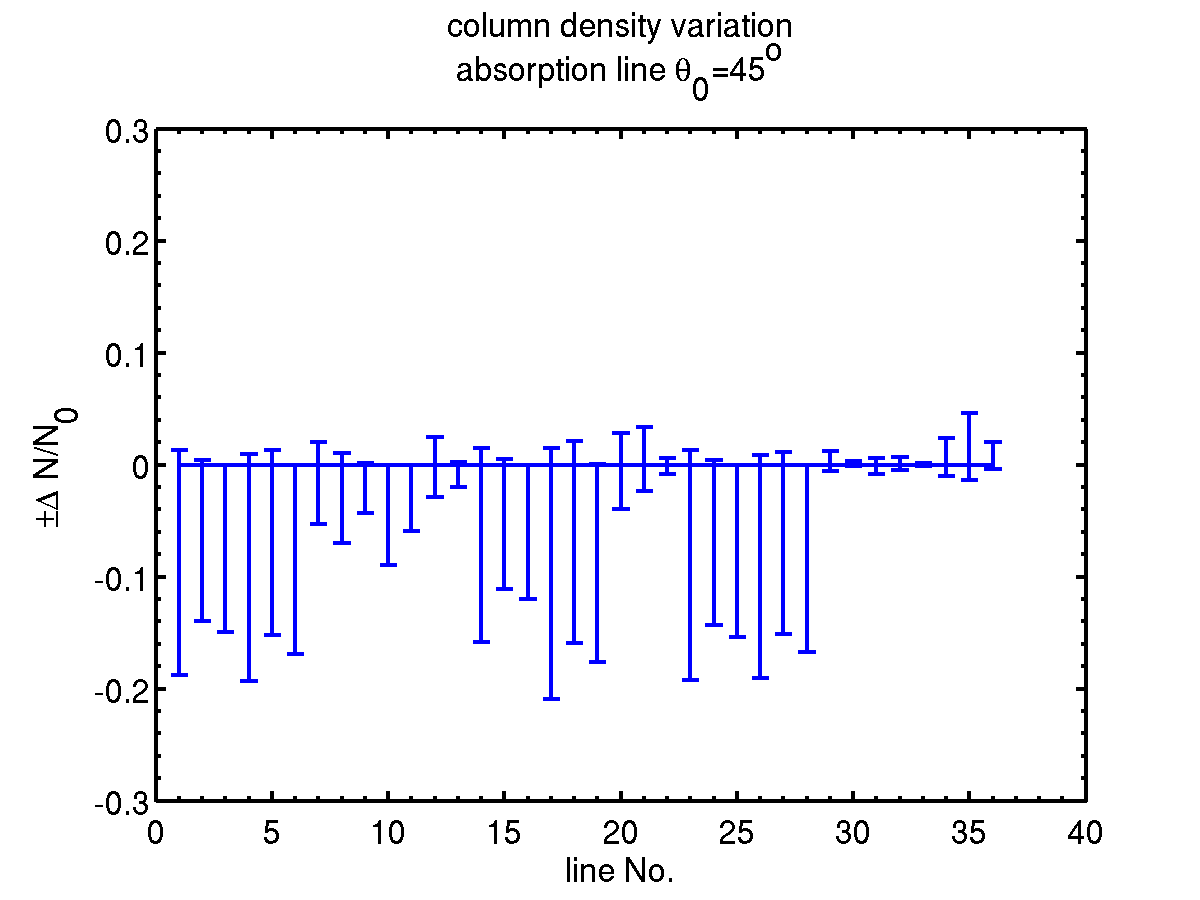}\label{varabc45}}
 \subfigure[]{
\includegraphics[width=0.47\columnwidth,
 height=0.35\textheight]{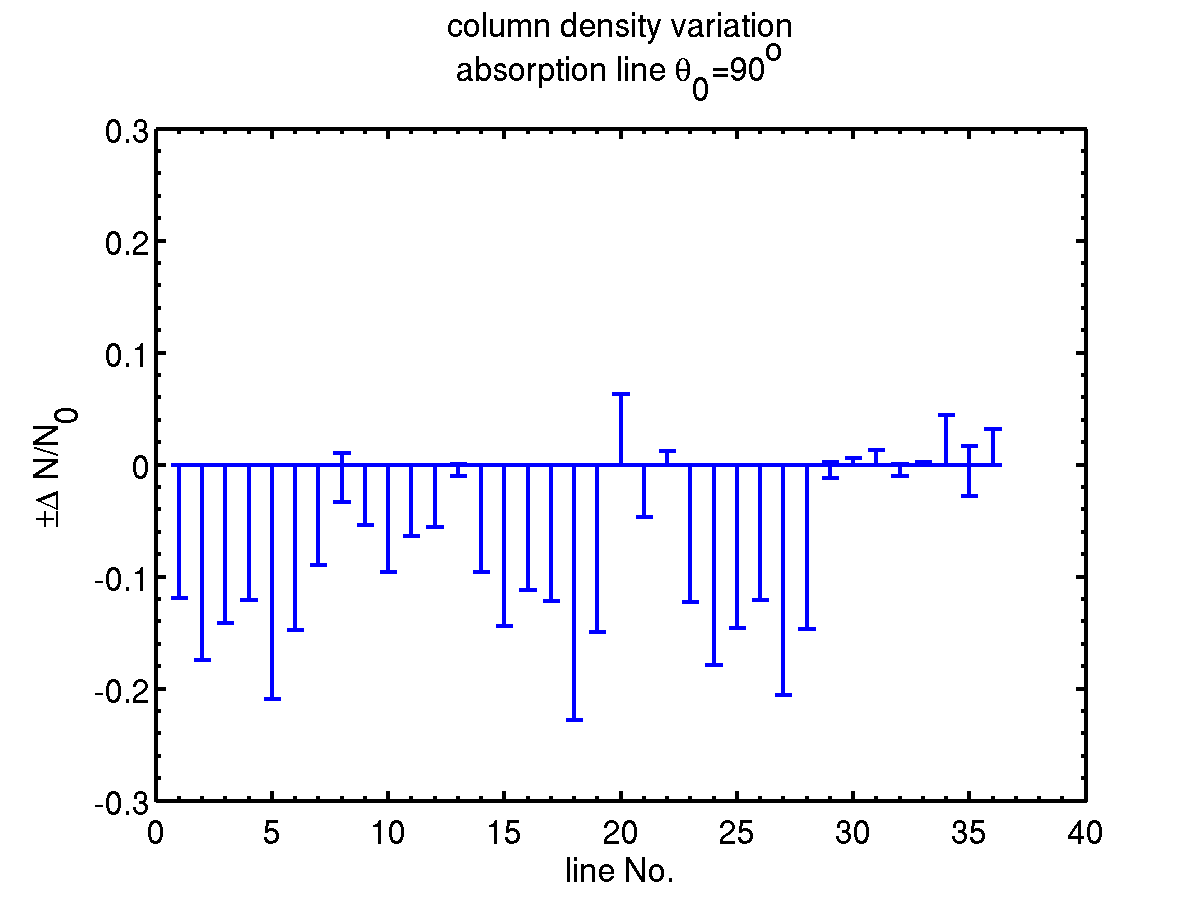}\label{varabc90}}
\caption{Examples on influence of atomic alignment on column density from absorption lines. Maximum and minimum column density variation deduced from absorption lines in blackbody radiation with $T_{source}=50000K$ for (a) $\theta_0=45^o$ and (b) $\theta_0=90^o$ are presented, respectively. Value above 0 means enrichment while below means depletion. Atoms are listed in the order of element number. The line numbers on $x-axis$ represent: $1\sim6$  C\,{\sc i} $\lambda\lambda1260.93, 1261.00, 1261.12, 1193.65, 1193.39, 1193.24\mbox{\AA}$; $7\sim9$, C\,{\sc ii} $\lambda\lambda1037.02, 1335.66, 1335.71\mbox{\AA}$; $10\sim11$, O\,{\sc i} $\lambda\lambda1025.76, 1027.43\mbox{\AA}$; $12\sim13$, Si\,{\sc ii} $\lambda\lambda1264.74, 1265.00\mbox{\AA}$;  $14\sim19$, S\,{\sc i} $\lambda\lambda1303.11, 1302.86, 1302.34, 1474.57, 1474.37, 1473.99\mbox{\AA}$; $20\sim22$, S\,{\sc ii} $\lambda\lambda1250.58,1253.81,1259.52\mbox{\AA}$; $23\sim28$, S\,{\sc iii} $\lambda\lambda1015.50, 1015.57, 1015.78, 1202.12, 1201.73, 1200.97\mbox{\AA}$; $29\sim30$, S \,{\sc iv} $\lambda\lambda1072.97, 1073.52\mbox{\AA}$;,$31\sim33$, Ti\,{\sc ii} $\lambda\lambda3072.97, 3241.98, 3229.19\mbox{\AA}$; $34\sim36$, Fe\,{\sc ii} $\lambda\lambda1142.36, 1143.22, 1144.923\mbox{\AA}$, respectively.\\
}\label{varabcp}
\end{figure*}

Fig. \ref{varabc45} and Fig. \ref{varabc90} provide an overview on the maximum and minimum column density variation from different absorption lines with (a) $\theta_0=45^o$ and (b) $\theta_0=90^o$, respectively. The more comprehensive results for maximum variation of column density for different absorption lines are presented in Table \ref{absorpretable}\footnote[7]{We present not only the absorption from the ground state, but also from some sublevels that have small energy difference with the ground state. Some of these lines can be detected at the places like galactic halos \citep{2013ApJ...772..111R,2014ApJ...787..147F} and circumburst medium in GRB\citep{2006A&A...451L..47F}. For example, according to Boltzmann distribution, when the medium temperature reaches the scale of a couple of hundred K, the metastable state C\,{\sc ii}$^{\ast}(J=3/2)$ will resides the same scale of atoms as the ground state C\,{\sc ii}$(J=1/2)$.}, where $\Delta N/N_{max}$ means enrichment and $\Delta N/N_{min}$ means depletion. Since chemical abundance studies always compare the log-scale of column density, the variations due to GSA on log-scale are also presented. For $\theta_0=90^o$,  S\,{\sc ii} $\lambda 1250.58$ can be enriched to the maximum of $7\%$ whereas S\,{\sc iv} $\lambda 1072.97$ can be depleted to the maximum of $27\%$; S\,{\sc i} $\lambda 1474.57$ can be depleted to the maximum of $22\%$ whereas the same line can be enriched to the maximum of $2\%$ (see Fig. \ref{varabc90}). Therefore it is obvious that magnetic fields have different influences on different atomic lines, as well as on the same atomic line with different direction of magnetic field.

\begin{figure*}
\centering
\includegraphics[width=0.7\columnwidth,
 height=0.5\textheight]{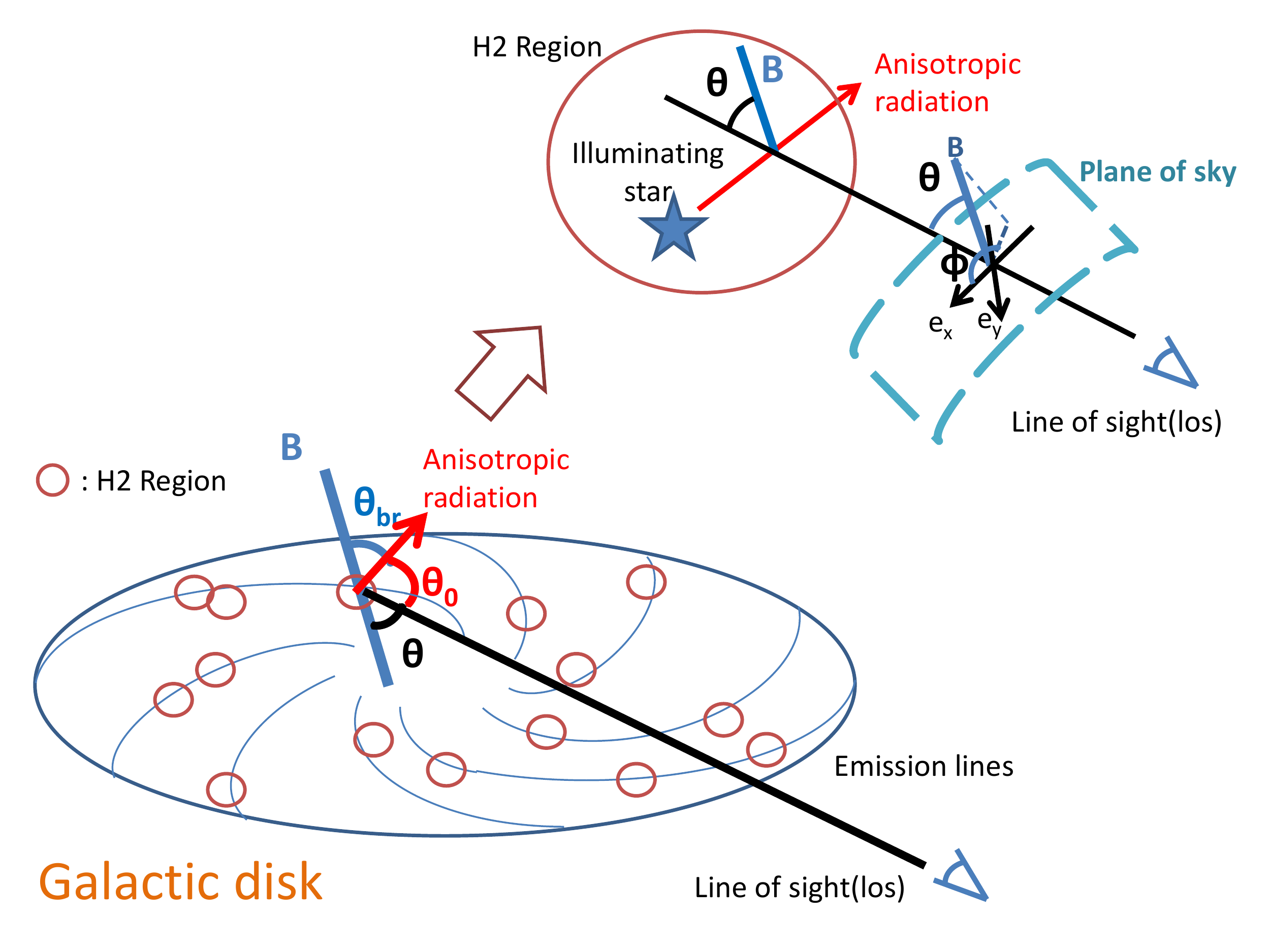}
\caption{Scenario for atomic alignment for emission lines in extragalactic H\,{\sc ii} region. Lower plot is a spiral galaxy with H\,{\sc ii} regions denoted in red circles. it is a typical environment where emission line is polarized owing to GSA. Alignment is produced by a pumping source, which both influence the polarization of scattering (emission) lines from the aligned medium. $\theta_{br},\theta$ are the polar angles of the pumping radiation and line of sight measured in reference to the magnetic field, respectively. $\theta_0$ is the angle between the anisotropic radiation and the line of sight. The upper plot depicts the influence of alignment on the scaterring emission lines from H\,{\sc ii} region in detail. Stars illuminate surrounding medium which gives emission lines. The whole process is influenced by the anisotropic radiation and magnetic field due to GSA.}
\label{scenaem}
\end{figure*}

\subsection{Magnetic influence on emission spectra}

Atomic emission spectra are very important in astrophysical studies. UV emission lines are used to model stellar wind \citep{2016ApJ...823...64L}. Optical recombination emission lines can help the analysis of chemical abundance in H\,{\sc ii} regions and Planetary Nebulae (PNe) (e.g., \citealt{2016ApJ...824L..27G,2006IAUS..234..227P}), and they have the advantages that they are not severely affected by the uncertainty of reddening correction as collision excited lines and they can directly analyze the ionized gas phase chemical (e.g., \citealt{2014MNRAS.443..624E}). Combined with absorption spectra, emission spectra can provide even more information, e.g., gas condition in DLAs (e.g., \citealt{2011ApJ...732...35M,2014ApJ...780...76K}).

\subsubsection{Scenario for emission spectra}

As an example, we show in Fig.\ref{scenaem} a typical scenario, where GSA affects the observational emission data. Presented is a typical spiral galaxy with diffuse emission nebulae (H\,{\sc ii} regions, which mostly reside on the spiral arms of the galaxies) denoted in red circle and an interstellar magnetic field on the disk. Stars in the H\,{\sc ii} region ionise and illuminate the surrounding medium, from which the emission lines can be observed. Radiation field here\footnote[8]{As illustrated in \citet{ZYD15}, alignment also exists in extended radiation source like cluster of stars. For simplicity, the radiation source considered here has been treated as a single pumping star.} is expected to have a certain degree of anisotropy. With the influence of magnetic fields and anisotropic radiation field, the ground state of the atoms in diffuse gas will be aligned and the alignment will be transferred to upper states due to optical excitation. Thus the atomic alignment will cause measurable changes in the emission spectra observed. Small density of the diffuse medium in H\,{\sc ii} region promises a much smaller collisional rate than Larmor precession rate, which makes atomic alignment effective. Latter part of this section will present the variation of the emission line intensity with the influence of GSA.

\begin{figure*}
\centering
\subfigure[]{
\includegraphics[width=0.32\columnwidth,
 height=0.25\textheight]{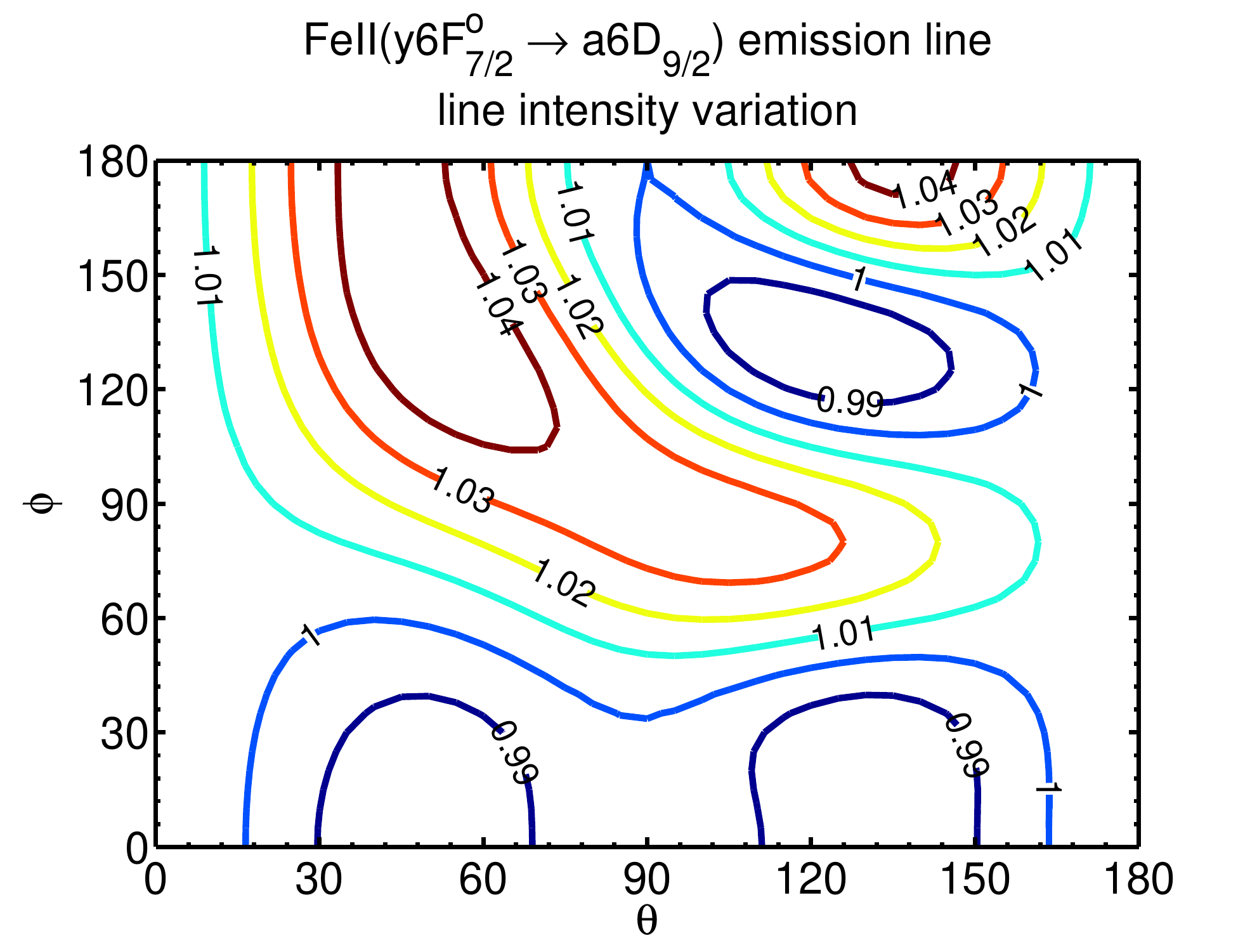}\label{Fe2bbremv7to9y6fc90}}
\subfigure[]{
\includegraphics[width=0.32\columnwidth,
 height=0.25\textheight]{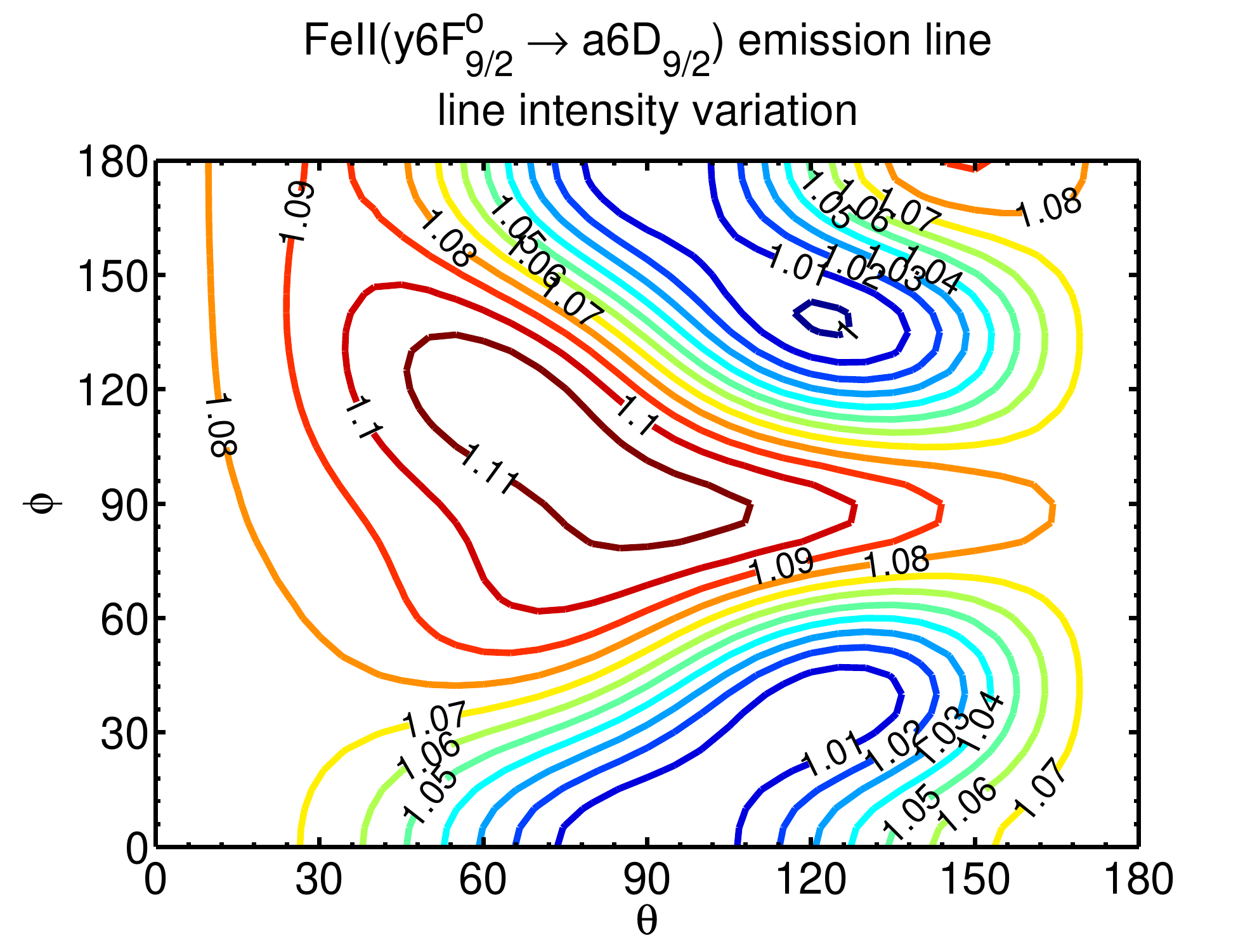}\label{Fe2bbremv9to9y6fc90}}
\subfigure[]{
\includegraphics[width=0.32\columnwidth,
 height=0.25\textheight]{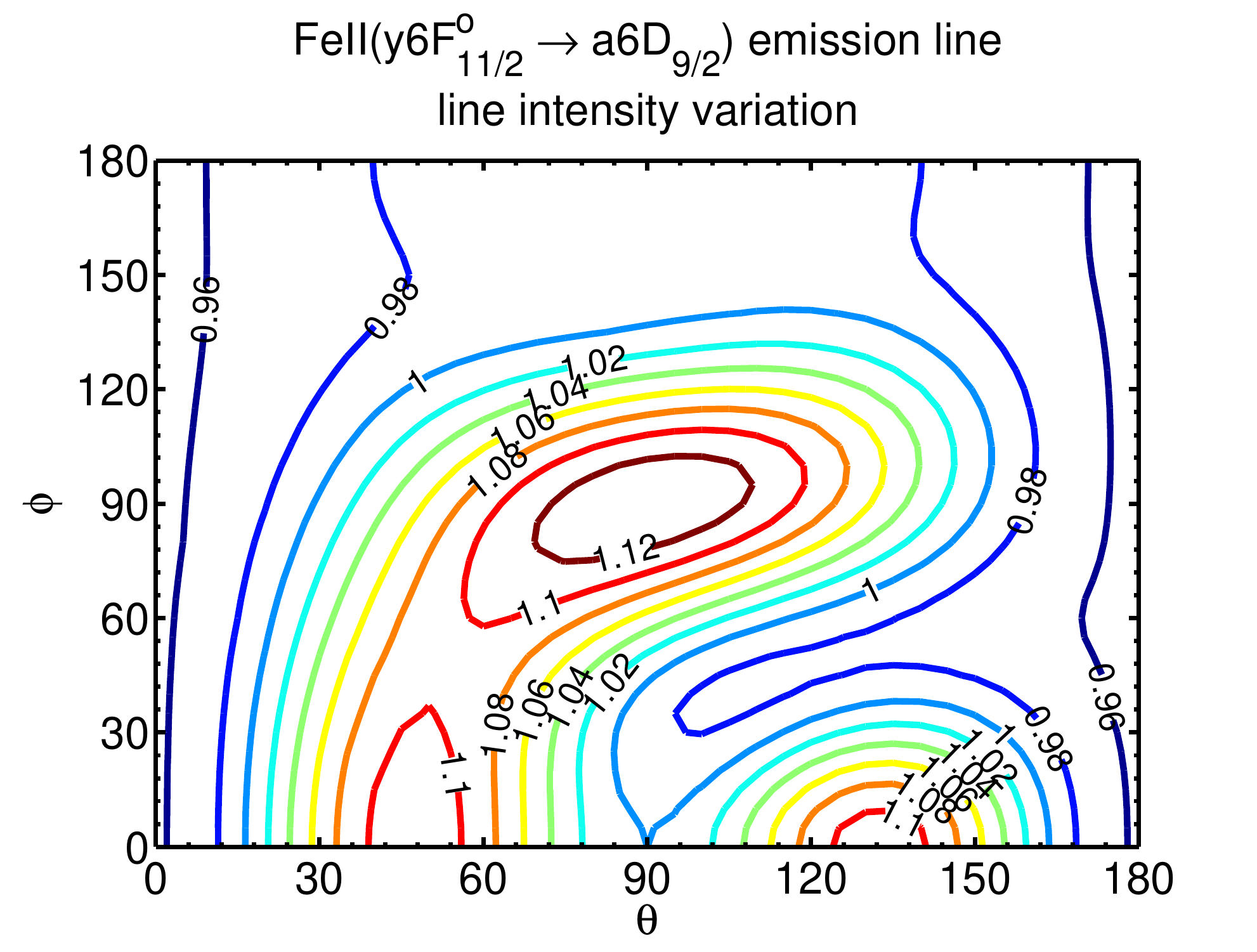}\label{Fe2bbremv11to9y6fc90}}
\caption{Fe\,{\sc ii} emission line in blackbody radiation with $T_{source}=50000K$. Different contour represents the wavelength (a)1142.36\mbox{\AA}; (b)1143.22\mbox{\AA}; (c)1144.93\mbox{\AA} in the case of line of sight vertical to radiation field($\theta_0=90^{\circ}$).}\label{Fe2bbremcp}
\end{figure*}

\subsubsection{Influence on emission lines}

Different from the case of absorption lines, atomic alignment for emission lines is a scattering process. The magnetic realignment on upper levels can be neglected when the precession rate of magnetic fields is smaller than the rate of emission from upper level. The alignment on the ground level can be transferred by optical excitation to upper levels, which then radiate spontaneous emission and the whole precess can be considered as a scattering process. Thus the observation of emission lines from these excited states is also influenced by the atomic alignment (see details in \citealt{YLhyf}). Given $i=0$ in Eq. \ref{epsiloni}, the intensity of the emission lines can be expressed by:
\begin{equation}\label{intenemi}
I_{em}(\nu,\Omega)=\frac{h\nu_0}{4\pi}An(J_u,\theta_{br})\Psi(\nu-\nu_0)\sum_{\substack{KQ}}\omega^{K}_{J_uJ_l}\sigma^K_Q(J_u,\theta_{br})\mathcal{J}^K_Q(0,\Omega).
\end{equation}

Let's denote the real intensity of emission line as $I^{em}_{re}$; and the observed intensity of emission line as $I^{em}_{ob}$. The ratio between observed line intensity and real line intensity for emission lines is defined by
\begin{equation}\label{ratioem1}
r^{em}(\lambda,\theta_0,\theta_{br},\theta)=I_{ob}/I_{re}\propto\left(\rho^0_0(J_u)\sqrt{2}+\sum_{\substack{Q}}\omega^2_{J_uJ_l}\rho^2_Q(J_u)\mathcal{J}^K_Q(0,\Omega)\right),
\end{equation}
where $\lambda$ is the wavelength of the line, $\theta_0,\theta_{br},\theta$ are defined in Fig. \ref{scea}, $Q=0,\pm1,\pm2$.

Fe\,{\sc ii} emission lines are presented as an example. The ground state of Fe\,{\sc ii} has 5 sublevels $a6D_{\frac{1}{2};\frac{3}{2};\frac{5}{2};\frac{7}{2};\frac{9}{2}}$, in which the ground level is $a6D_{J_l=\frac{9}{2}}$. The upper states of Fe\,{\sc ii} are $y6F^o_{\frac{1}{2};\frac{3}{2};\frac{5}{2};\frac{7}{2};\frac{9}{2};\frac{11}{2}}$, $z6D^o_{\frac{1}{2};\frac{3}{2};\frac{5}{2};\frac{7}{2};\frac{9}{2}}$, $z6F^o_{\frac{1}{2};\frac{3}{2};\frac{5}{2};\frac{7}{2};\frac{9}{2};\frac{11}{2}}$, $z6P^o_{\frac{3}{2};\frac{5}{2};\frac{7}{2}}$, $6D^o_{\frac{3}{2};\frac{5}{2};\frac{7}{2};\frac{9}{2}}$, etc. Two coordinate systems based on line of sight and radiation as z-axis respectively are compared in Fig. \ref{Fe2bbremcp}. The corresponding Fe\,{\sc ii} emission line variation with the wavelength $\lambda\lambda1142.36, 1143.22, 1144.93$ when $\theta_0=90^\circ$ radiation source temperature $T_{source}=50000K$ are compared in Fig. \ref{Fe2bbremv7to9y6fc90}, Fig. \ref{Fe2bbremv9to9y6fc90}, and Fig. \ref{Fe2bbremv11to9y6fc90}. These lines are from the excited state $y6F^o$ with different sublevels $y6F^o_{\frac{7}{2};\frac{9}{2};\frac{11}{2}}$ to the ground state $a6D_{\frac{9}{2}}$. It is obvious that with the same direction of incidental radiation and line of sight, the influence of the magnetic fields on different emission spectra is different: when $\theta=15^\circ$, $\phi=120^\circ$, GSA enriches $\lambda\lambda1142.36$ by $1\%$, enriches $\lambda\lambda1143.22$ by $8\%$ and depletes $\lambda\lambda1144.93$ by $2\%$.

\begin{figure*}
\centering
\subfigure[]{
\includegraphics[width=0.47\columnwidth,
 height=0.35\textheight]{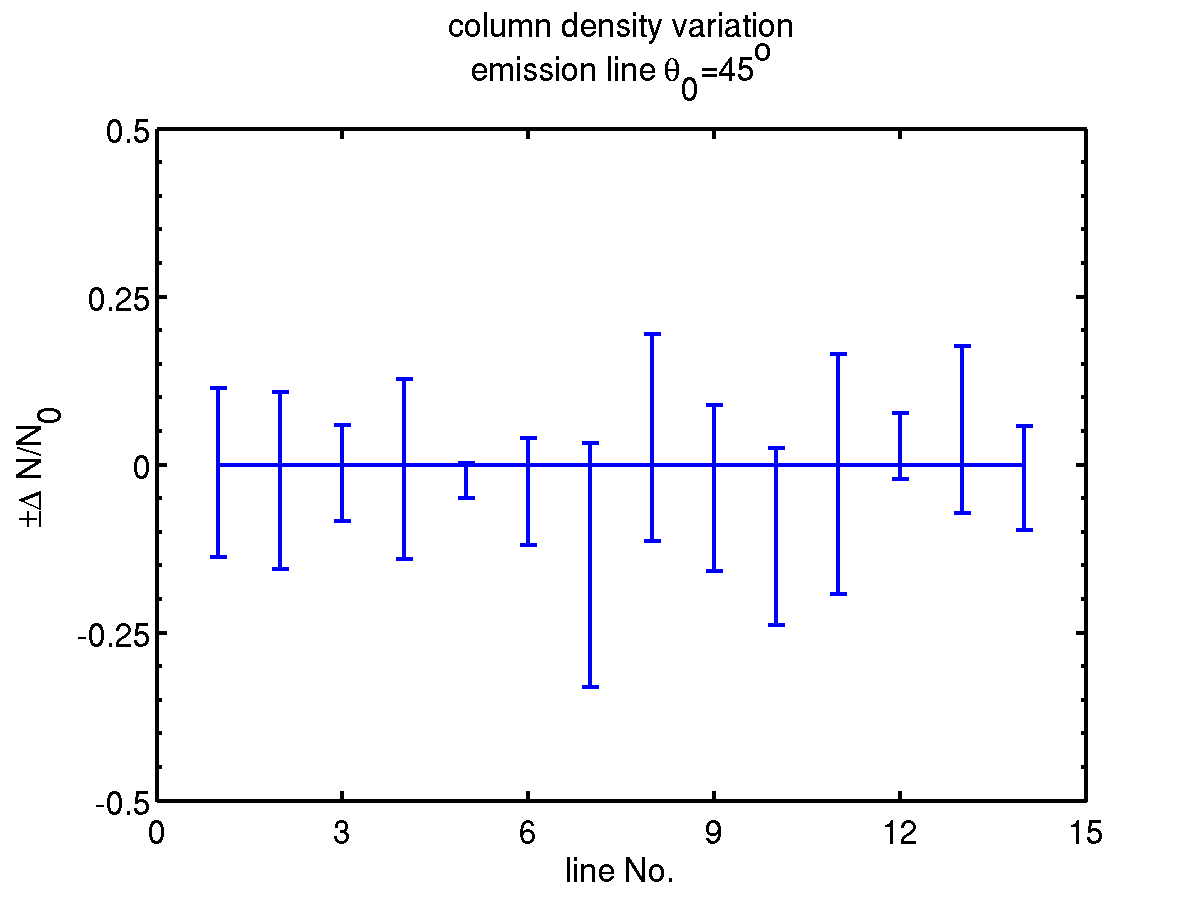}\label{varemc45}}
 \subfigure[]{
\includegraphics[width=0.47\columnwidth,
 height=0.35\textheight]{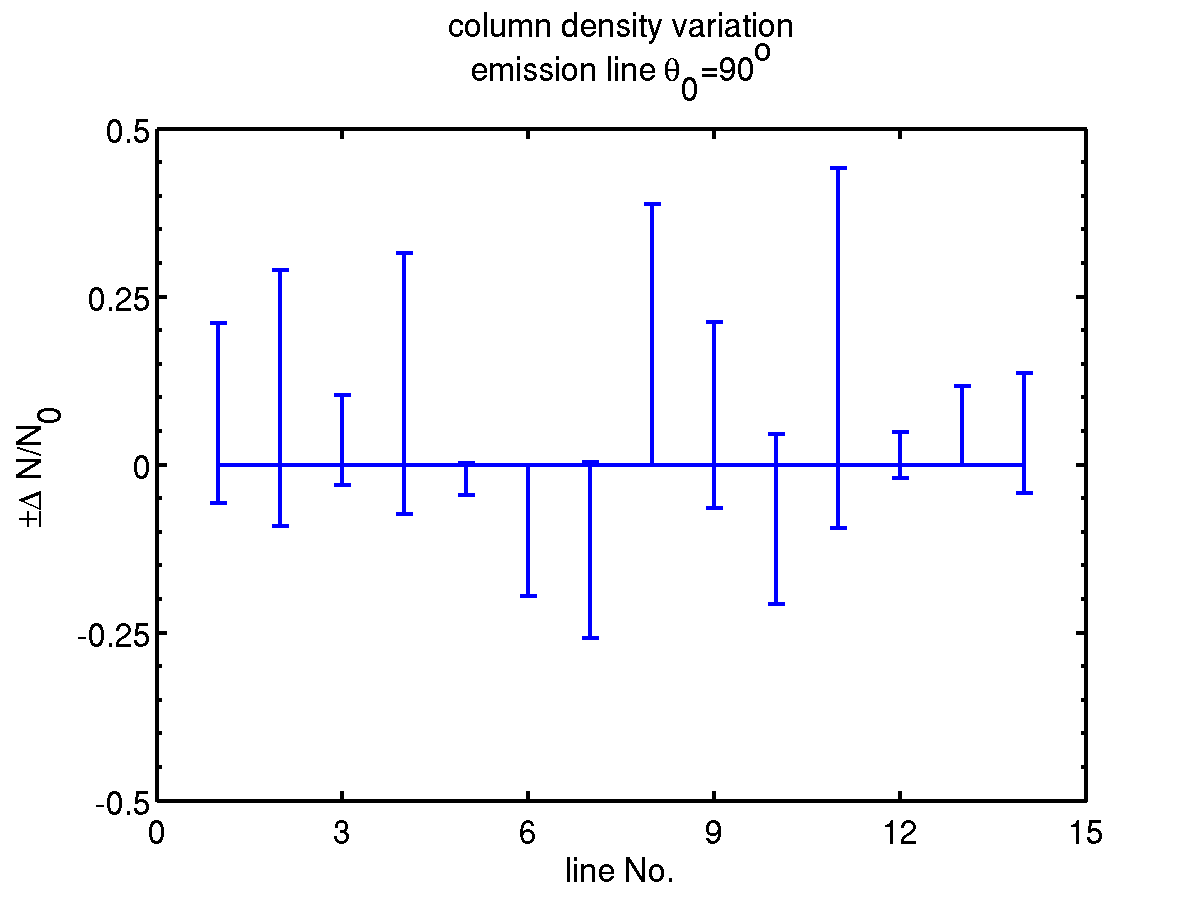}\label{varemc90}}
\caption{Examples on influence of atomic alignment on column density from emission lines. Maximum and minimum column density variations deduced from emission lines in blackbody radiation with $T_{source}=50000K$ for (a) $\theta_0=45^o$ and (b) $\theta_0=90^o$ are presented, respectively. Value above 0 means enrichment while below means depletion. Atoms are listed in the order of element number. Only transitions that have higher probability (hereby, to the ground state mostly) are considered. The line numbers on $x-axis$ represent: $1$, C\,{\sc i} $\lambda 1656.93\mbox{\AA}$; $2$, C\,{\sc ii} $\lambda 1334.53\mbox{\AA}$; $3$, O\,{\sc i} $\lambda 1025.76\mbox{\AA}$; $4$, Si\,{\sc ii} $\lambda 1190.42\mbox{\AA}$; $5\sim7$, S\,{\sc i} $\lambda\lambda1474.57, 1474.37, 1473.99\mbox{\AA}$; $8\sim9$, S\,{\sc ii} $\lambda\lambda1253.81, 1259.52\mbox{\AA}$;  $10$, S\,{\sc iii} $\lambda 1012.49\mbox{\AA}$; $11$, S\,{\sc iv} $\lambda 1062.66\mbox{\AA}$; $12\sim14$, Fe\,{\sc ii} $\lambda\lambda1142.36, 1143.22, 1144.93\mbox{\AA}$, respectively.\\}\label{varemcp}
\end{figure*}

Moreover, even for the same spectrum with the same radiation field observed from the same line of sight (hereby, the same $\theta_0$), the line intensity observed can be different only due to the change of magnetic fields. For example, as demonstrated in Fig. \ref{Fe2bbremv9to9y6fc90}, the magnetic field almost does not influence the spectrum $\theta=120^{\circ}, \phi=120^\circ$, whereas enriches more than $10\%$ of the spectrum when the direction of the magnetic field is perpendicular to both the line of sight and incident radiation ($\theta=90^{\circ}, \phi=90^{\circ}$). Therefore,taking into consideration of GSA makes the spectrum observed vary with respect to different direction of magnetic field.

Fig. \ref{varemc45} and Fig. \ref{varemc90} provide an overview on the maximum and minimum intensity variation of different emission lines with  (a) $\theta_0=45^o$ and (b) $\theta_0=90^o$, respectively. The more comprehensive results for maximum variation of intensity for defferent lines are presented in Table \ref{emiretable}, where $\Delta I/I_{max}$ means enrichment and $\Delta I/I_{min}$ means depletion. Variations on log-scale are also presented. For $\theta_0=90^o$, S\,{\sc ii} $\lambda\lambda 1259.52$ can be depleted to the maximum of more than $6\%$ whereas the same line can be enriched to the maximum of $22\%$;  S\,{\sc i} $\lambda\lambda 1473.99$ can be depleted to the maximum of more than $25\%$ whereas the line \,{\sc ii} $\lambda\lambda 1253.81$ can be enriched to the maximum of $38\%$ , as demonstrated in Fig. \ref{varemc90}. Thus it is clear that magnetic fields have different influences on different atomic lines, as well as on the same atomic line with different direction of magnetic field.

\begin{table}[!hbp]
\centering
\caption{LINE INTENSITY VARIATION FOR EMISSION LINE INDUCED BY GSA}\label{emiretable}
\begin{tabular}{|c|c|c|c|c|c|c|}
\hline \hline
Species & Transition & Wavelength & $\Delta I/I_{min}$ & $\Delta I/I_{max}$ & $\Delta$ log$(N/N_0)_{min}$ & $\Delta$ log$(N/N_0)_{max}$ \\
\hline
C\,{\sc i} & $3D^{\circ}_{1}\rightarrow 3P_{0}$ & $1193.03\mbox{\AA}$ & $-20.85\%$ & $+21.00\%$ & $-1.02\times10^{-1}$ & $8.28\times10^{-2}$ \\
\hline
C\,{\sc ii} & $2S_{\frac{3}{2}}\rightarrow 2P^{\circ}_{\frac{1}{2}}$ & $1334.53\mbox{\AA}$ & $-29.17\%$ & $+28.90\%$ & $-1.50\times10^{-1}$ & $1.1\times10^{-1}$ \\
\hline
O\,{\sc i} & $3D^{\circ}\rightarrow 3P_{2}$ & $1025.76\mbox{\AA}$ & $-14.30\%$ & $+10.44\%$ & $-6.70\times10^{-2}$ & $4.31\times10^{-2}$ \\
\hline
Si\,{\sc ii} & $2P_{\frac{3}{2}}\rightarrow 2P^{\circ}_{\frac{1}{2}}$ & $1190.42\mbox{\AA}$ & $-27.20\%$ & $+31.49\%$ & $-1.38\times10^{-1}$ & $1.19\times10^{-1}$ \\
\hline
\multirow{3}{*}{S\,{\sc i}} & $3D^{\circ}_{1}\rightarrow 3P_{2}$ & $1474.57\mbox{\AA}$ & $-6.32\%$ & $+0.50\%$ & $-2.84\times10^{-2}$ & $2.17\times10^{-3}$ \\
\cline{2-7}
& $3D^{\circ}_{2}\rightarrow 3P_{2}$ & $1474.38\mbox{\AA}$ & $-19.62\%$ & $+7.49\%$ & $-9.49\times10^{-2}$ & $3.13\times10^{-2}$ \\
\cline{2-7}
& $3D^{\circ}_{3}\rightarrow 3P_{2}$ & $1473.99\mbox{\AA}$ & $-38.67\%$ & $+3.89\%$ & $-2.12\times10^{-1}$ & $1.66\times10^{-2}$ \\
\hline
\multirow{3}{*}{S\,{\sc ii}} & $4P_{\frac{3}{2}}\rightarrow 4S^{\circ}_{\frac{3}{2}}$ & $1253.81\mbox{\AA}$ & $-20.51\%$ & $+38.79\%$ & $-9.97\times10^{-2}$ & $1.42\times10^{-1}$ \\
\cline{2-7}
& $4P_{\frac{5}{2}}\rightarrow 4S^{\circ}_{\frac{3}{2}}$ & $1259.52\mbox{\AA}$ & $-23.04\%$ & $+21.27\%$ & $-1.14\times10^{-1}$ & $8.38\times10^{-2}$ \\
\hline
S\,{\sc iii} & $3P^{\circ}_{1}\rightarrow 3P_{0}$ & $1012.50\mbox{\AA}$ & $-32.02\%$ & $+6.01\%$ & $-1.68\times10^{-1}$ & $2.54\times10^{-2}$ \\
\hline
S\,{\sc iv} & $2D_{\frac{3}{2}}\rightarrow 2P^{\circ}_{\frac{1}{2}}$ & $1062.66\mbox{\AA}$ & $-35.44\%$ & $+44.09\%$ & $-1.90\times10^{-1}$ & $1.59\times10^{-1}$ \\
\hline
\multirow{3}{*}{Fe\,{\sc ii}} & $y6F^{\circ}_{\frac{7}{2}}\rightarrow a6D_{\frac{9}{2}}$ & $1142.36\mbox{\AA}$ & $-2.90\%$ & $+12.10\%$ & $-1.28\times10^{-2}$ & $4.96\times10^{-2}$ \\
\cline{2-7}
& $y6F^{\circ}_{\frac{9}{2}}\rightarrow a6D_{\frac{9}{2}}$ & $1143.22\mbox{\AA}$ & $-9.37\%$ & $+22.44\%$ & $-4.27\times10^{-2}$ & $8.79\times10^{-2}$ \\
\cline{2-7}
& $y6F^{\circ}_{\frac{11}{2}}\rightarrow a6D_{\frac{9}{2}}$ & $1144.93\mbox{\AA}$ & $-14.38\%$ & $+13.64\%$ & $-6.74\times10^{-2}$ & $5.55\times10^{-2}$ \\
\hline
\end{tabular}
\end{table}

\subsection{Magnetic influence on submilimetre fine structure lines}

Submillimeter fine structure spectra, which arise from magnetic dipole transitions, have a broad applicability in astrophysics. Submillimeter lines can be used to decide chemical properties of distant star-forming galaxies and derive electron and ionization temperature of galaxy \citep{1999ApJ...514..544K}. They can be applied to predicting star burst size \citep{2013ApJ...774...68D}. They also help the modelling of flaring Herbig Ae/Be disks \citep{2013A&A...555A..64M}.

However, previous studies do not consider atomic alignment, which in two aspects influences the results. First, in real circumstances, the assumption that radiation field is isotropic is problematic. For example, in protoplanetary disks,  dominant radiation source could be localized (see Appendix A in \citealt{2016arXiv160101449K} for details). Moreover, the magnetic realignment has to be taken into consideration. Thus, as discussed in \S 2, emission and absorption coefficients for Stokes parameters will be modified due to atomic alignment.

\begin{figure*}
\centering
\subfigure[]{
\includegraphics[width=0.47\columnwidth,
 height=0.35\textheight]{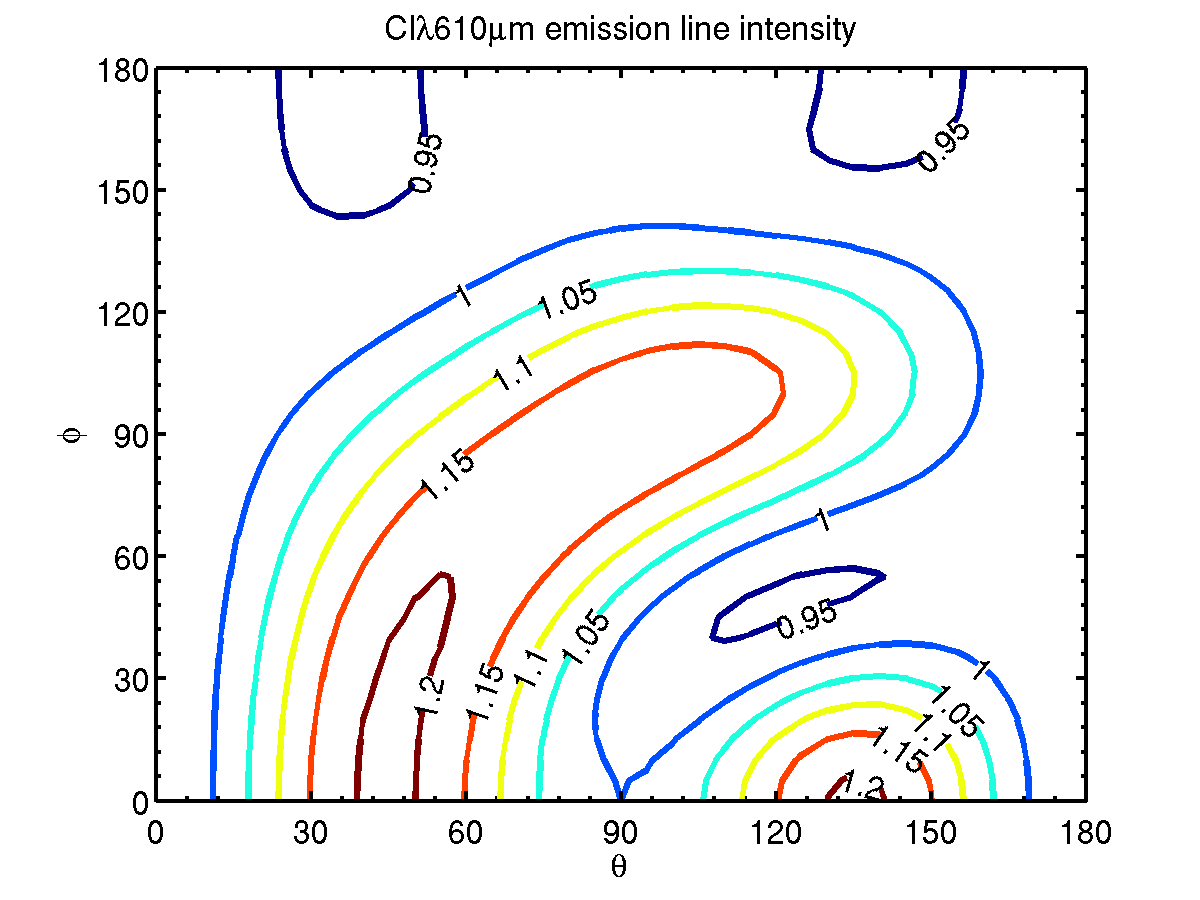}\label{C1bbrsubem1c90}}
\subfigure[]{
\includegraphics[width=0.47\columnwidth,
 height=0.35\textheight]{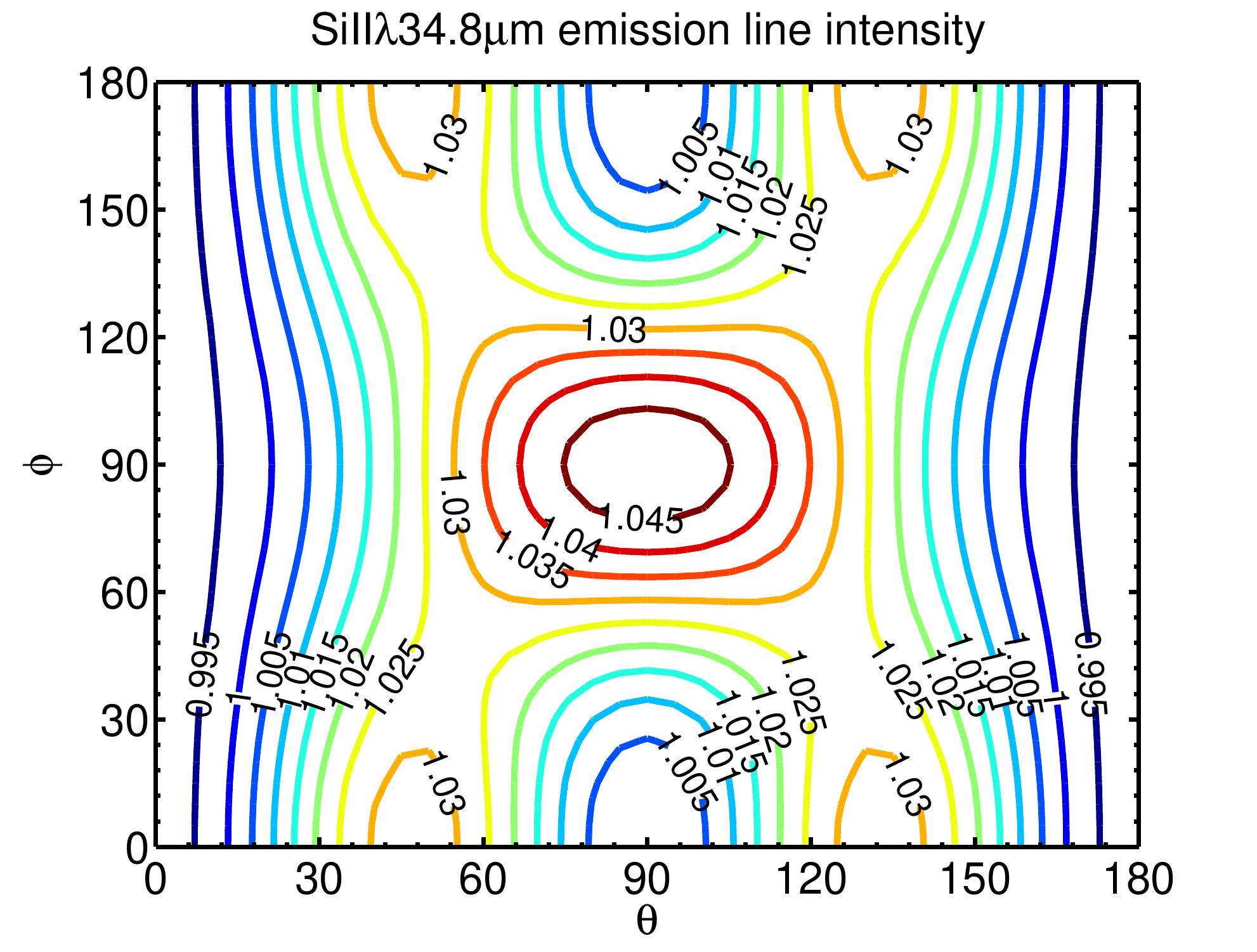}\label{Si2bbrsubem1c90}}
\caption{The variation of line intensity for submillimeter lines with respect to the direction of the magnetic field with $T_{source}=50000K$. Contours represent the line intensity variation of (a) C\,{\sc i} $\lambda610{\mu}m$ and (b) Si\,{\sc ii} $\lambda34.8{\mu}m$ with the change of magnetic field due to GSA when $\theta_0=90^\circ$.}\label{subC1}
\end{figure*}

For example, the influence of atomic alignment on C\,{\sc i}  $\lambda610{\mu}m$ and  Si\,{\sc ii} $\lambda34.8{\mu}m$ lines are presented in Fig. \ref{subC1}. The ground state of C\,{\sc i} has 3 sublevels $3P_{0;1;2}$, in which the ground level is $3P_{J_l=0}$. C\,{\sc i}  $\lambda610{\mu}m$ line represents the transition within lower sublevels $3P_{J_l=1}$ and $3P_{J_l=0}$. The ground state of Si\,{\sc ii} has 2 sublevels $2P^o_{\frac{1}{2};\frac{3}{2}}$, in which the ground level is $2P_{J_l=\frac{1}{2}}^o$. Si\,{\sc ii} $\lambda34.8{\mu}m$ line represents the transition within lower sublevels $2P_{J_l=\frac{3}{2}}^o$ and $2P_{J_l=\frac{1}{2}}^o$. As demonstrated in Fig. \ref{C1bbrsubem1c90}, atomic alignment can enrich C\,{\sc i}$\lambda\lambda610{\mu}m$ by $20\%$ when $\theta=45^{\circ}, \phi=0^{\circ}$, while deplete the same line by $5\%$ when $\theta=120^{\circ}, \phi=45^{\circ}$. In addition, by the comparison between Fig. \ref{C1bbrsubem1c90} and Fig. \ref{Si2bbrsubem1c90}, atomic alignment can enrich C\,{\sc i}  $\lambda610{\mu}m$ line by a factor of more than $20\%$ and enrich Si\,{\sc ii} $\lambda34.8{\mu}m$ line by a factor of around $2\%$ when $\theta=45^{\circ}, \phi=45^{\circ}$. Thus, atomic alignment not only has an influence on submillimeter lines obtained, but also influences different lines differently.

\begin{figure*}
\centering
\subfigure[]{
\includegraphics[width=0.47\columnwidth,
 height=0.35\textheight]{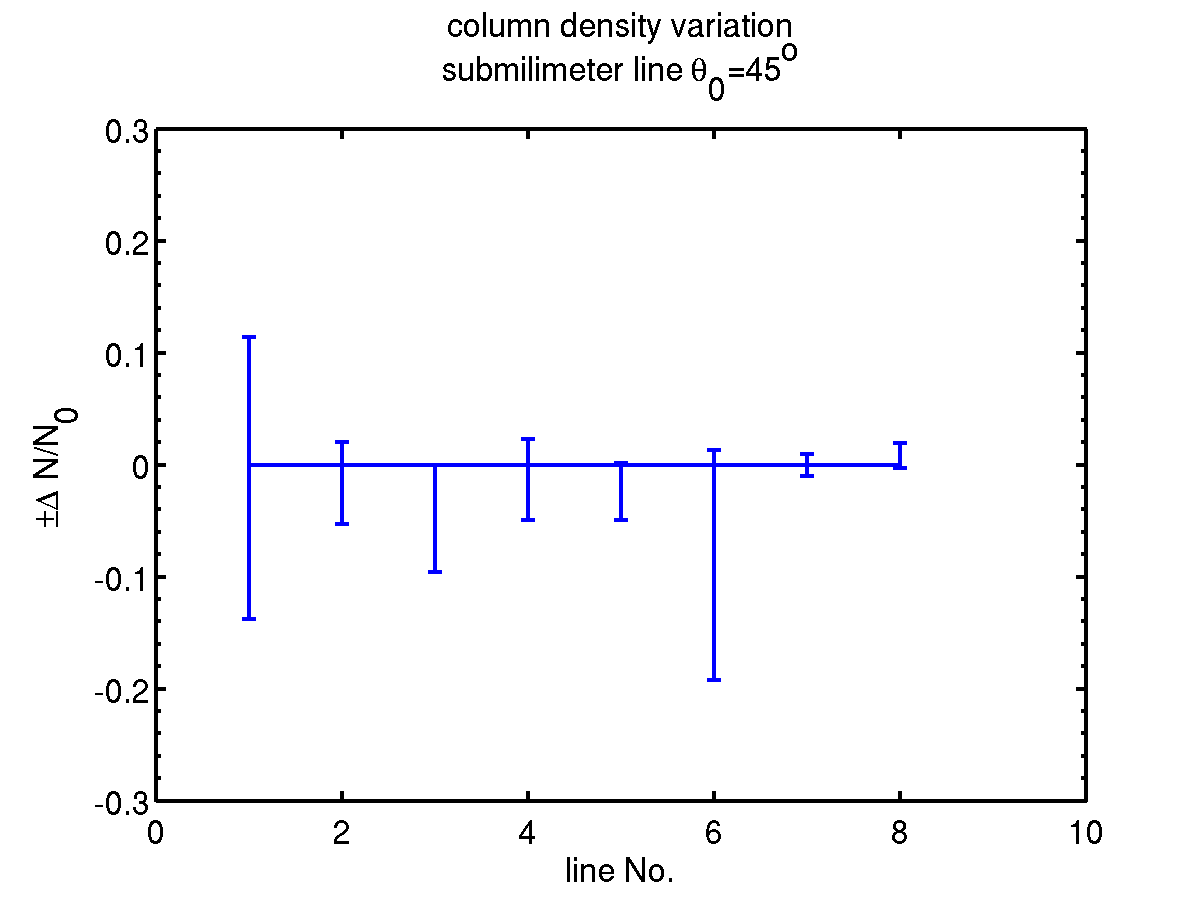}\label{varsubc45}}
 \subfigure[]{
\includegraphics[width=0.47\columnwidth,
 height=0.35\textheight]{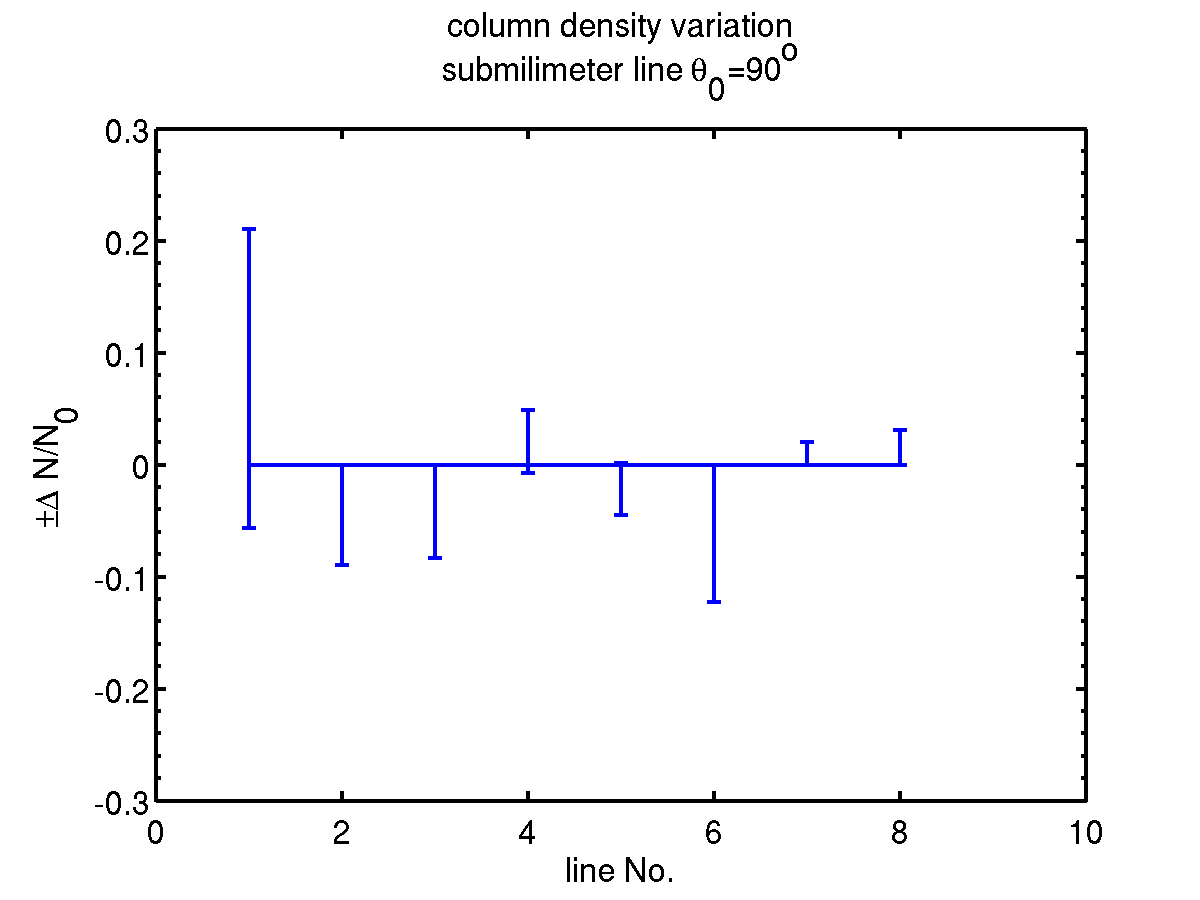}\label{varsubc90}}
\caption{Examples on influence of atomic alignment on column density from submillimeter lines. Maximum and minimum column density variation deduced from submillimeter lines in blackbody radiation with $T_{source}=50000K$ for (a) $\theta_0=45^o$ and (b) $\theta_0=90^o$ are presented, respectively.
Value above 0 means enrichment while below means depletion. Atoms are listed in the order of element number. Only transitions to the ground level is considered. The line numbers on $x-axis$ represent: $1$, C\,{\sc i} $\lambda 610\mu m$; $2$, C\,{\sc ii} $\lambda 157.7\mu m$; $3$, O\,{\sc i} $\lambda 63.2\mu m$; $4$, Si\,{\sc ii} $\lambda 34.8\mu m$; $5$, S\,{\sc i} $\lambda 25.2\mu m$; $6$, S\,{\sc iii} $\lambda 33.5\mu m$; $7$, S\,{\sc iv} $\lambda 10.5\mu m$; $8$, Fe\,{\sc ii} $\lambda 26.0\mu m$, respectively.}\label{varsubcp}
\end{figure*}

Fig. \ref{varsubc45} and Fig. \ref{varsubc90} provide an overview on the maximum and minimum column density  variation from different submillimeter lines with  (a) $\theta_0=45^o$ and (b) $\theta_0=90^o$, respectively .Table \ref{submiretable} is provided to present the influence of atomic alignment on different submillimeter lines, where $\Delta I/I_{max}$ means enrichment and $\Delta I/I_{min}$ means depletion. It's clear that magnetic fields have different influences on different atomic lines, as well as on the same atomic line with different direction of magnetic field. For example, for  $\theta_0=45^o$,  S\,{\sc iii} $\lambda 33.5{\mu}m$ can be enriched to the maximum of more than $2\%$ whereas the same line can be depleted to the maximum of almost $20\%$ only because of the different direction of the magnetic field; C\,{\sc i} $\lambda 610{\mu}m$ can be enriched to the maximum of more than $11\%$ whereas O\,{\sc i} $\lambda 63.2{\mu}m$ can be depleted to the maximum of almost $10\%$, as illustrated in Fig. \ref{varsubc45}.

\begin{table}[!hbp]
\centering
\caption{LINE INTENSITY VARIATION FOR submillimeter LINE INDUCED BY GSA}\label{submiretable}
\begin{tabular}{|c|c|c|c|c|c|c|}
\hline \hline
Species & Transition & Wavelength & $\Delta I/I_{min}$ & $\Delta I/I_{max}$ & $\Delta$ log$(N/N_0)_{min}$ & $\Delta$ log$(N/N_0)_{max}$ \\
\hline
[C\,{\sc i}] & $3P_{1}\rightarrow 3P_{0}$ & $610\mu m$ & $-20.85\%$ & $+22.00\%$ & $1.02\times10^{-1}$ & $8.28\times10^{-2}$ \\
\hline
[C\,{\sc ii}] & $2P^{\circ}_{3/2}\rightarrow 2P^{\circ}_{1/2}$ & $157.7\mu m$ & $-9.02\%$ & $+3.68\%$ & $4.10\times10^{-2}$ & $1.57\times10^{-2}$ \\
\hline
[O\,{\sc i}] & $3P_{1}\rightarrow 3P_{2}$ & $63.2\mu m$ & $-10.77\%$ & $+0\%$ & $4.95\times10^{-2}$ & $0$ \\
\hline
[Si\,{\sc ii}] & $2P^{\circ}_{3/2}\rightarrow 2P^{\circ}_{1/2}$ & $34.8\mu m$ & $-8.76\%$ & $+4.90\%$ & $3.98\times10^{-2}$ & $2.08\times10^{-2}$ \\
\hline
[S\,{\sc i}] & $3P_{1}\rightarrow 3P_{2}$ & $25.2\mu m$ & $-6.32\%$ & $+4.90\%$ & $2.84\times10^{-2}$ & $2.2\times10^{-3}$ \\
\hline
[S\,{\sc iii}] & $3P^{\circ}_{1}\rightarrow 3P_{0}$ & $33.5\mu m$ & $-24.67\%$ & $+1.53\%$ & $1.23\times10^{-1}$ & $6.58\times10^{-3}$ \\
\hline
[S\,{\sc iv}] & $2P^{\circ}_{3/2}\rightarrow 2P^{\circ}_{1/2}$ & $10.5\mu m$ & $-2.04\%$ & $+2.00\%$ & $8.96\times10^{-3}$ & $8.59\times10^{-3}$ \\
\hline
[Fe\,{\sc ii}] & $a6D_{7/2}\rightarrow a6D_{9/2}$ & $26.0\mu m$ & $-0.39\%$ & $+3.11\%$ & $1.71\times10^{-3}$ & $1.33\times10^{-2}$ \\
\hline
\end{tabular}
\end{table}

\section{Influence of GSA on physical parameters derived from line ratio}

Many important physical parameters in astronomy are derived from spectral line ratio. For example, The metastable level of C\,{\sc ii}$^{\ast}$ has an angular momentum $J=3/2$ and an energy $7.8\times10^{-3}$ eV above the ground the level and the ratio between the absorption line from  C\,{\sc ii}$^{\ast}(J=3/2)$ (e.g., $\lambda 1036.34$ or $\lambda 1335.66$) and from the ground level C\,{\sc ii}$(J=1/2)$ (e.g., $\lambda 1334.53$) are used to estimate the electron density in different astrophysical environments (see \citealt{2004ApJ...615..767L,2008ApJ...679..460Z,2013ApJ...772..111R} for details). According to Eq. \eqref{ratioab1}, the ratio of line intensity between two different lines observed from one direction for absorption line is defined by
\begin{equation}\label{ratioab2}
\begin{split}
\mathcal{R}^{ab}_{\lambda_1,\lambda_2}(\theta_0,\theta_{br},\theta)&=r^{ab}(\lambda_1,\theta_0,\theta_{br},\theta)/r^{ab}(\lambda_2,\theta_0,\theta_{br},\theta),\\
&\propto\frac{\rho^0_0(J'_l)\sqrt{2}+\omega^2_{J'_lJ'_u}\rho^2_0(J'_l)\left(1-1.5\sin^2\theta'\right)}{\rho^0_0(J_l)\sqrt{2}+\omega^2_{J_lJ_u}\rho^2_0(J_l)\left(1-1.5\sin^2\theta\right)}.
\end{split}
\end{equation}

In addition, the ratio of line intensity of emission line between two different lines observed from one direction can be obtained from Eq. \eqref{ratioem1}:
\begin{equation}\label{ratioem2}
\begin{split}
\mathcal{R}^{em}_{\lambda_1,\lambda_2}(\theta_0,\theta_{br},\theta)&=r^{em}(\lambda_1,\theta_0,\theta_{br},\theta)/r^{em}(\lambda_2,\theta_0,\theta_{br},\theta),\\
&\propto\frac{\left(\rho^0_0(J'_u)\sqrt{2}+\sum_{\substack{Q}}\omega^2_{J'_uJ'_l}\rho^2_Q(J'_u)\mathcal{J}^K_Q(0,\Omega)\right)}{\left(\rho^0_0(J_u)\sqrt{2}+\sum_{\substack{Q}}\omega^2_{J_uJ_l}\rho^2_Q(J_u)\mathcal{J}^K_Q(0,\Omega)\right)}.
\end{split}
\end{equation}
Since the depletion and enrichment of different spectral lines vary with the same magnetic field, the influence on atomic line ratio can be expected to be more significant with one line depleted and the other enriched. A few examples will be presented in the following subsections.

\begin{figure*}
\centering
\subfigure[]{
\includegraphics[width=0.32\columnwidth,
 height=0.25\textheight]{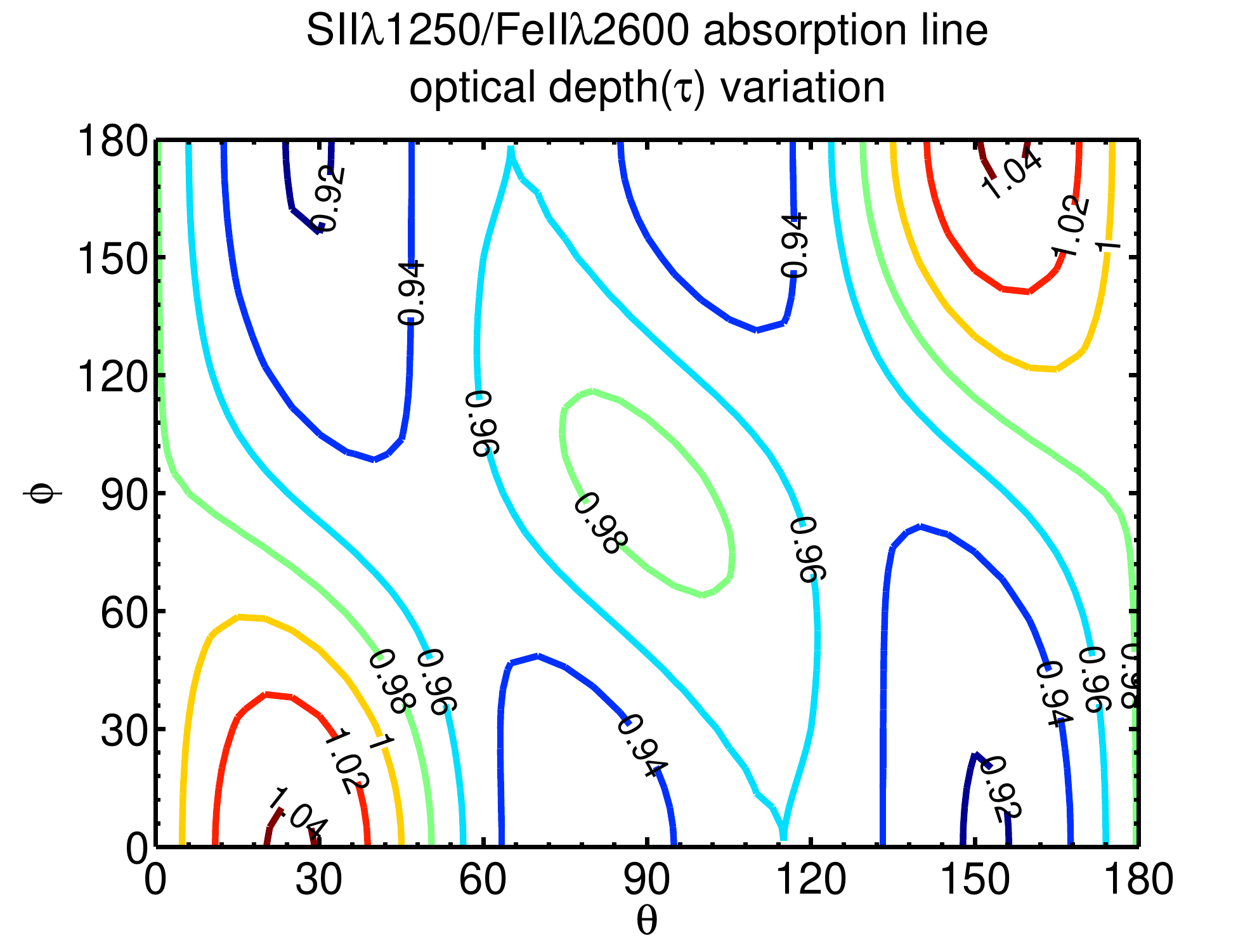}\label{S2Fe2bbrab9to9z6dr1c45}}
\subfigure[]{
\includegraphics[width=0.32\columnwidth,
 height=0.25\textheight]{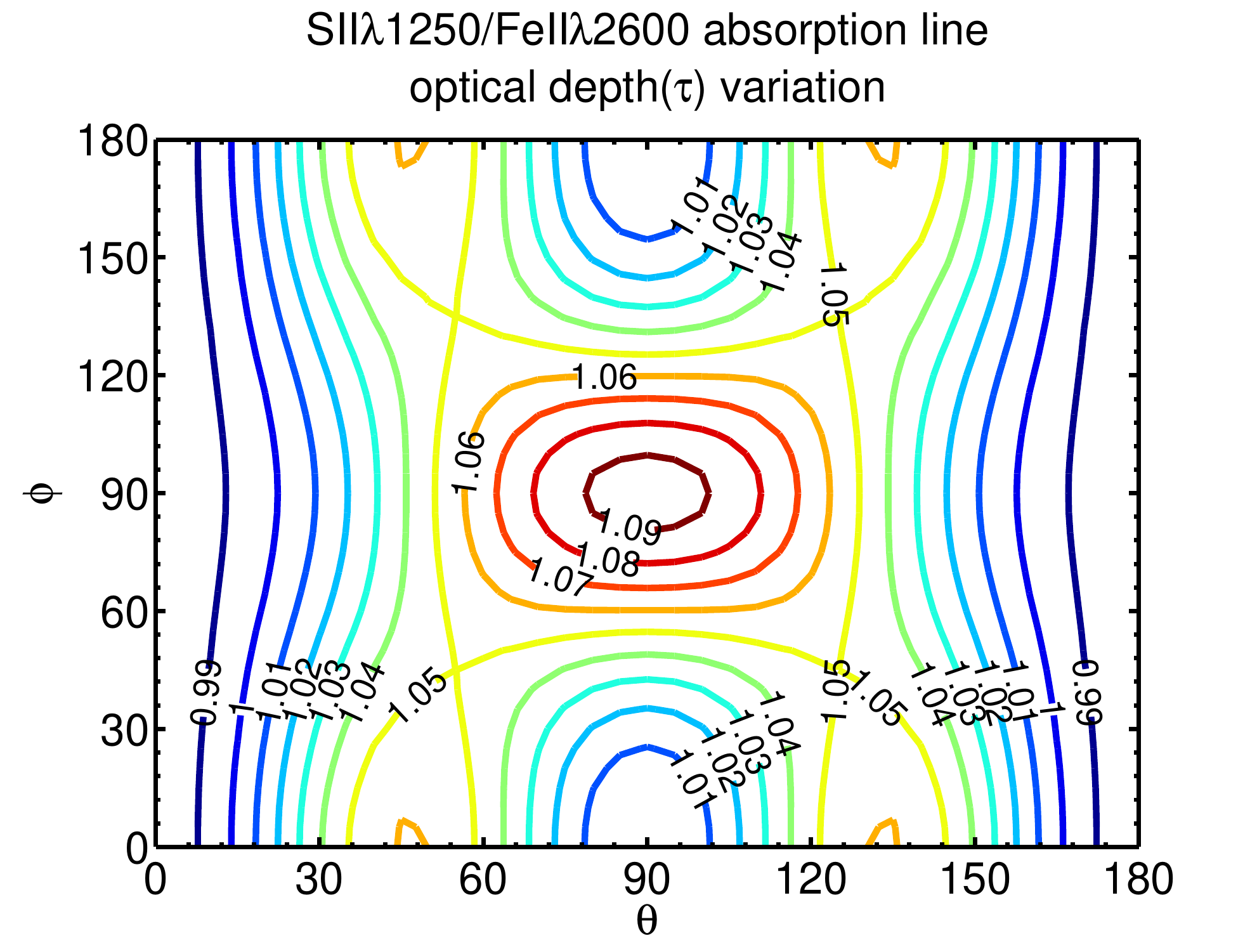}\label{S2Fe2bbrab9to9z6dr1c90}}
 \subfigure[]{
\includegraphics[width=0.32\columnwidth,
 height=0.25\textheight]{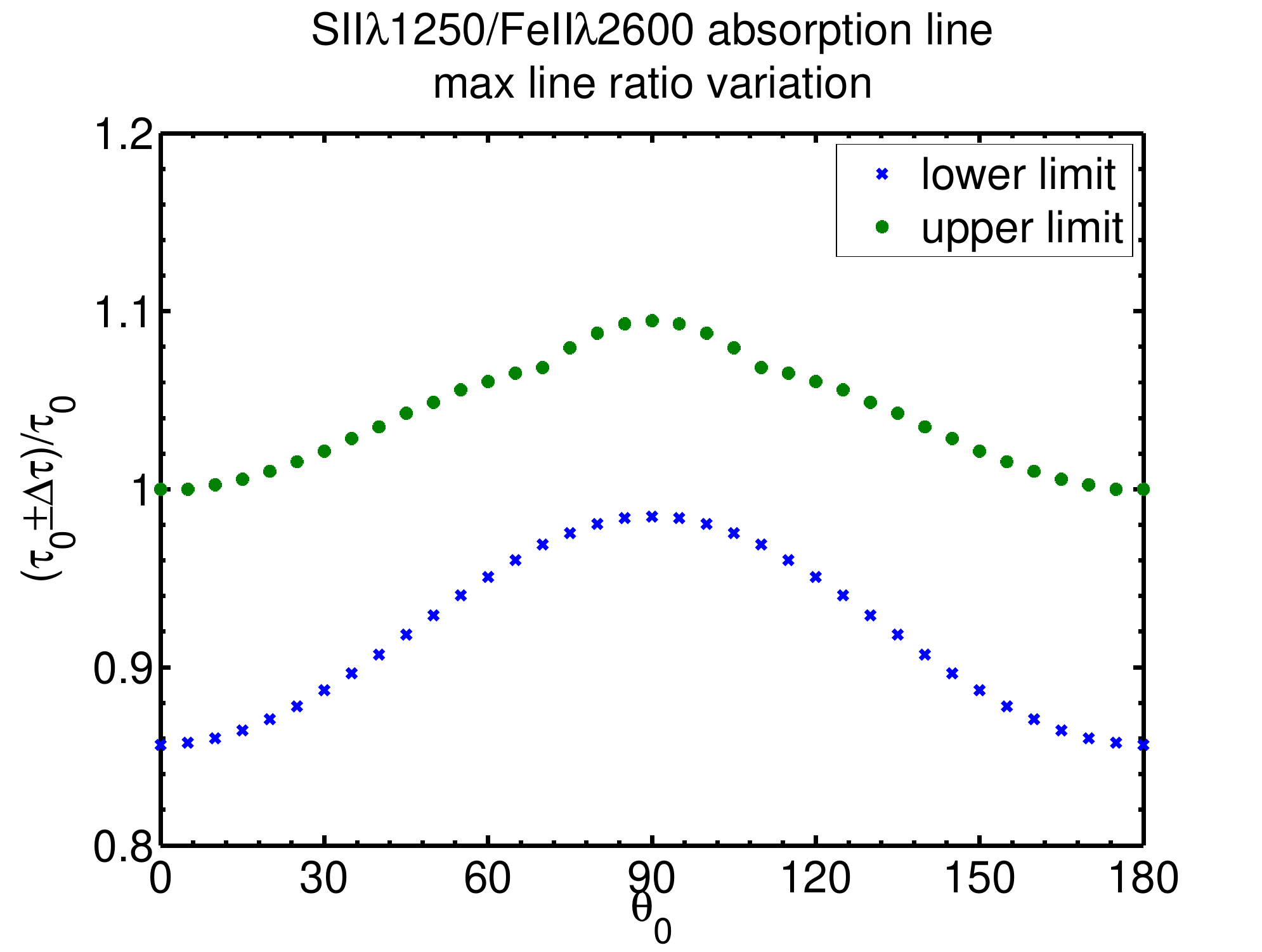}\label{S2Fe2bbrab9to9z6dr1}}
\subfigure[]{
\includegraphics[width=0.32\columnwidth,
 height=0.25\textheight]{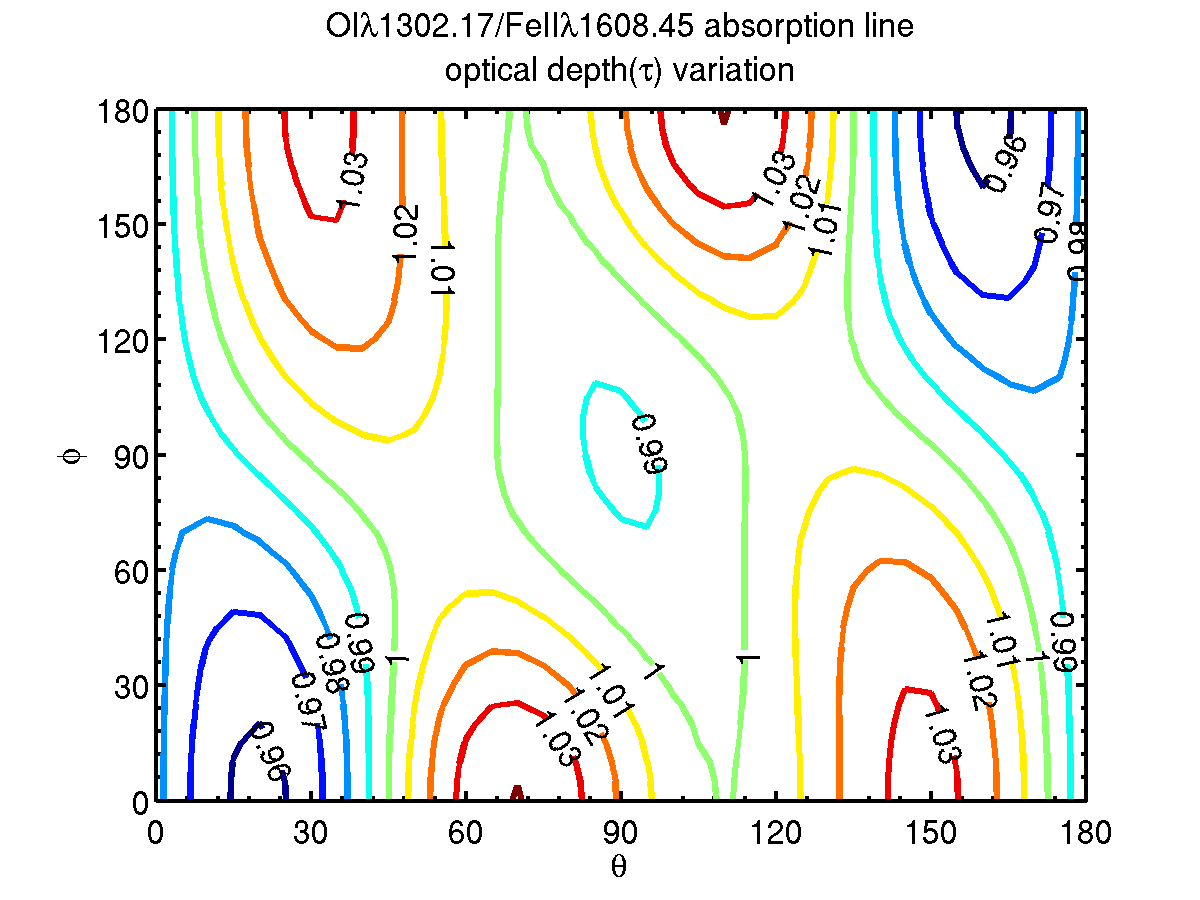}\label{O1Fe2bbrab2to1r9to7c45}}
\subfigure[]{
\includegraphics[width=0.32\columnwidth,
 height=0.25\textheight]{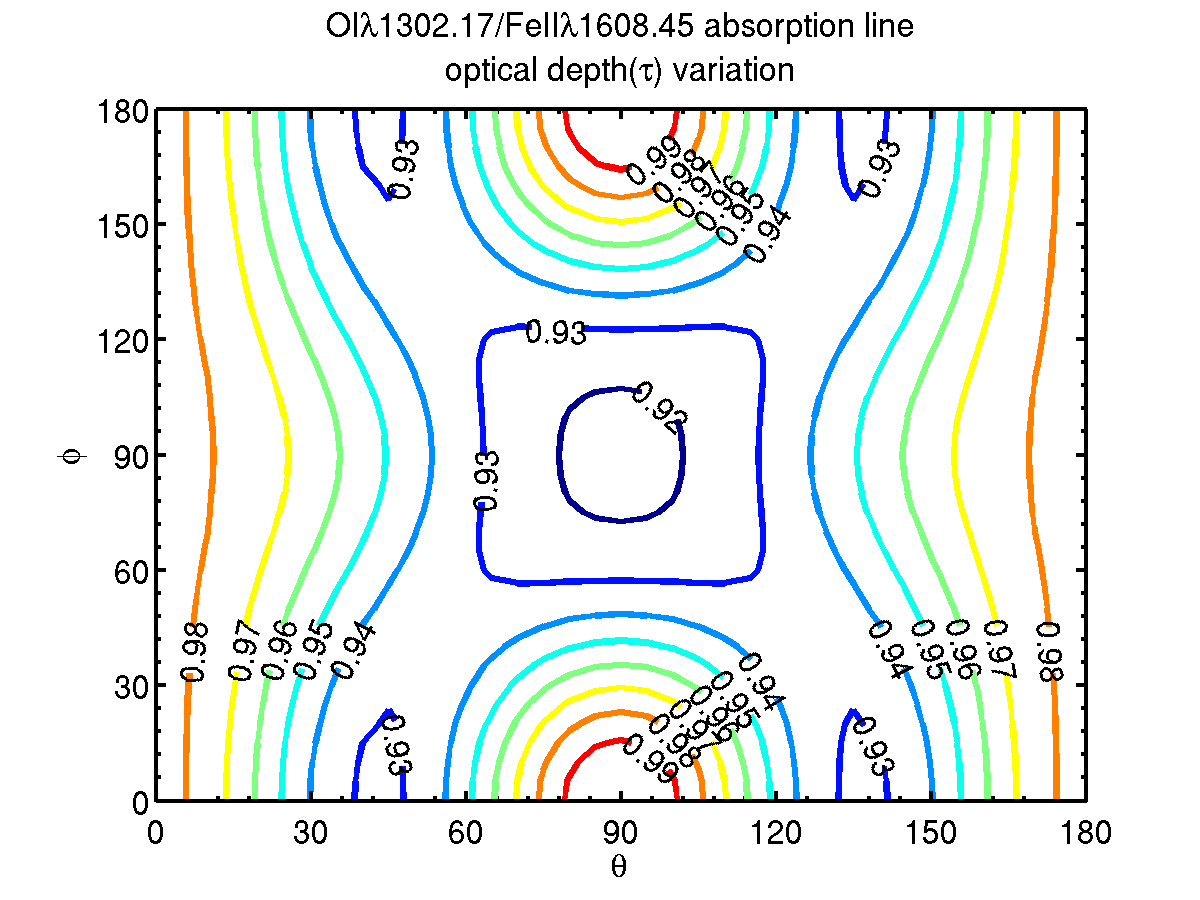}\label{O1Fe2bbrab2to1r9to7c90}}
 \subfigure[]{
\includegraphics[width=0.32\columnwidth,
 height=0.25\textheight]{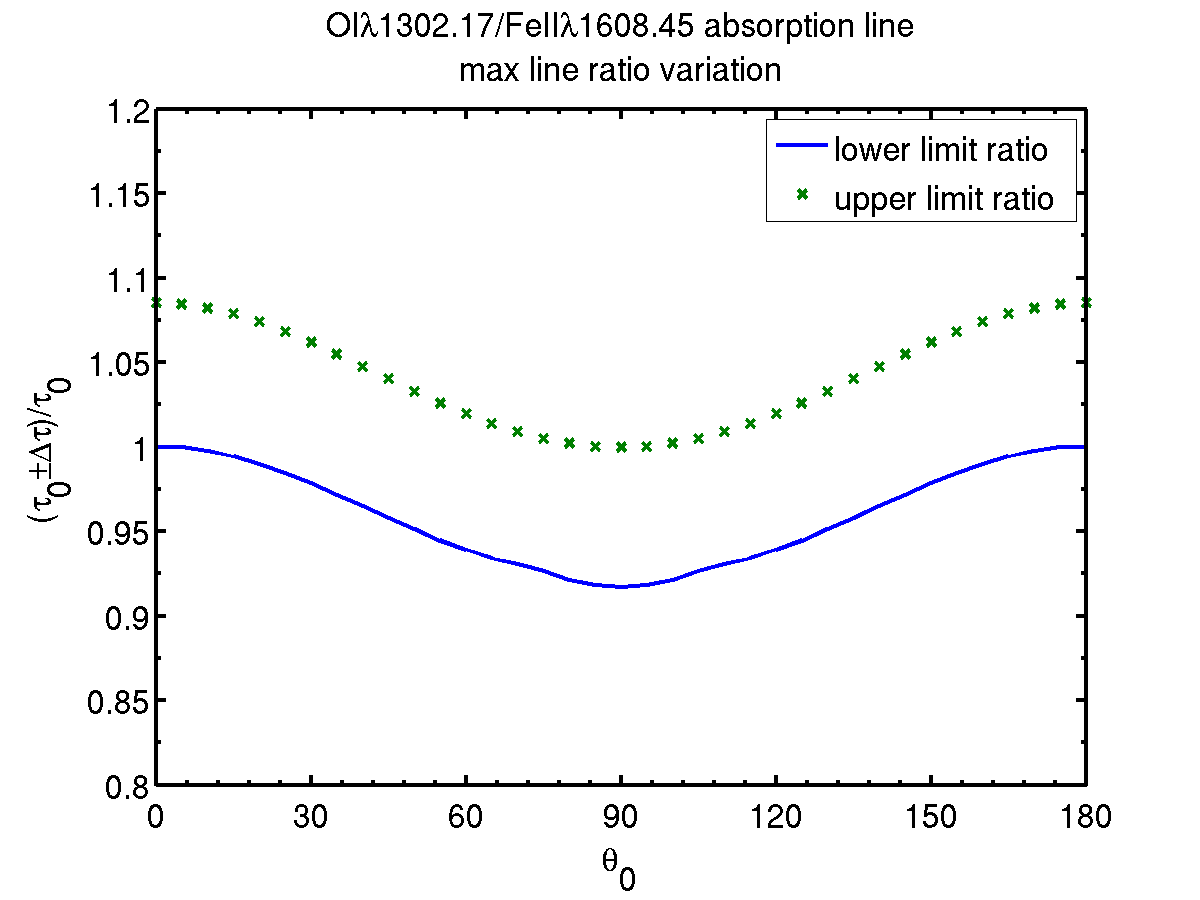}\label{O1Fe2bbrab2to1r9to7}}
\caption{[$\alpha$/Fe] ratio using S\,{\sc ii} $\lambda 1250$/Fe\,{\sc ii} $\lambda 2599$ and O\,{\sc i} $\lambda 1302.17$/ Fe\,{\sc ii} $\lambda 1608.45$ absorption line ratios in blackbody radiation with $T_{source}=50000K$. Line ratio with the case when the magnetic field is parallel to the direction of incidental radiation is set as 1. (a),(b) The variation of line ratio S\,{\sc ii} $\lambda 1250$/Fe\,{\sc ii} $\lambda 2599$ with respect to the direction of the magnetic field when $\theta_0=45^{\circ}$ and $\theta_0=90^\circ$ respectively; (c) The maximum and minimum ratio S\,{\sc ii} $\lambda 1250$/Fe\,{\sc ii} $\lambda 2599$ with respect to different $\theta_0$; (e),(f) The variation of line ratio O\,{\sc i} $\lambda 1302.17$/ Fe\,{\sc ii} $\lambda 1608.45$ with respect to the direction of the magnetic field in the case when $\theta_0=45^{\circ}$ and $\theta_0=90^\circ$ respectively; (g) The maximum and minimum ratio O\,{\sc i} $\lambda 1302.17$/ Fe\,{\sc ii} $\lambda 1608.45$ with respect to different $\theta_0$.}
\end{figure*}

\subsection{Influence on nucleosynthetic studies in DLAs}

The alpha-to-iron ratio is widely used in spectral analysis. This line ratio comparison reflects the nucleosynthetic processes in star forming region (e.g., \citealt{2001ApJS..137...21P,2002ApJ...566...68P}). The $\alpha$ elements can be measured by elements like O, Si, S, Ti etc, whereas Fe refers to Fe peak elements like Cr, Mn, Co, Fe etc. $\alpha$ elements like O and Fe in the medium of the galaxy with low metallicity ([Fe/H]$\lesssim$-1.0) are produced exclusively by Type-II supernovae(SNe II), but the [$\alpha$/Fe] ratio suffers a drop when the delayed contribution of Fe from Type-Ia supernovae (SNe Ia) is effective \citep{1997ARA&A..35..503M, 2011MNRAS.417.1534C}. The alpha-to-iron ratio [X/Fe] (X represents the chemical elements for $\alpha$ elements) is defined by the ratio of abundance observed from the medium in comparison with that from the sun:
\begin{equation}\label{alphafe}
[X/Fe]\equiv log[N(X)/N(Fe)]-log[N(X)/N(Fe)]_{\odot}
\end{equation}

The abundance of the elements is assumed to be equal to the column density of the dominant ionization state. For example, Fe\,{\sc ii} lines are used for Fe abundance and S\,{\sc ii} for S since in DLAs these elements are mostly single ionized. Thus, the ratio N(S)/N(Fe) can be obtained by measuring $\mathcal{R}^{ab}_{S II, Fe II}$. When GSA is taken into consideration, the variation of the ratio S\,{\sc ii}/Fe\,{\sc ii} with respect to different direction of magnetic fields when $\theta_0=45^\circ$ and $\theta_0=90^\circ$ are presented in Fig. \ref{S2Fe2bbrab9to9z6dr1c45} and Fig. \ref{S2Fe2bbrab9to9z6dr1c90}, respectively. As demonstrated in Fig. \ref{S2Fe2bbrab9to9z6dr1}, the maximum variation for N(S)/N(Fe) with the influence of GSA is [$-14\%,+10\%$], which makes the maximum variation for log[N(S)/N(Fe)] [-0.04,+0.07]. If log[N(X)/N(Fe)]$_{\odot}$ obtained from solar is considered to be fixed, the influence of GSA on [S/Fe] is [-0.07,+0.06].

In comparison, another $\alpha$ element O is presented as an example. Most O in DLAs are neutral and hence the ratio N(O)/N(Fe) can be represented by  $\mathcal{R}^{ab}_{O I, Fe II}$. The influence of GSA on the line ratio O\,{\sc i}$\lambda1302.17$/Fe\,{\sc ii}$\lambda1608.45$ when $\theta_0=45^\circ$ and $\theta_0=90^\circ$ are presented in Fig. \ref{O1Fe2bbrab2to1r9to7c45} and Fig. \ref{O1Fe2bbrab2to1r9to7c90}, respectively. By comparing Fig. \ref{S2Fe2bbrab9to9z6dr1c45} and Fig. \ref{O1Fe2bbrab2to1r9to7c90}, it is obvious that when magnetic field is parallel to both line of sight and incidental radiation, the alignment enriches N(S)/N(Fe) by $10\%$ whereas depletes N(O)/N(Fe) by $9\%$. As a result, with the same direction of magnetic field and the same line of sight, [S/Fe] observed varies $+0.06$ while [O/Fe] observed varies $-0.06$. Hence, GSA will affect results from nucleosynthetic studies, although the effect is small, in particular when compared to possible ionization effect.

\begin{figure*}
\centering
\subfigure[]{
\includegraphics[width=0.47\columnwidth,
 height=0.35\textheight]{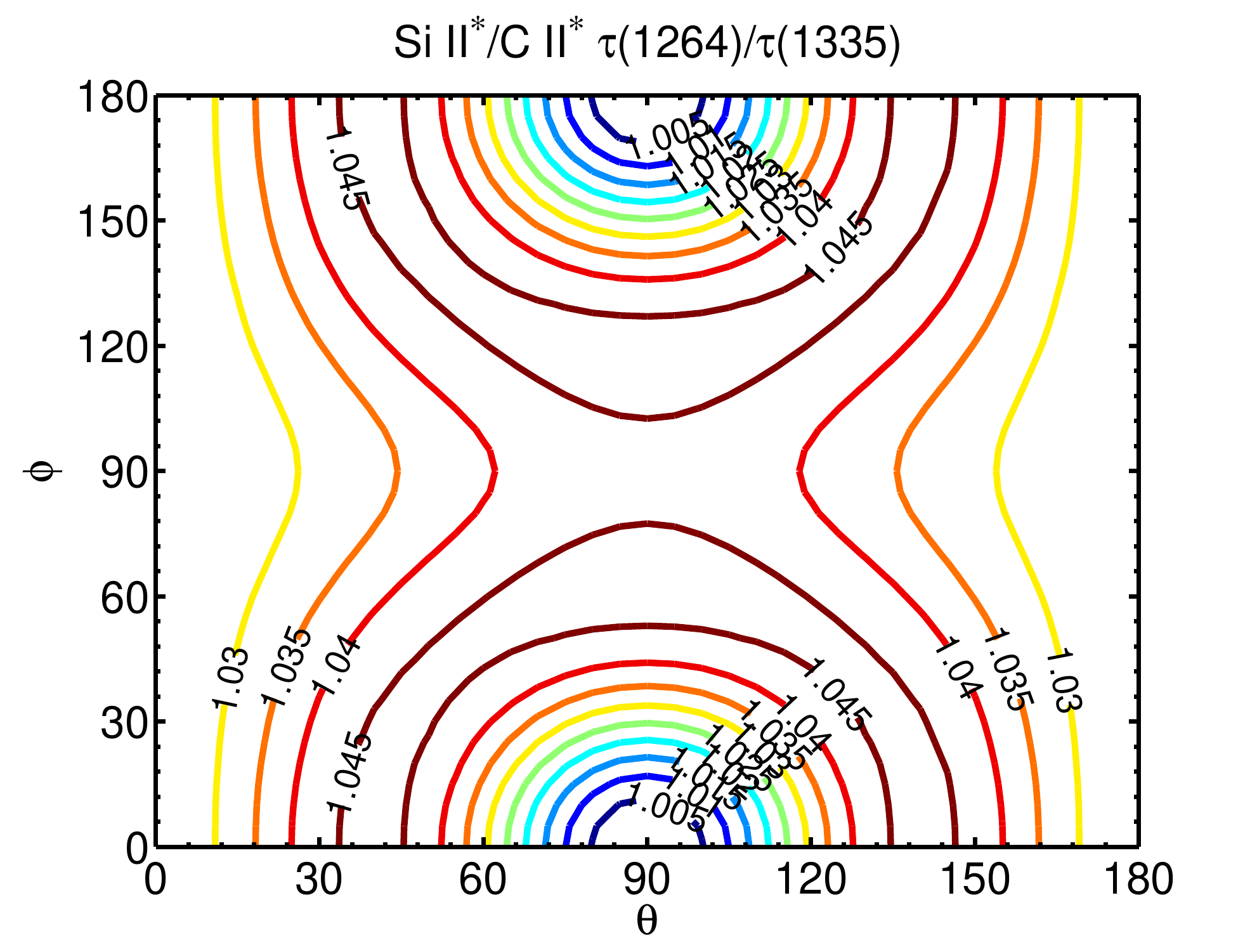}\label{Si2C2bbrab5r3p5c90}}
 \subfigure[]{
\includegraphics[width=0.47\columnwidth,
 height=0.35\textheight]{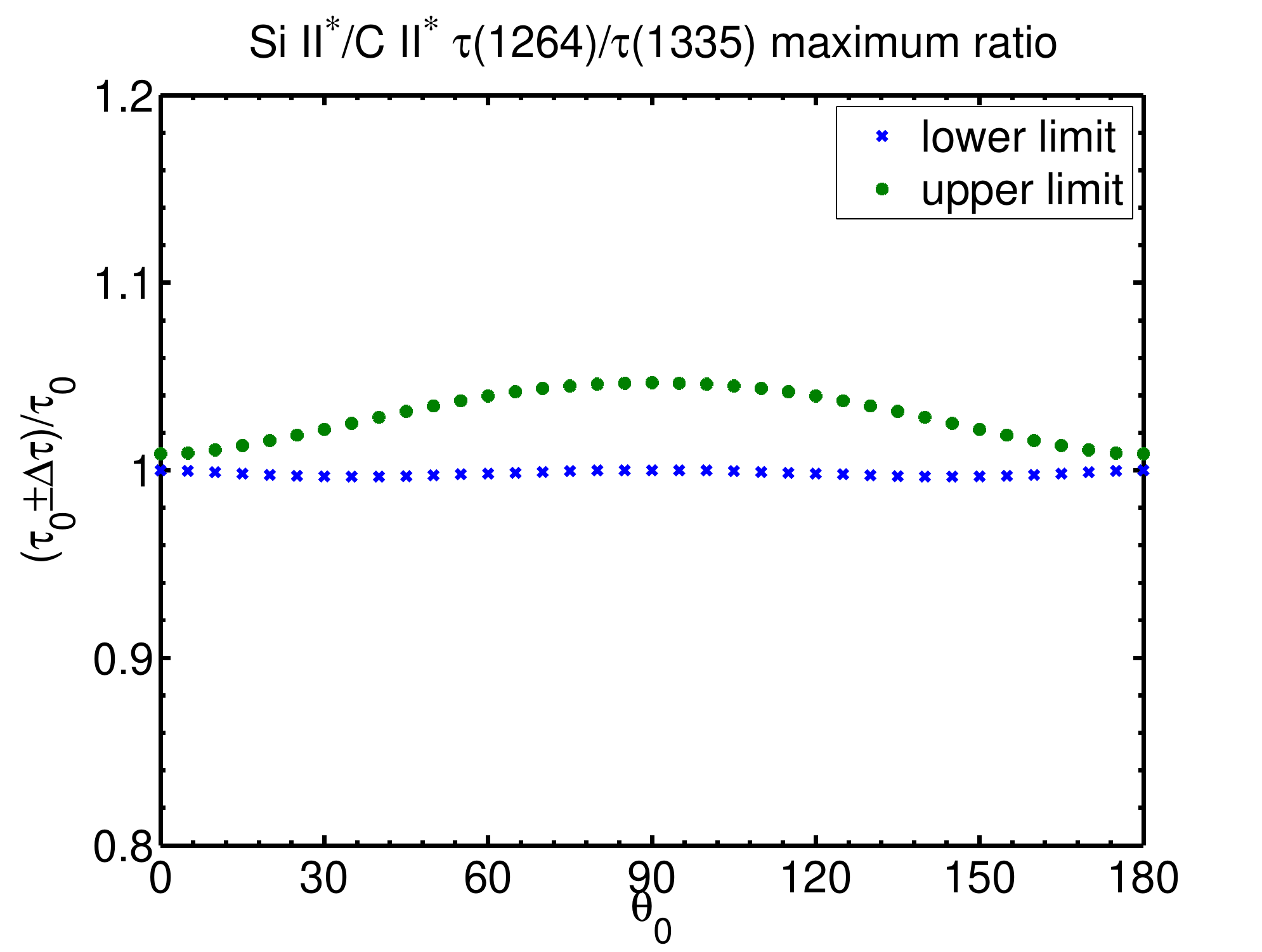}\label{Si2C2bbrab5r3p5}}
\caption{Line ratio of Si\,{\sc ii} $\lambda 1264.74$ and C\,{\sc ii} $\lambda 1335.71$ doublet absorption lines in blackbody radiation with $T_{source}=50000K$. Line ratio with the case when the magnetic field is parallel to the direction of incident radiation is set as 1.
(a) The variation of line ratio with respect to the direction of the magnetic field in the case of line of sight vertical to radiation field($\theta_0=90^{\circ}$);
(b) The maximum and minimum ratio with respect to different $\theta_0$.}
\end{figure*}

\subsection{Influence on environmental temperature studies in the interstellar medium}

As discussed in \S 3.1, the energy difference between some low metastable levels and the ground level on the ground state for some elements(e.g., C\,{\sc ii} and Si\,{\sc ii}) is relatively small. Thus these metastable levels need to be taken into consideration in the study of GSA. The ratio of absorption out of Si\,{\sc ii} $^2P_{3/2}$ and C\,{\sc ii} $^2P_{3/2}$ proves to be a good tool to determine interstellar temperature (\citealt{2005ApJ...622L..81H}, see also \citealt{2015ApJ...800....7N}). The absorption line Si\,{\sc ii}$^{\ast} \lambda\lambda1264.74$ can be used to analyze N(Si\,{\sc ii}$^\ast$), whereas the unresolved doublet C\,{\sc ii}$^\ast \lambda\lambda 1335.66, 1335.71$ can be used to decide N(C\,{\sc ii}$^\ast$). For example, Fig. 1 in \citep{2005ApJ...622L..81H} demonstrated that with the temperature $T\lesssim$ 1000 K, log[N(Si\,{\sc ii}$^\ast$)/N(C\,{\sc ii}$^\ast$)]$\lesssim$-2.8.  The influence of GSA on the corresponding optical depth ratio $\tau(1264)/\tau(1335)$ is presented in Fig.\ref{Si2C2bbrab5r3p5c90}. As demonstrated in Fig. \ref{Si2C2bbrab5r3p5}, the maximum variation of N(Si\,{\sc ii}$^\ast$)/N(C\,{\sc ii}$^\ast$) is $5\%$, which means a variation of 0.02 on log[N(Si\,{\sc ii}$^\ast$)/N(C\,{\sc ii}$^\ast$)]. Since this technique of measuring the interstellar temperature requires high resolution, high signal-to-noise(S/N) spectra (see \citealt{2005ApJ...622L..81H}), the influence of GSA is measurable, in principle.

\begin{figure*}
\centering
\subfigure[]{
\includegraphics[width=0.32\columnwidth,
 height=0.25\textheight]{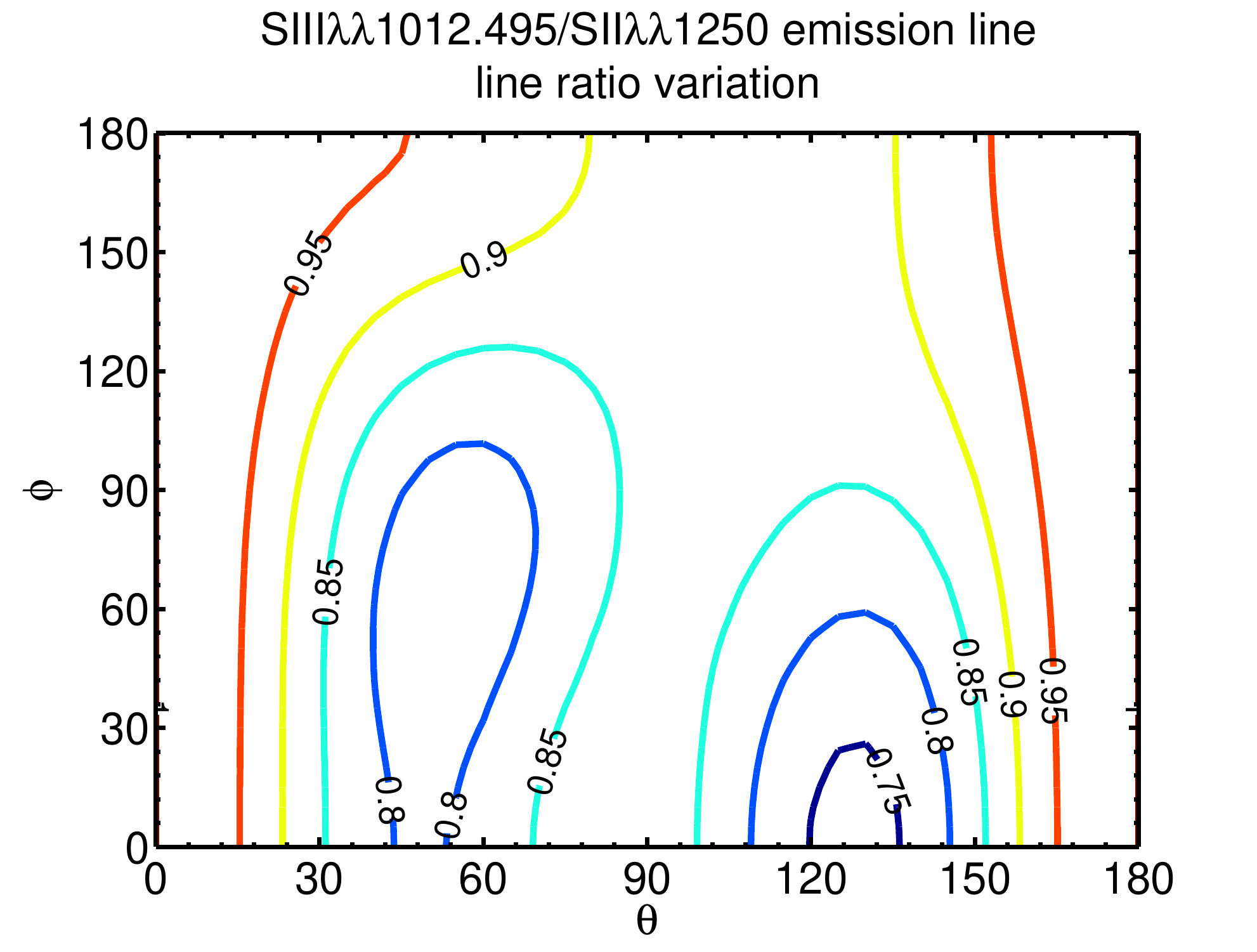}\label{S3S2bbrem1to0pr1c0}}
\subfigure[]{
\includegraphics[width=0.32\columnwidth,
 height=0.25\textheight]{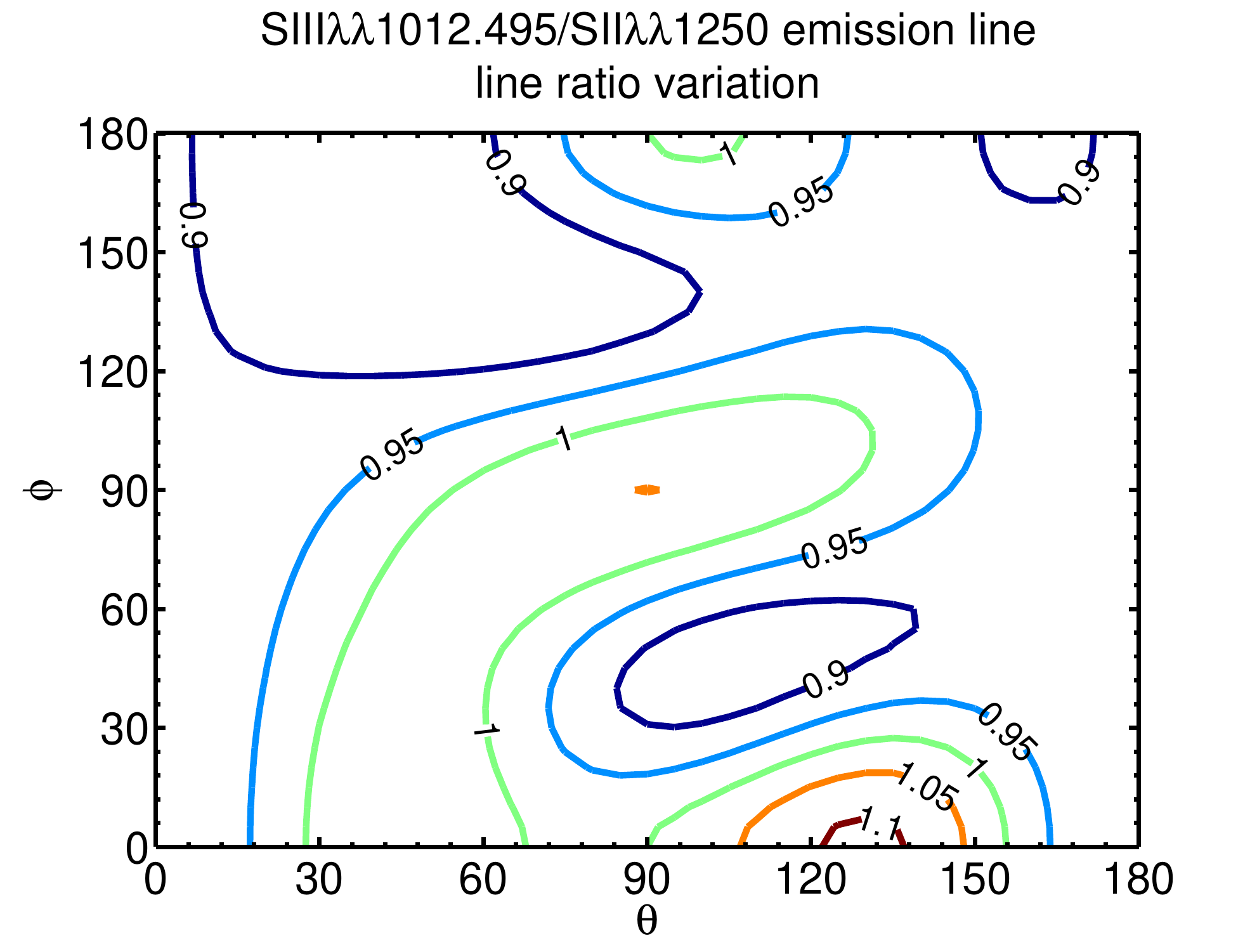}\label{S3S2bbrem1to0pr1c90}}
 \subfigure[]{
\includegraphics[width=0.32\columnwidth,
 height=0.25\textheight]{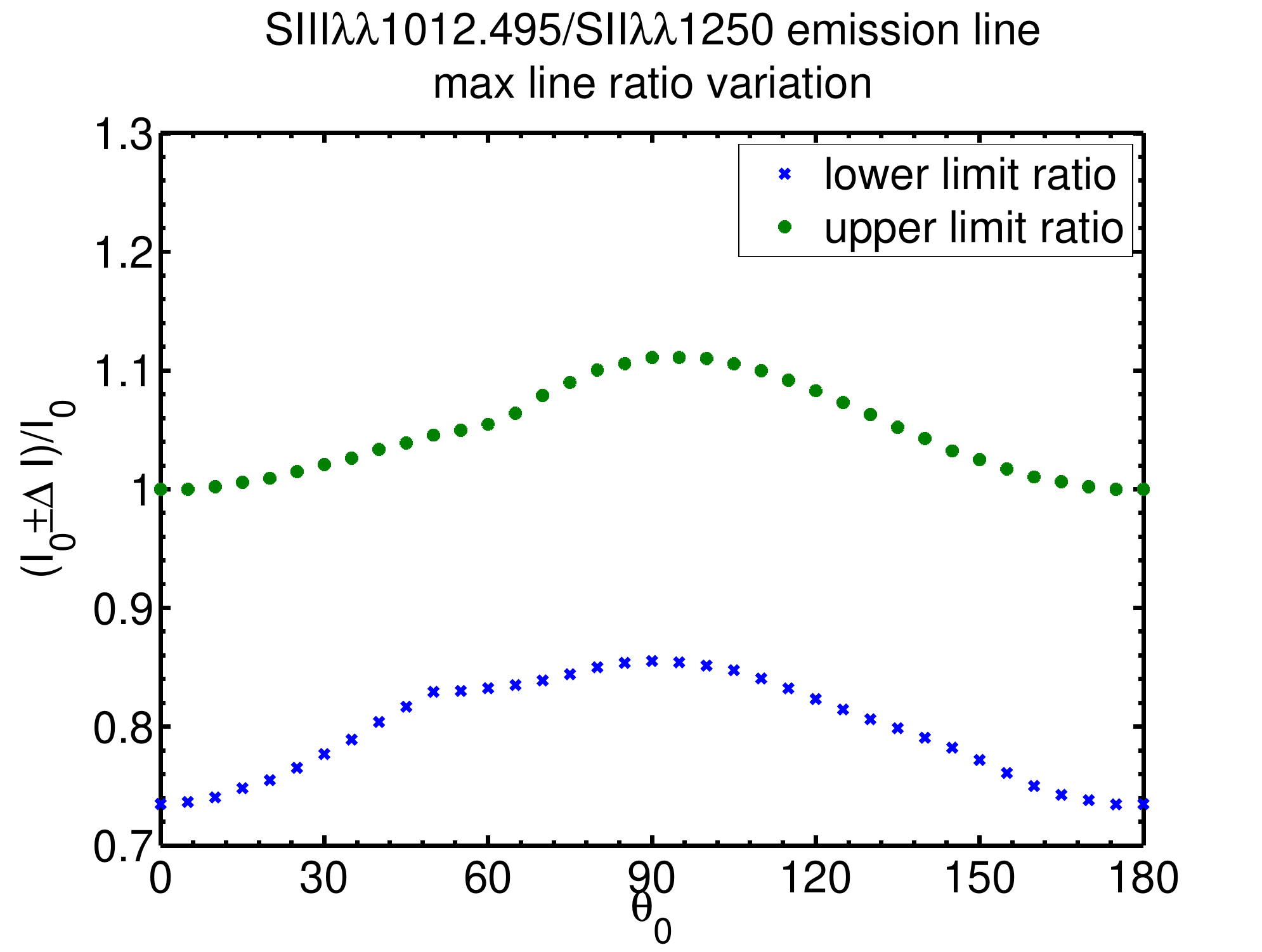}\label{S3S2bbrem1to0pr1}}
\caption{Line ratio of S\,{\sc iii} $\lambda\lambda1012.49$ and S\,{\sc ii} $\lambda\lambda1250.58$ absorption line in blackbody radiation with $T_{source}=50000K$. Line ratio with the case when the magnetic field is parallel to the direction of incident radiation is set as 1.
(a),(b) The variation of line ratio with respect to the direction of the magnetic field in the case of line of sight parallel and vertical to radiation field($\theta_0=0^{\circ}, 90^\circ$, respectively);
(c) The maximum and minimum ratio with respect to different $\theta_0$.}
\end{figure*}

\subsection{Influence on ionization studies in diffuse gas}

The ratio of the abundance\footnote[9]{The abundance ratio can be analyzed from both absorption ratios and emission ratios} of different ionization states of the same element is often used to depict the ionization rate (see e.g., \citealt{2016A&A...590A..68R}). As an example, (S\,{\sc iii})/(S\,{\sc ii}) emission line ratio can be applied to decide the ionization rate in extragalactic H\,{\sc ii} region (see e.g., \citealt{1988MNRAS.235..633V}), because higher ionized Sulphur is not important in most of these  H\,{\sc ii} regions \citep{1982ApJ...261..195M, 1985ApJ...291..247M}. To illustrate the influence of GSA, the line ratio S\,{\sc iii} $\lambda\lambda1012.49$/S\,{\sc ii} $\lambda\lambda1250.58$ is used to represent (S\,{\sc iii})/(S\,{\sc ii}). Fig. \ref{S3S2bbrem1to0pr1c0} and Fig. \ref{S3S2bbrem1to0pr1c90} demonstrate the influence of GSA with different direction of magnetic fields when $\theta_0=0^\circ$ and $90^\circ$. As presented in Fig. \ref{S3S2bbrem1to0pr1}, the maximum influence of GSA on different $\theta_0$ is [$-27\%,+12\%$], which means a variation of [-0.14,+0.05] on log[N(S\,{\sc iii})/N(S\,{\sc ii})]. GSA will affect the error budget of ionization rate when high resolution and high signal-to-noise(S/N) ratio spectra is used.

\section{Summary $\And$ Discussion}
\subsection{Conclusions}
GSA is a well-studied effect in laboratory, whilst its importance still requires attention in astrophysical community. In this paper, we generate synthetic spectra according to a scenario about DLAs in galaxy to present  the change of observed spectrum with the influence of magnetic fields due to GSA. We demonstrate the variation of emission lines with the influence of GSA in a scenario of extragalactic H\,{\sc ii} Region. The influence of GSA on submillimeter fine-structure lines is discussed under a scenario for star forming regions. The variations GSA causes to physical parameters derived from atomic line ratios are analysed. It is concluded that:

\begin{enumerate}

\item Due to GSA, magnetic fields will have a small but measurable influence on the atomic spectra.
\item The depletion and enrichment of the same spectrum due to GSA can be different given a different direction of the magnetic field.
\item The influence of GSA on different spectrum observed from the same line of sight with the same magnetic field is different.
\item The influence of magnetic fields on line ratio varies due to GSA. Sometimes the influence can be significant due to Conclusion 2 and 3.
\item Due to GSA, magnetic fields have a small but measurable influence absorption spectra from absorbers' continuum (QSOs) to analysis diffuse medium like DLAs and galactic halos.
\item The observation for atomic emission spectra from diffuse medium like extragalactic H\,{\sc ii} Regions, circumburst medium near GRBs, and supernova remnants is influenced by GSA through a scattering process.
\item GSA can affect submillimeter fine-structure lines from diffuse medium like PDR regions, star forming regions and Herbig Ae/Be disks, etc.
\item GSA will affect the analysis of all the physical parameters derived from atomic spectra and atomic line ratios. Though the effect can be small, it should be taken into account in the error budget.
\end{enumerate}

\subsection{Further discussion}
We have demonstrated the main idea the influence of magnetic fields due to GSA on spectroscopic studies. Real observational data with high resolution and high signal-to-noise(S/N) ratio can be compared with the theoretical predictions to trace the effect. Nevertheless, the effect is not limited to the examples we provide. Meanwhile, some physical processes that may influence the alignment can be taken into account in further studies. We present some future outlooks as follows:

\subsubsection{Observation of the effect and Hyperfine splitting}

As illustrated in this paper, GSA has a small but measurable influence in spectroscopic studies. Thus, we can use observational signals with high resolution and high S/N ratio to prove the effect in future work. Furthermore, as illustrated in \citep{YLhyf}, hyperfine splitting of atoms and ions will also influence the alignment. Even with the same fine structure, the atoms with hyperfine structure (the nuclear spin is not 0) is affected by GSA differently than the atoms without (see the comparison between the alignment of S\,{\sc ii} and N\,{\sc i} in Fig. 19 in \citealt{YLhyf}, in which the two elements share the same atomic fine structure, but due to a different nuclear spin, the alignment on the ground state of the two elements are totally different). The alignment on different hyperfine sublevels of the same fine structure level can be different. Hence, future observation with higher resolution that can separate transitions to different hyperfine sublevels will be perfect for analyzing the influence of magnetic fields because the span of the hyperfine lines are so small that they would share the same signal-to-noise(S/N) ratio. Note that hyperfine splitting also has to be considered when studying the GSA effect on fine structure transitions of the atoms with nuclear spin.

\subsubsection{fluorescence lines}

As illustrated in \S 3.2, the alignment on the ground state can be transferred to the upper state through absorption process. Atoms in upper states quickly decay by radiative transitions and hence photons are emitted. In real circumstances, the excited atoms may experience several decays and photons emitted may be scattered several times before they actually escape, which can be considered in future work. The intensity of the lines derived from successive decays to different levels of atoms are thus dependent on the density of the initial upper levels, which can be influenced by GSA by scattering process demonstrated in Fig. \ref{scenaem}.

\subsubsection{The influence of strong magnetic fields}

In most interstellar medium where the magnetic field is relatively weak (1 Gauss$\gtrsim$ B $\gtrsim$ $10^{-15}$ Gauss), the GSA is applicable for the influence of magnetic field on atomic spectra. Strong magnetic field can also influence the spectra observed (e.g., spectra from stars) due to the upper level Hanle effect (see \citealt{landi2004}). The magnetic fields can affect the analysis of spectra observed in all circumstances.

\subsubsection{The influence of collision}

The influence of collision is negligible in our discussion, which applies to diffuse interstellar medium. However, collision can become important in the case of higher density and higher temperature medium. The threshold for collision being more important than optical pumping is when the collision rate $\tau_C^{-1}$(either inelastic collision rate or Van der Waals collision rate) is larger than optical pumping rate $B_{J_lJ_u}I$ (see \citealt{YLfine} for details). Collisions can also redistribute atoms to different sublevels, yet a reduced efficiency is expected (see \citealt{Hawkins:1955fv}). In higher temperature H\,{\sc ii} Regions, collisional energy may be enough to excite atoms partially to the metastable levels from the ground level. Hence, the recombination lines and collisional excited lines can be combined together to evaluate the influence of GSA in future work.

\subsection{summary}
{\em We emphasize that in spectroscopic studies GSA will have a measurable effect in the error budget of observations with high resolution and high signal-to-noise(S/N) ratio. Although the effect can be small, the physical parameters derived from spectral line ratio can have a better accuracy when we take into account the measurable influence of magnetic fields due to GSA.}

\begin{figure*}
\centering
\subfigure[]{
 \includegraphics[width=0.47\columnwidth,
 height=0.32\textheight]{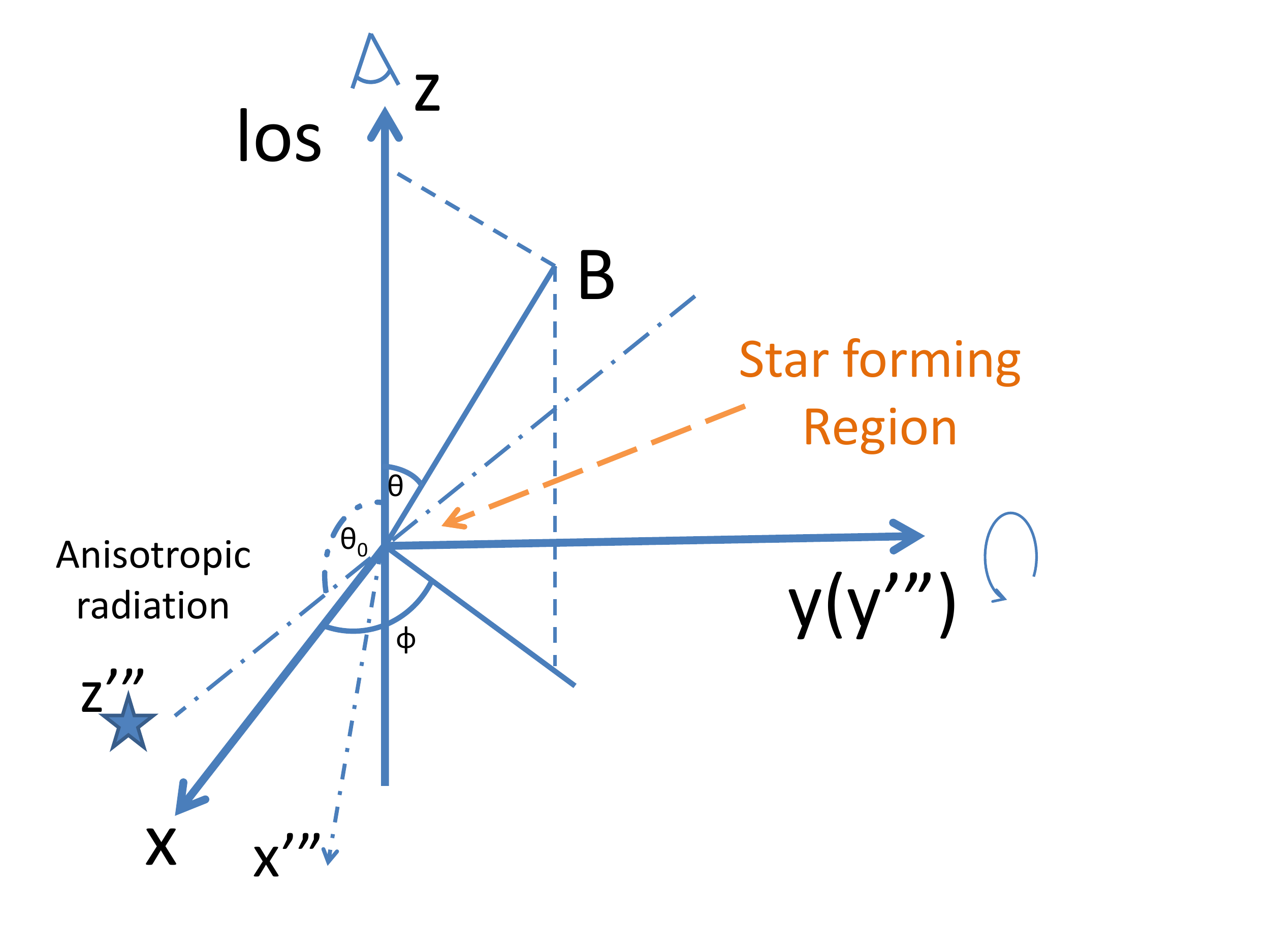}\label{sceemb}}
\subfigure[]{
 \includegraphics[width=0.47\columnwidth,
 height=0.32\textheight]{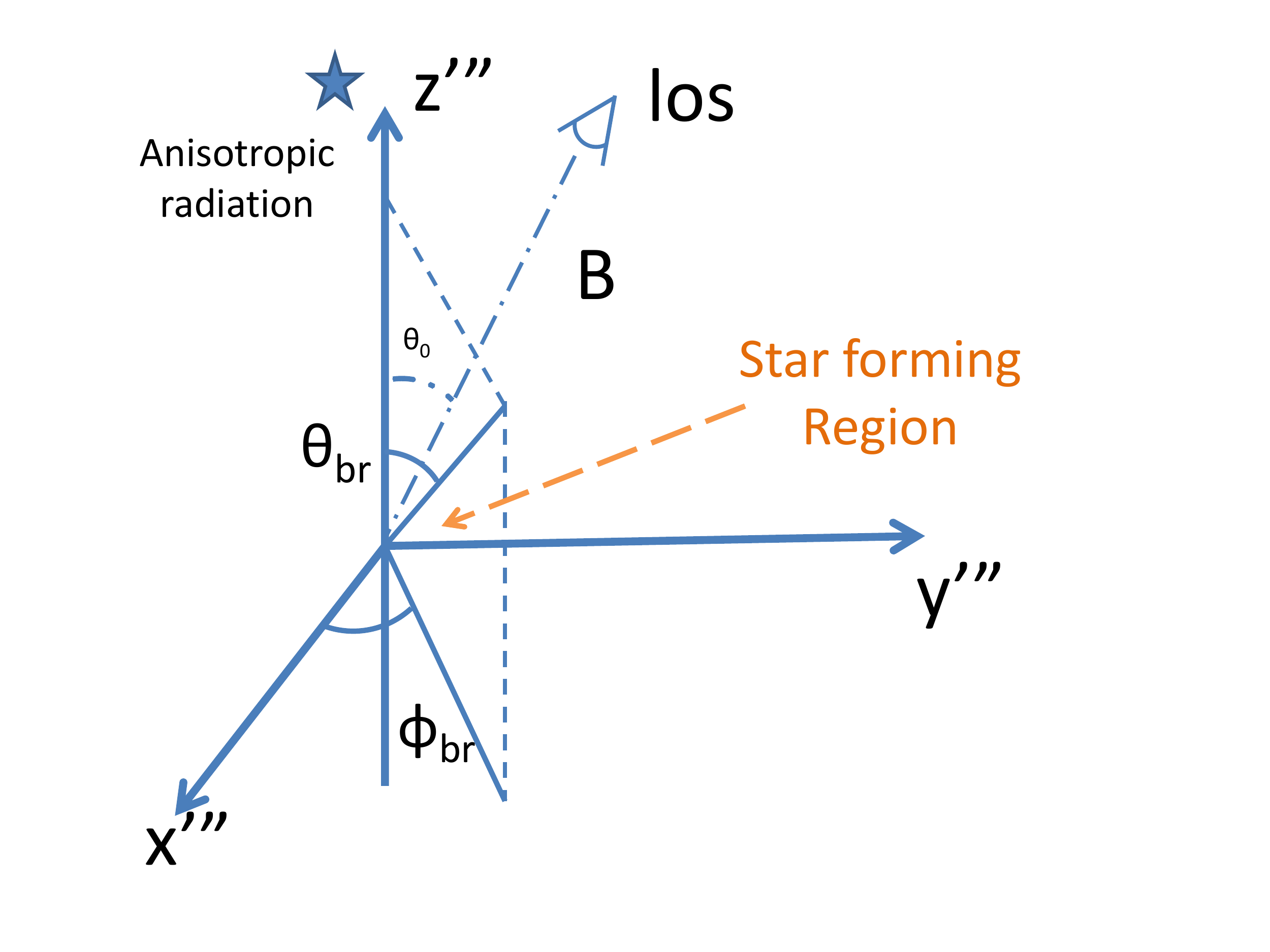}\label{sceemc}}
 \caption{Different coordinate system for comparison. The variation of line intensity with respect to the direction of the magnetic field is compared to the spectrum when magnetic field is parallel to the incidental radiation. (a) xyz-coordinate system with line of sight being z-axis; $\theta$ and $\phi$ are the polar and azimuth angle of the magnetic field in line of sight coordinate; (b) x'''y'''z'''-coordinate system with radiation field being z'''-axis;  $\theta_{br}$ and $\phi_{br}$ are the polar and azimuth angle of the magnetic field in radiation coordinate.
}\label{subscemm}
\end{figure*}
\begin{figure*}
\centering
\subfigure[]{
\includegraphics[width=0.47\columnwidth,
 height=0.35\textheight]{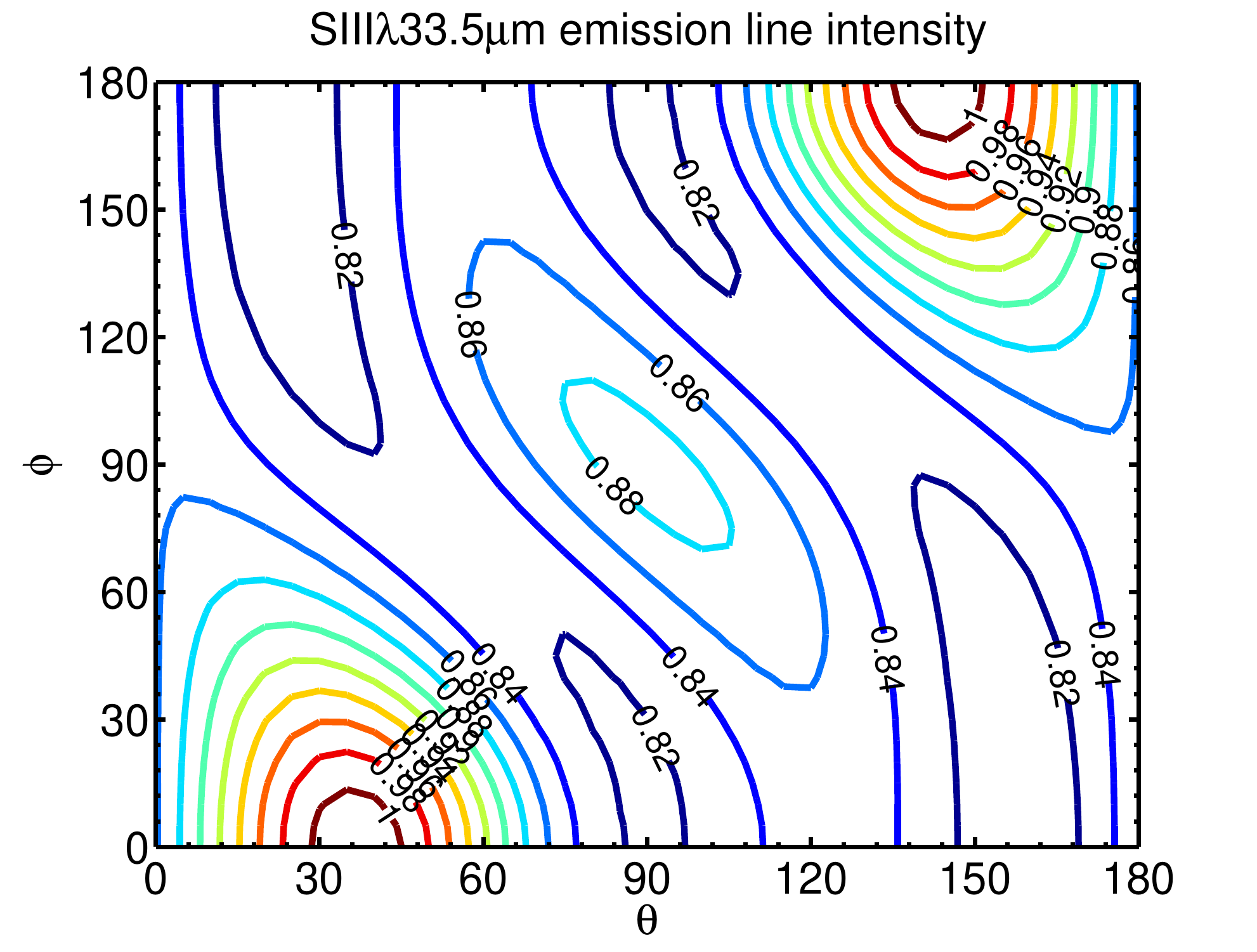}\label{S3bbrsubem1c45}}
\subfigure[]{
\includegraphics[width=0.47\columnwidth,
 height=0.35\textheight]{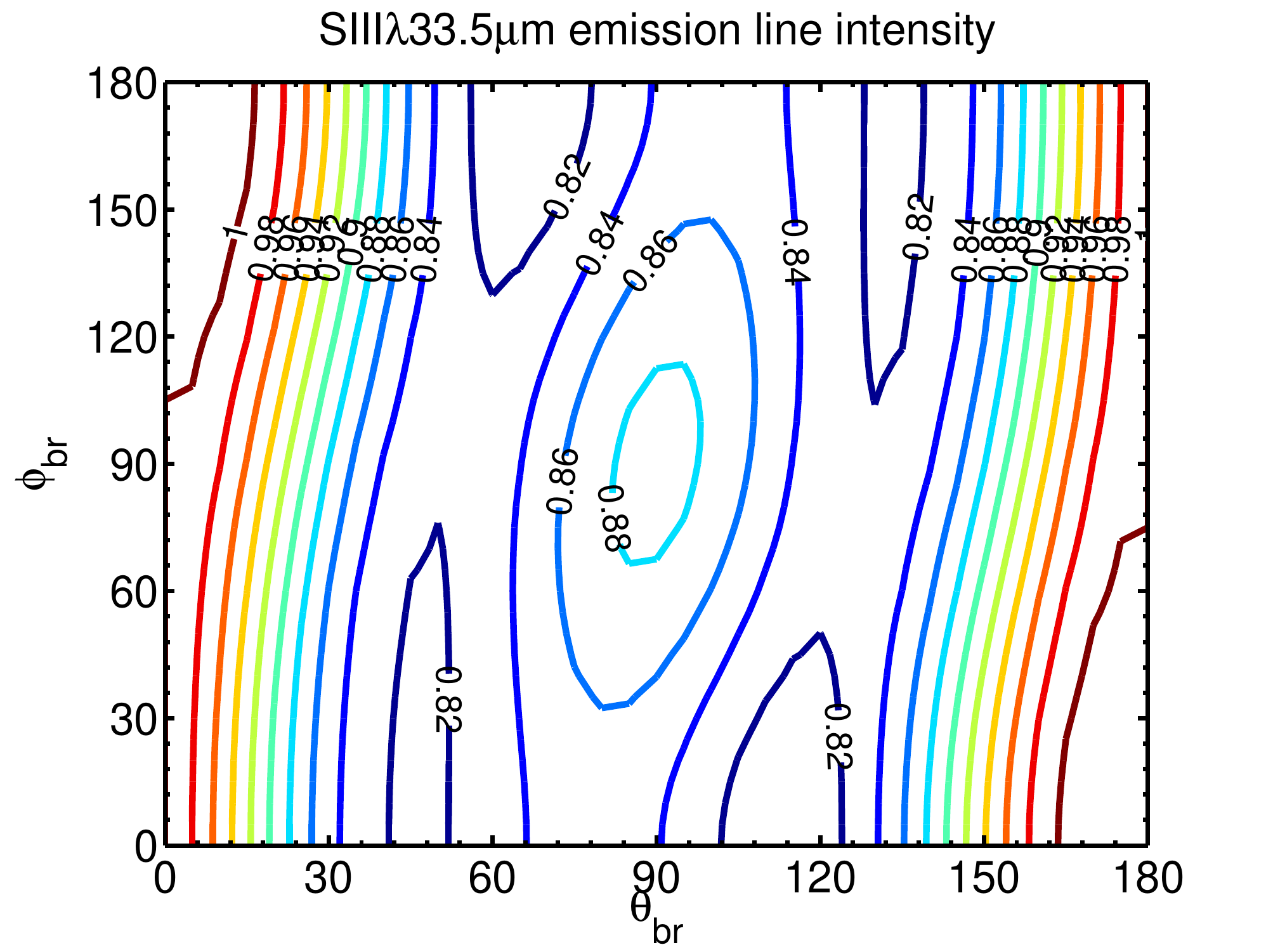}\label{S3bbrsubem1c45a}}
\caption{The variation of line intensity for submillimeter lines with respect to the direction of the magnetic field with $T_{source}=50000K$. Contours are presented in different coordinate system defined in Fig. \ref{subscemm} for the line intensity variation of S\,{\sc iii} $\lambda\lambda33.5{\mu}m$ when $\theta_0=45^\circ$, respectively. (a) $\theta$ and $\phi$ are the polar and azimuth angle of the magnetic field in line of sight coordinate defined in Fig. \ref{sceemb}; (b) $\theta_{br}$ and $\phi_{br}$ are the polar and azimuth angle of the magnetic field in line of sight coordinate defined in Fig. \ref{sceemc}.}\label{submm}
\end{figure*}

\begin{appendix}
\section{A. COMPARISON OF DIFFERENT COORDINATE SYSTEM}\label{coordinatescompare}

Throughout the paper, in order to show the GSA effect with the change of magnetic fields, we fix the direction of line of sight and the direction of incidental radiation for all the contour plots. However, with this setting, two coordinate systems can be used to depict the change of magnetic fields. xyz-coordinate system with the line of sight as z-axis is presented in Fig. \ref{sceemb}, which is considered as the real observation structure. The coordinate system with radiation as z'''-axis and the corresponding spectral variation are presented in  Fig. \ref{sceemc}.

As an example to compare the two coordinate systems, the influence of atomic alignment on S\,{\sc iii}  $\lambda33.5{\mu}m$ is presented in Fig. \ref{submm}. The ground state of S\,{\sc iii} has 3 sublevels $3P_{0;1;2}$, in which the ground level is $3P_{J_l=0}$. S\,{\sc iii}  $\lambda\lambda33.5{\mu}m$ line represents the transition within lower sublevels $3P_{J_l=1}$ and $3P_{J_l=0}$. Fig. \ref{S3bbrsubem1c45} shows the variation of the line intensity S\,{\sc iii}$\lambda\lambda33.5{\mu}m$ when $\theta_0=45^\circ$ in xyz-coordinate system in Fig. \ref{sceemb}. Fig. \ref{S3bbrsubem1c45a} shows the variation of the line intensity S\,{\sc iii}$\lambda\lambda33.5{\mu}m$ when $\theta_0=45^\circ$ in x'''y'''z'''-coordinate system in Fig. \ref{sceemc}. As illustrated in Eq. \eqref{groundoccu}, as the angle $\theta_{br}$ is fixed, the alignment parameter $\sigma^2_0$ is the same. As a result, in Fig. \ref{S3bbrsubem1c45a}, the change of $\phi_{br}$ under the same $\theta_{br}$ means the spectra with the same alignment observed from different line of sight.(For example, by comparing Fig. \ref{S3bbrsubem1c45} and Fig. \ref{S3bbrsubem1c45a}, it is obvious that spectra with the same alignment (same $\theta_{br}$ in Fig. \ref{S3bbrsubem1c45a}) have a smaller variation.) Thus x'''y'''z'''-coordinate system exhibits a more explicit physical meaning than xyz-coordinate system. However, by comparing the same situation in different coordinate systems, it is obvious that the maximum and minimum variations of the two plots remain the same. For the sake of observation, we present our results through the paper with the actual observation system: xyz-coordinate system with the line of sight magnetic angle $\theta$ and the plane of sky magnetic angle $\phi$, as well defined in the scenarios Fig. \ref{scea} and Fig. \ref{scenaem}.

\end{appendix}
\label{lastpage}
 \bibliographystyle{aasjournal}
 \setcitestyle{square,aysep={},yysep={;}}
 \bibliography{Zhan0306}

\begin{thebibliography}{}
\expandafter\ifx\csname natexlab\endcsname\relax\def\natexlab#1{#1}\fi

\bibitem[{{Asplund} {et~al.}(2009){Asplund}, {Grevesse}, {Sauval}, \&
  {Scott}}]{Asplund2009}
{Asplund}, M., {Grevesse}, N., {Sauval}, A.~J., \& {Scott}, P. 2009, \araa, 47,
  481

\bibitem[{{Ceccarelli} {et~al.}(1996){Ceccarelli}, {Hollenbach}, \&
  {Tielens}}]{1996ApJ...471..400C}
{Ceccarelli}, C., {Hollenbach}, D.~J., \& {Tielens}, A.~G.~G.~M. 1996, \apj,
  471, 400

\bibitem[{{Cooke} {et~al.}(2011){Cooke}, {Pettini}, {Steidel}, {Rudie}, \&
  {Nissen}}]{2011MNRAS.417.1534C}
{Cooke}, R., {Pettini}, M., {Steidel}, C.~C., {Rudie}, G.~C., \& {Nissen},
  P.~E. 2011, \mnras, 417, 1534

\bibitem[{{D{\'{\i}}az-Santos} {et~al.}(2013){D{\'{\i}}az-Santos}, {Armus},
  {Charmandaris}, {Stierwalt}, {Murphy}, {Haan}, {Inami}, {Malhotra},
  {Meijerink}, {Stacey}, {Petric}, {Evans}, {Veilleux}, {van der Werf}, {Lord},
  {Lu}, {Howell}, {Appleton}, {Mazzarella}, {Surace}, {Xu}, {Schulz},
  {Sanders}, {Bridge}, {Chan}, {Frayer}, {Iwasawa}, {Melbourne}, \&
  {Sturm}}]{2013ApJ...774...68D}
{D{\'{\i}}az-Santos}, T., {Armus}, L., {Charmandaris}, V., {et~al.} 2013, \apj,
  774, 68

\bibitem[{{Esteban} {et~al.}(2014){Esteban}, {Garc{\'{\i}}a-Rojas}, {Carigi},
  {Peimbert}, {Bresolin}, {L{\'o}pez-S{\'a}nchez}, \&
  {Mesa-Delgado}}]{2014MNRAS.443..624E}
{Esteban}, C., {Garc{\'{\i}}a-Rojas}, J., {Carigi}, L., {et~al.} 2014, \mnras,
  443, 624

\bibitem[{{Fontana} \& {Ballester}(1995)}]{Fontana1995}
{Fontana}, A., \& {Ballester}, P. 1995, The Messenger, 80, 37

\bibitem[{{Fox} {et~al.}(2014{\natexlab{a}}){Fox}, {Richter}, \&
  {Fechner}}]{2014A&A...572A.102F}
{Fox}, A., {Richter}, P., \& {Fechner}, C. 2014{\natexlab{a}}, \aap, 572, A102

\bibitem[{{Fox} {et~al.}(2013){Fox}, {Richter}, {Wakker}, {Lehner}, {Howk},
  {Ben Bekhti}, {Bland-Hawthorn}, \& {Lucas}}]{2013ApJ...772..110F}
{Fox}, A.~J., {Richter}, P., {Wakker}, B.~P., {et~al.} 2013, \apj, 772, 110

\bibitem[{{Fox} {et~al.}(2014{\natexlab{b}}){Fox}, {Wakker}, {Barger},
  {Hernandez}, {Richter}, {Lehner}, {Bland-Hawthorn}, {Charlton}, {Westmeier},
  {Thom}, {Tumlinson}, {Misawa}, {Howk}, {Haffner}, {Ely}, {Rodriguez-Hidalgo},
  \& {Kumari}}]{2014ApJ...787..147F}
{Fox}, A.~J., {Wakker}, B.~P., {Barger}, K.~A., {et~al.} 2014{\natexlab{b}},
  \apj, 787, 147

\bibitem[{{Fynbo} {et~al.}(2006){Fynbo}, {Starling}, {Ledoux}, {Wiersema},
  {Th{\"o}ne}, {Sollerman}, {Jakobsson}, {Hjorth}, {Watson}, {Vreeswijk},
  {M{\o}ller}, {Rol}, {Gorosabel}, {N{\"a}r{\"a}nen}, {Wijers},
  {Bj{\"o}rnsson}, {Castro Cer{\'o}n}, {Curran}, {Hartmann}, {Holland},
  {Jensen}, {Levan}, {Limousin}, {Kouveliotou}, {Nelemans}, {Pedersen},
  {Priddey}, \& {Tanvir}}]{2006A&A...451L..47F}
{Fynbo}, J.~P.~U., {Starling}, R.~L.~C., {Ledoux}, C., {et~al.} 2006, \aap,
  451, L47

\bibitem[{{Garc{\'{\i}}a-Rojas} {et~al.}(2016){Garc{\'{\i}}a-Rojas}, {Corradi},
  {Monteiro}, {Jones}, {Rodr{\'{\i}}guez-Gil}, \&
  {Cabrera-Lavers}}]{2016ApJ...824L..27G}
{Garc{\'{\i}}a-Rojas}, J., {Corradi}, R.~L.~M., {Monteiro}, H., {et~al.} 2016,
  \apjl, 824, L27

\bibitem[{{Hawkins}(1955)}]{Hawkins:1955fv}
{Hawkins}, W.~B. 1955, Physical Review, 98, 478

\bibitem[{{Howk} {et~al.}(2005){Howk}, {Wolfe}, \&
  {Prochaska}}]{2005ApJ...622L..81H}
{Howk}, J.~C., {Wolfe}, A.~M., \& {Prochaska}, J.~X. 2005, \apjl, 622, L81

\bibitem[{{Kama} {et~al.}(2016){Kama}, {Bruderer}, {Carney}, {Hogerheijde},
  {van Dishoeck}, {Fedele}, {Baryshev}, {Boland}, {G{\"u}sten}, {Aikutalp},
  {Choi}, {Endo}, {Frieswijk}, {Karska}, {Klaassen}, {Koumpia}, {Kristensen},
  {Leurini}, {Nagy}, {Perez Beaupuits}, {Risacher}, {van der Marel}, {van
  Kempen}, {van Weeren}, {Wyrowski}, \& {Y{\i}ld{\i}z}}]{2016arXiv160101449K}
{Kama}, M., {Bruderer}, S., {Carney}, M., {et~al.} 2016, ArXiv e-prints,
  arXiv:1601.01449

\bibitem[{Kastler(1950)}]{KASTLER-1950-234250}
Kastler, A. 1950, J. Phys. Radium, 11, 255

\bibitem[{{Kisielius} {et~al.}(2014){Kisielius}, {Kulkarni}, {Ferland},
  {Bogdanovich}, \& {Lykins}}]{2014ApJ...780...76K}
{Kisielius}, R., {Kulkarni}, V.~P., {Ferland}, G.~J., {Bogdanovich}, P., \&
  {Lykins}, M.~L. 2014, \apj, 780, 76

\bibitem[{{Kobulnicky} {et~al.}(1999){Kobulnicky}, {Kennicutt}, \&
  {Pizagno}}]{1999ApJ...514..544K}
{Kobulnicky}, H.~A., {Kennicutt}, Jr., R.~C., \& {Pizagno}, J.~L. 1999, \apj,
  514, 544

\bibitem[{{Landi Degl'Innocenti}(1984)}]{Landi-DeglInnocenti:1984kl}
{Landi Degl'Innocenti}, E. 1984, \solphys, 91, 1

\bibitem[{{Landi Degl'Innocenti} \& {Landolfi}(2004)}]{landi2004}
{Landi Degl'Innocenti}, E., \& {Landolfi}, M., eds. 2004, Astrophysics and
  Space Science Library, Vol. 307, {Polarization in Spectral Lines}

\bibitem[{{Landolfi} \& {Landi Degl'Innocenti}(1986)}]{Landolfi:1986lh}
{Landolfi}, M., \& {Landi Degl'Innocenti}, E. 1986, \aap, 167, 200

\bibitem[{{Lehner} {et~al.}(2004){Lehner}, {Wakker}, \&
  {Savage}}]{2004ApJ...615..767L}
{Lehner}, N., {Wakker}, B.~P., \& {Savage}, B.~D. 2004, \apj, 615, 767

\bibitem[{{Leitherer} {et~al.}(2016){Leitherer}, {Hernandez}, {Lee}, \&
  {Oey}}]{2016ApJ...823...64L}
{Leitherer}, C., {Hernandez}, S., {Lee}, J.~C., \& {Oey}, M.~S. 2016, \apj,
  823, 64

\bibitem[{{Maaskant} {et~al.}(2013){Maaskant}, {Honda}, {Waters}, {Tielens},
  {Dominik}, {Min}, {Verhoeff}, {Meeus}, \& {van den
  Ancker}}]{2013A&A...555A..64M}
{Maaskant}, K.~M., {Honda}, M., {Waters}, L.~B.~F.~M., {et~al.} 2013, \aap,
  555, A64

\bibitem[{{Mathis}(1982)}]{1982ApJ...261..195M}
{Mathis}, J.~S. 1982, \apj, 261, 195

\bibitem[{{Mathis}(1985)}]{1985ApJ...291..247M}
---. 1985, \apj, 291, 247

\bibitem[{{McWilliam}(1997)}]{1997ARA&A..35..503M}
{McWilliam}, A. 1997, \araa, 35, 503

\bibitem[{{Meiring} {et~al.}(2011){Meiring}, {Tripp}, {Prochaska}, {Tumlinson},
  {Werk}, {Jenkins}, {Thom}, {O'Meara}, \& {Sembach}}]{2011ApJ...732...35M}
{Meiring}, J.~D., {Tripp}, T.~M., {Prochaska}, J.~X., {et~al.} 2011, \apj, 732,
  35

\bibitem[{{Miller} \& {Bregman}(2015)}]{2015ApJ...800...14M}
{Miller}, M.~J., \& {Bregman}, J.~N. 2015, \apj, 800, 14

\bibitem[{{Morton}(2003)}]{Morton2003}
{Morton}, D.~C. 2003, \apjs, 149, 205

\bibitem[{{Neeleman} {et~al.}(2015){Neeleman}, {Prochaska}, \&
  {Wolfe}}]{2015ApJ...800....7N}
{Neeleman}, M., {Prochaska}, J.~X., \& {Wolfe}, A.~M. 2015, \apj, 800, 7

\bibitem[{{Peimbert} \& {Peimbert}(2006)}]{2006IAUS..234..227P}
{Peimbert}, M., \& {Peimbert}, A. 2006, in IAU Symposium, Vol. 234, Planetary
  Nebulae in our Galaxy and Beyond, ed. M.~J. {Barlow} \& R.~H. {M{\'e}ndez},
  227--234

\bibitem[{{Pickering} {et~al.}(2001){Pickering}, {Thorne}, \&
  {Perez}}]{2001ApJS..132..403P}
{Pickering}, J.~C., {Thorne}, A.~P., \& {Perez}, R. 2001, \apjs, 132, 403

\bibitem[{{Prochaska} \& {Wolfe}(2002)}]{2002ApJ...566...68P}
{Prochaska}, J.~X., \& {Wolfe}, A.~M. 2002, \apj, 566, 68

\bibitem[{{Prochaska} {et~al.}(2001){Prochaska}, {Wolfe}, {Tytler}, {Burles},
  {Cooke}, {Gawiser}, {Kirkman}, {O'Meara}, \&
  {Storrie-Lombardi}}]{2001ApJS..137...21P}
{Prochaska}, J.~X., {Wolfe}, A.~M., {Tytler}, D., {et~al.} 2001, \apjs, 137, 21

\bibitem[{{Richter} {et~al.}(2013){Richter}, {Fox}, {Wakker}, {Lehner}, {Howk},
  {Bland-Hawthorn}, {Ben Bekhti}, \& {Fechner}}]{2013ApJ...772..111R}
{Richter}, P., {Fox}, A.~J., {Wakker}, B.~P., {et~al.} 2013, \apj, 772, 111

\bibitem[{{Richter} {et~al.}(2001){Richter}, {Sembach}, {Wakker}, {Savage},
  {Tripp}, {Murphy}, {Kalberla}, \& {Jenkins}}]{Richter2001}
{Richter}, P., {Sembach}, K.~R., {Wakker}, B.~P., {et~al.} 2001, \apj, 559, 318

\bibitem[{{Richter} {et~al.}(2016){Richter}, {Wakker}, {Fechner}, {Herenz},
  {Tepper-Garc{\'{\i}}a}, \& {Fox}}]{2016A&A...590A..68R}
{Richter}, P., {Wakker}, B.~P., {Fechner}, C., {et~al.} 2016, \aap, 590, A68

\bibitem[{{Savage} \& {Sembach}(1996)}]{Savage1996}
{Savage}, B.~D., \& {Sembach}, K.~R. 1996, \araa, 34, 279

\bibitem[{{Stutzki} {et~al.}(1988){Stutzki}, {Stacey}, {Genzel}, {Harris},
  {Jaffe}, \& {Lugten}}]{1988ApJ...332..379S}
{Stutzki}, J., {Stacey}, G.~J., {Genzel}, R., {et~al.} 1988, \apj, 332, 379

\bibitem[{{Varshalovich}(1968)}]{Varshalovich:1968qc}
{Varshalovich}, D.~A. 1968, Astrofizika, 4, 519

\bibitem[{{Varshalovich}(1971)}]{Varshalovich:1971mw}
---. 1971, Soviet Physics Uspekhi, 13, 429

\bibitem[{{Vilchez} {et~al.}(1988){Vilchez}, {Pagel}, {Diaz}, {Terlevich}, \&
  {Edmunds}}]{1988MNRAS.235..633V}
{Vilchez}, J.~M., {Pagel}, B.~E.~J., {Diaz}, A.~I., {Terlevich}, E., \&
  {Edmunds}, M.~G. 1988, \mnras, 235, 633

\bibitem[{{Welsh} \& {Lallement}(2012{\natexlab{a}})}]{2012PASP..124..566W}
{Welsh}, B.~Y., \& {Lallement}, R. 2012{\natexlab{a}}, \pasp, 124, 566

\bibitem[{{Welsh} \& {Lallement}(2012{\natexlab{b}})}]{Welsh2012}
---. 2012{\natexlab{b}}, \pasp, 124, 566

\bibitem[{{Yan} \& {Lazarian}(2006)}]{YLfine}
{Yan}, H., \& {Lazarian}, A. 2006, \apj, 653, 1292

\bibitem[{{Yan} \& {Lazarian}(2007)}]{YLhyf}
---. 2007, \apj, 657, 618

\bibitem[{{Yan} \& {Lazarian}(2008)}]{YLhanle}
---. 2008, \apj, 677, 1401

\bibitem[{{Yan} \& {Lazarian}(2012)}]{2012JQSRT.113.1409Y}
---. 2012, \jqsrt, 113, 1409

\bibitem[{{Yan} \& {Lazarian}(2013)}]{2013arXiv1302.3264Y}
---. 2013, ArXiv e-prints, arXiv:1302.3264

\bibitem[{{Zare} \& {Harter}(1989)}]{1989PhT....42l..68Z}
{Zare}, R.~N., \& {Harter}, W.~G. 1989, Physics Today, 42, 68

\bibitem[{{Zech} {et~al.}(2008){Zech}, {Lehner}, {Howk}, {Van Dyke Dixon}, \&
  {Brown}}]{2008ApJ...679..460Z}
{Zech}, W.~F., {Lehner}, N., {Howk}, J.~C., {Van Dyke Dixon}, W., \& {Brown},
  T.~M. 2008, \apj, 679, 460

\bibitem[{{Zhang} {et~al.}(2015){Zhang}, {Yan}, \& {Dong}}]{ZYD15}
{Zhang}, H., {Yan}, H., \& {Dong}, L. 2015, \apj, 804, 142

\end{thebibliography}
\end{document}